\documentclass[12pt]{article}
\usepackage{graphicx}

\input{epsf}

\def\rf#1{(\ref{eq:#1})}
\def\lab#1{\label{eq:#1}}

\def\br{\begin{eqnarray}}
\def\er{\end{eqnarray}}
\def\be{\begin{equation}}
\def\ee{\end{equation}}
\def\({\left(}
\def\){\right)}
\def\pa{\partial}
\def\rlx{\relax\leavevmode}
\def\IR{\rlx\hbox{\rm I\kern-.18em R}}
\def\vp{\varphi}
\def\ve{\varepsilon}
\def\psib{{\bar \psi}}

\newcommand{\sbr}[2]{\left\lbrack\,{#1}\, ,\,{#2}\,\right\rbrack}

\def\IZ{\rlx\hbox{\sf Z\kern-.4em Z}}
\def\IR{\rlx\hbox{\rm I\kern-.18em R}}
\def\IC{\rlx\hbox{\,$\inbar\kern-.3em{\rm C}$}}
\def\one{\hbox{{1}\kern-.25em\hbox{l}}}

\begin{document}

\begin{titlepage}
\vspace*{-1cm}

\vskip 2cm

\vspace{.2in}
\begin{center}
{\large\bf The concept of quasi-integrability for  modified non-linear Schr\"odinger models}
\end{center}

\vspace{.5cm}

\begin{center}
L. A. Ferreira~$^{\star}$, G. Luchini~$^{\star}$ and Wojtek J. Zakrzewski~$^{\dagger}$

\vspace{.3 in}
\small

\par \vskip .2in \noindent
$^{(\star)}$Instituto de F\'\i sica de S\~ao Carlos; IFSC/USP;\\
Universidade de S\~ao Paulo  \\ 
Caixa Postal 369, CEP 13560-970, S\~ao Carlos-SP, Brazil\\
email: laf@ifsc.usp.br\\
email: gabriel.luchini@gmail.com

\par \vskip .2in \noindent
$^{(\dagger)}$~Department of Mathematical Sciences,\\
 University of Durham, Durham DH1 3LE, U.K.\\
email: W.J.Zakrzewski@durham.ac.uk

\normalsize
\end{center}


\begin{abstract}
We consider modifications of the nonlinear Schr\"odinger model (NLS) to look at the recently introduced concept of quasi-integrability.
We show that such models possess an infinite number of quasi-conserved charges which present intriguing properties in relation to very 
specific space-time parity transformations. For the case of two-soliton solutions where the fields are eigenstates of this parity, those charges are asymptotically conserved in the scattering process of the solitons. Even though the charges vary in time their values in the far past and the far future are the same. Such results are obtained through analytical and numerical methods, and employ adaptations of algebraic techniques used in integrable field theories. Our findings may have important consequences on the applications of these models in several areas of non-linear science. We make a detailed numerical study of the modified NLS potential of the form $V\sim \(\mid \psi\mid^2\)^{2+\ve}$, with $\ve$ being a perturbation parameter. We perform numerical simulations of the scattering of solitons for this model and find a good agreement with the results predicted by the analytical considerations. Our paper shows that the  quasi-integrability concepts recently proposed in the context of modifications of the sine-Gordon model remain valid for perturbations of the NLS model.

\end{abstract} 
\end{titlepage}

\section{Introduction}
\label{sec:intro}
\setcounter{equation}{0}

The concept of a soliton, introduced half a century ago by Zabusky and Kruskal \cite{zabusky}, was based by the seminal work of Fermi, Pasta and Ulam \cite{fermi}. Solitons are special solutions of non-linear evolution equations that propagate without changing their shapes and without dissipating their energies. The solitons interact among themselves but the special property they possess is, that after a long time after their scattering, the only effect of it is a shift in their position (the so-called time delay or time advance) w.r.t. the values they would have had had the scattering have not taken place. There is no emission of radiation during their interaction and, well after  the scattering process, their  shape  and other physical properties like energy,  are preserved.  For the case of $(1+1)$ dimensional theories this behaviour of solitons has been understood in the context of integrable field theories. Indeed, it has been observed that (practically) all models possessing soliton solutions admit a representation of their equations of motion in terms of the so-called Lax-Zakharov-Shabat (LZS) equation or zero curvature condition \cite{lax}, where the Lax potential or connection lives in an infinite dimensional Kac-Moody algebra. The LZS equation has led to the development of many exact and non-perturbative methods to study such $(1+1)$ dimensional theories, including the construction of exact solutions and of an infinite number of conservation laws \cite{faddeev,babelon}. In the context of such a {\em  soliton theory}, the above mentioned  striking properties of solitons  can be credited to the constraints on their dynamics imposed by the infinite number of exactly conserved charges coming from the LZS equation. 

Of course, the class of $(1+1)$ dimensional  integrable field theories, admitting the LZS equation, is not very large. Indeed, most of the two dimensional physical non-linear phenomena are described by theories that do not belong to that class. In many cases however, integrable models can be used as approximations to more realistic theories, and many interesting developments have been done in that direction. In fact, the literature on applications of perturbations around integrable theories is quite vast and diverse, and we shall not attempt to quote the many interesting and important results obtained. We shall concentrate, however, on the fact that many non-integrable theories possess solutions that behave much like solitons despite the lack of a large number of conservation laws. 

In this context, two of us \cite{us} have recently  looked at a class of models which generalizes the integrable sine-Gordon model  and used it to introduce the concept of {\em quasi-integrability}. According to \cite{us} a $(1+1)$ dimensional field theory is quasi-integrable if although it {\em does not} admit a representation of its equations of motion in terms of the LZS equation, it does possess soliton like solutions that which, when they undergo a scattering process, preserve their basic physical properties like mass, topological charges, etc.  It is also required that the theory should possess an infinity number of quasi-conservation laws with the property that the  corresponding charges are conserved when evaluated on the one-soliton solutions, and are asymptotically conserved in the scattering of these solitons. In other words, during the scattering of the solitons the charges do vary in time, but they return to their original values (in the far past), when the solitons are well separated after the collision (in the far future). Essentially, the theory possesses  anomalous conservation laws of the form
\be
\frac{d\,Q^{(n)}}{d\,t}= \beta_n \(t\) 
\lab{introquasicharges}
\ee
with the label $n$  being an integer.  For instance, in the scattering of two solitons one has 
\be
Q^{(n)}\(t\rightarrow \infty\)- Q^{(n)}\(t\rightarrow -\infty\)=\int_{-\infty}^{\infty} dt\, \beta_n = 0 .
\lab{introquasiconserv}
\ee
For breather like solutions it was shown in \cite{us} that in many special cases the vanishing in \rf{introquasiconserv} occurs when the time integral is performed over a period $T$ determined by the breather, {\it i.e.} the charges are periodic in time, $Q^{(n)}\(t+T\)= Q^{(n)}\(t\)$.  

Reference  \cite{us} has also considered  particular modifications \cite{bazeia} of the sine-Gordon model that admit topological soliton like solutions (kinks), and a representation of their equations of motion in terms of an anomalous (non-zero) LZS equation.  Adapting  techniques of integrable field theories to this anomalous equation, an infinite set of quasi-conserved charges was constructed. Employing both analytical and numerical techniques the scattering of solitons was studied and it was verified that for some special solutions the asymptotic conservation of charges does take place. The key observation of \cite{us} was based on
the fact that  the two-soliton solutions  satisfying \rf{introquasiconserv} had the property that their fields were eigenstates of a very special space-time parity transformation 
\be
P:\qquad \({\tilde x},{\tilde t}\)\rightarrow \(-{\tilde x},-{\tilde t}\) \qquad\qquad {\rm with} \qquad \quad{\tilde x}= x-x_{\Delta} \qquad \quad{\tilde t}=t-t_{\Delta}.
\lab{introparitydef}
\ee
where the point $\(x_{\Delta},t_{\Delta}\)$ in space-time, depends upon the parameters of the solution. Since the charges are obtained from some densities, {\it i.e.} $Q^{(n)}=\int_{-\infty}^{\infty}dx\, j_0^{(n)}$, so are the anomalies $\beta_n=\int_{-\infty}^{\infty}dx\,\gamma_n$. Therefore, the vanishing of $\int_{-\infty}^{\infty} dt\,\int_{-\infty}^{\infty}dx\,\gamma_n$, follows from the properties of $\gamma_n$ under \rf{introparitydef}. Note that the solutions for which the fields are eigenstates of the parity \rf{introparitydef} cannot be selected by  choosing appropriate initial boundary conditions. The reason for this is simple:
 the boundary conditions are set at a given initial time and the transformation  \rf{introparitydef}  relates the past and the future of the solutions. In other words, boundary conditions are kinematical statements, and the fact that a field is an eigenstate under  \rf{introparitydef}  is a dynamical statement. For these reasons, the physical mechanism that guarantees that such special solutions have the required parity properties is not clear yet.

The models studied in \cite{us} were perturbed sine-Gordon models; {\it i.e.} Lorentz covariant mo\-dels with topological solitons. Thus it would be interesting to see whether similar phenomena hold in other models, with other symmetries. Hence in this paper we look at the nonlinear Schr\"odinger (NLS) model and its perturbations. Even though this model is also integrable, it differs from the sine-Gordon in the sense that 
 it possesses solitons which are not topological, their dynamics is governed by a first order (in time derivatives) equation and it does not possess any breather like structures. However, this model is probably even more important than the sine-Gordon model in its applications, which are abundant in all areas of nonlinear science. Hence the understanding of quasi-integrability in this context would have very important implications. The modifications of the NLS  model we consider in this paper have equations of motion of the form 
\br
i\, \partial_t\psi&=&-\partial^2_x\psi+\frac{\partial\,V}{\partial \mid\psi\mid^2}\, \psi,
\lab{introeqofmot}
\er
where $\psi$ is a complex scalar field and $V$ is a potential dependent only on the modulus of $\psi$. The unperturbed NLS equation corresponds to $V\sim \mid \psi \mid^4$. We start our analysis of such models by writing the equations of motion \rf{introeqofmot} as an anomalous LZS equation of the form
\be
\partial_t A_x-\partial_x A_t + \sbr{A_x}{A_t}={\cal X},
\lab{introlzs}
\ee
where the connection $A_{\mu}$ is a functional of $\psi$ and its derivatives, and takes values in the $SL(2)$ loop algebra  (Kac-Moody algebra with vanishing central element), and ${\cal X}$ is the anomaly that vanishes when $V$ is the NLS potential.  

We construct the infinite set of quasi-conserved charges by employing the standard techniques of integrable field theories known as Drinfeld-Sokolov  reduction \cite{drinfeld}, or abelianization procedure \cite{oliveabelian,afgzcharges,simple}.  Using these techniques we gauge transform the $A_x$ component of the connection into an infinite dimensional abelian subalgebra  of the loop algebra, generated by $T_3^n\equiv \lambda^n\,T_3$. Even though the anomaly ${\cal X}$ prevents the gauge transformation  to rotate the $A_t$ component into the same abelian subalgebra, the component of the transformed curvature \rf{introlzs} in that subalgebra, leads to an infinite set of  quasi-conservation laws, $\partial^{\mu}j_{\mu}^{(n)}=\gamma_n$, or equivalently leads to \rf{introquasicharges} with $Q^{(n)}=\int_{-\infty}^{\infty}dx\, j_0^{(n)}$ and $\beta_n=\int_{-\infty}^{\infty}dx\,\gamma_n$.

Next we employ a more refined algebraic technique, involving two $\IZ_2$ transformations, to understand the conditions for the vanishing of the integrated anomalies. The first $\IZ_2$ is an order two automorphism of the $SL(2)$ loop algebra and the second is the parity transformation \rf{introparitydef}. For the solutions for which the field $\psi$ transforms under \rf{introparitydef} as
\be
\psi \rightarrow e^{i\,\alpha}\, \psi^* \qquad\qquad\qquad \qquad \mbox{\rm with $\alpha$ constant}
\lab{intropsitransf}
\ee
we show that $\int_{-{\tilde t}_0}^{{\tilde t}_0} dt\,\int_{-{\tilde x}_0}^{{\tilde x}_0} dx\,\gamma_n=0$, where ${\tilde t}_0$ and ${\tilde x}_0$ are any given fixed values of the space-time coordinates ${\tilde t}$ and ${\tilde x}$, respectively, introduced in \rf{introparitydef}. 
This shows that 
\be
Q^{(n)}\(t={\tilde t}_0 +t_{\Delta}\)=Q^{(n)}\(t=-{\tilde t}_0 +t_{\Delta}\)
\lab{intromirror}
\ee
 which is a type of a mirror symmetry for the charges. Therefore, for a two-soliton solution satisfying \rf{intropsitransf},  the asymptotic conservation of the charges \rf{introquasiconserv} follows from such stronger result. 

Such results certainly unravel important structures responsible for the phenomena that we have called quasi-integrability. They involve an anomalous LZS equation, internal and external $\IZ_2$  symmetries, and algebraic techniques borrowed from integrable field theories. However, they rely on the assumption \rf{intropsitransf} which is, as we have argued above, a dynamical statement since it relates the past and the future of the solutions. In order to shed more light on this issue we study the relation between \rf{intropsitransf} and the dynamics defined by 
\rf{introeqofmot}. 

It is easier to work with the modulus and phase of $\psi$, and so we parametrize the fields as $\psi=\sqrt{R}\, e^{i\frac{\vp}{2}}$, with $R$ and $\vp$ being real scalars fields. We split them into their eigen-components under the parity \rf{introparitydef}, as $R=R^{(+)}+R^{(-)}$, and $\vp=\vp^{(+)}+\vp^{(-)}$. The assumption  \rf{intropsitransf} implies that the solution should contain only the components $\(R^{(+)}, \vp^{(-)}\)$, and  nothing of the pair $\(R^{(-)}, \vp^{(+)}\)$. By splitting  the equations of motion \rf{introeqofmot} into their even and odd components under  \rf{introparitydef}, we show that there cannot exist non-trivial solutions carrying only the pair $\(R^{(-)}, \vp^{(+)}\)$. In addition, if the potential $V$ in \rf{introeqofmot} is a deformation of the NLS potential, in the sense that we can expand it as 
\be
V=V_{{\rm NLS}}+ \ve\, V_1+\ve^2\, V_2+\ldots
\ee
 with $\ve$ being a deformation parameter, then we can make even stronger statements.  In such a case we expand the equations of motion and the solutions into power series in $\ve$, as 
 \be
 R^{(\pm)}= R^{(\pm)}_0 + \ve\, R^{(\pm)}_1+ \ve^2\, R^{(\pm)}_2 +\ldots\; ; \qquad\qquad\qquad 
 \vp^{(\pm)}= \vp^{(\pm)}_0 + \ve\, \vp^{(\pm)}_1+ \ve^2\, \vp^{(\pm)}_2 +\ldots
 \ee
 If we select a zero order solution, {\it i.e.} a solution of the NLS equation, satisfying \rf{intropsitransf}, {\it i.e.} carrying only the pair $\(R^{(+)}_0, \vp^{(-)}_0\)$, then the equations for the first order fields, which are obviously linear in them, are such that the pair $\(R^{(+)}_1, \vp^{(-)}_1\)$  satisfies inhomogeneous equations, while the pair $\(R^{(-)}_1, \vp^{(+)}_1\)$, satisfies homogeneous ones. Therefore,  $\(R^{(-)}_1, \vp^{(+)}_1\)=\(0,{\rm const.}\)$, is a solution of the equations of motion, but $\(R^{(+)}_1, \vp^{(-)}_1\)=\(0,{\rm const.}\)$, is not.
 By selecting the first order solution such that the pair $\(R^{(-)}_1, \vp^{(+)}_1\)$ is absent, we see that the same happens in second order, {\it i.e.} that the pair $\(R^{(+)}_2, \vp^{(-)}_2\)$ also satisfies inhomogeneous equations, and the pair $\(R^{(-)}_2, \vp^{(+)}_2\)$
the homogeneous ones. By repeating this procedure, order by order, one can build a perturbative solution which satisfies \rf{intropsitransf}, and so has charges satisfying \rf{intromirror}. Note that the converse could not be done, {\it i.e.} we cannot construct a solution involving only the pair $\(R^{(-)}, \vp^{(+)}\)$. So, the dynamics dictated by \rf{introeqofmot} favours solutions of the type \rf{intropsitransf}. 

Finally we show that the one-bright-soliton and the one-dark-soliton solutions of the NLS equation satisfy the condition \rf{intropsitransf}, and that not all two-bright-soliton solutions satisfy it. However, one can choose the parameters of the general solution so that the corresponding two-bright-soliton solutions  do satisfy \rf{intropsitransf}. This involves a choice of the relative phase between the two one-bright-solitons forming the two-soliton solution. We do not analyze in this paper the two-dark soliton solutions of the NLS equation. Therefore, our perturbative expansion explained above can be used to build a sub-sector of two-bright-soliton solutions of  \rf{introeqofmot} that obeys \rf{intropsitransf} and so has charges satisfying  \rf{intromirror}. This would constitute our quasi-integrable sub-model of \rf{introeqofmot}.

Despite the fact that the equations of motion satisfied by the $n$-order fields $\(R^{(\pm)}_n, \vp^{(\pm)}_n\)$ are linear, the coefficients are highly non-linear in the lower order fields and so, unfortunately, these  equations are not easy to solve. We then use numerical methods to study the properties of our solutions. In addition, such numerical analysis can clarify possible convergence issues of our perturbative expansions. We chose to perform our numerical simulations  for a potential of the form 
\be
V = \frac{\eta}{2+\ve}\, \(\mid \psi\mid^{2}\)^{2+\ve} \qquad\qquad \qquad \qquad \eta<0
\ee

 We performed several simulations using the 4th order Runge Kutta method of simulating the time evolution.
These simulations involved the NLS case with the two bright solitons 
sent towards each other with different values of velocity (including $v=0$) and for various values of the relative phase. We then repeated that for the modified models. We looked at various values of $\epsilon$
and have found that the numerical results were reliable for only a small range of $\epsilon$ around 0. For very small values we saw no difference 
from the results for the NLS model but for $\vert \epsilon\vert \sim 0.1$ or $\sim0.2$ the results of the simulations became less reliable. Hence, 
we are quite confident of our results for $\vert \epsilon\vert<0.1$
and in the numerical section we present the results for $\epsilon=\pm 0.06$.

We also present the results for the first anomaly as seen in our simulations. We find that our results confirm our expectations.

The paper is organized as follows: in section \ref{sec:model} we describe in detail the models to be studied, construct the anomalous LZS equation, the quasi-conserved charges and establish the conditions, that have to be satisfied by the solutions, for the integrated anomalies to vanish. We also give an argument, valid in a space-time of any dimension, for a field theory to possess charges satisfying symmetries of the type given in \rf{intromirror}. In section \ref{sec:deformnls} we discuss how the  dynamics of the model favours solutions satisfying \rf{intropsitransf}. We also discuss further the relation between the dynamics and parity for the case when the potential is a deformation  of the NLS potential. In section \ref{sec:nlssoliton} we discuss the parity properties of the one and two-soliton solutions of the NLS theory and show how to select those that satisfy \rf{intropsitransf}. We then present, in section \ref{sec:nsupport},   the results of our numerical simulations which support the analytical results discussed in the previous sections. The conclusions are given in section  \ref{sec:conclusions}, and in the appendix \ref{sec:rotatedpot} we present the details of the calculation used in section \ref{sec:model}, and in appendix \ref{sec:hirota} we use  the Hirota method to construct one and two-bright-soliton solutions of the NLS theory.

\section{The model}
\label{sec:model}
\setcounter{equation}{0}

We consider a non-relativistic complex scalar field in ($1+1$) dimensions with
the  Lagrangian given by  
\be
{\cal L}= \frac{i}{2}\(\psib\,\partial_t \psi-\psi\,\partial_t\psib\) 
- \partial_x\psib\,\partial_x\psi-V\(\mid\psi\mid^2\),
\lab{lagrangian}
\ee
where $\psib$ is the complex conjugate of $\psi$. The equations of motion are 
\br
i\, \partial_t\psi&=&-\partial^2_x\psi+\frac{\partial\,V}{\partial \mid\psi\mid^2}\, \psi
\lab{eqofmot}
\er
together with its complex conjugate. The corresponding Hamiltonian is given by
\be
{\cal H}=\mid\partial_x\psi\mid^2+V\(\mid\psi\mid^2\).
\ee
We shall consider solutions of \rf{eqofmot} satisfying the following boundary conditions
\be
\mid\psi\mid_{x=-\infty}=\mid\psi\mid_{x=\infty}\; ;\qquad\qquad \qquad \qquad
\partial_x\psi \rightarrow 0 \qquad {\rm for} \qquad x\rightarrow \pm \infty. 
\lab{bc}
\ee
It is easy to check that  the energy $E$, momentum $P$ and normalization $N$ of the solutions of the equations of motion\rf{eqofmot} satisfying \rf{bc}, as defined below, are conserved in time.
\br
E&=&\int_{-\infty}^{\infty}dx\,\(\mid\partial_x\psi\mid^2+V\),\\
P&=&i\int_{-\infty}^{\infty}dx\, \(\psib\,\partial_x\psi-\psi\,\partial_x\psib\),\\
N&=&\int_{-\infty}^{\infty}dx\, \mid\psi\mid^2.
\lab{conserved}
\er
In fact, these conserved quantities correspond to the Noether charges of the model. The energy $E$ is connected with the invariance of  \rf{lagrangian} under time translations, the momentum $P$ under the space translations, and $N$ is related to the following internal symmetry of the Lagrangian \rf{lagrangian}
\be
\psi \rightarrow e^{i\,\alpha}\, \psi\qquad\qquad\qquad \qquad \alpha \equiv {\rm const.}
\lab{phasetransf}
\ee

The integrable Non-Linear Schr\"odinger theory (NLS) corresponds to the potential 
\be
V_{\rm NLS}= \eta\, \mid\psi_0\mid^4,
\lab{nlspot}
\ee
which leads to the NLS equation 
\be
i\, \partial_t\psi_0=-\partial^2_x\psi_0+2\,\eta\, \mid\psi_0\mid^2\, \psi_0.
\lab{nlseq}
\ee
The sign of the parameter $\eta$ plays an important role in the properties of the solutions. Indeed, for $\eta<0$ we have the so-called bright soliton solutions given by
\be
\psi_0= \frac{\mid \rho\mid}{\sqrt{\mid \eta\mid}}\, 
\frac{e^{i\left[\(\rho^2-\frac{v^2}{4}\)\,t+ \frac{v}{2}\, x\right]}}{\cosh\left[\rho\,\(x-v\,t-x_0\)\right]}
\lab{brightsol}
\ee
with $\rho$, $v$ and $x_0$ being real parameters of the solution.  For $\eta>0$ we have the dark soliton solution given by
 \be
 \psi_0=\frac{\mid \rho\mid}{\sqrt{\eta}}\,\tanh\left[\rho\(x-v\,t-x_0\)\right]\, e^{i\,\left[\frac{v}{2}\,x-\(2\,\rho^2+\frac{v^2}{4}\)\,t\right]}.
 \lab{darkonesol}
 \ee
Note, that the solutions are defined up to an overall constant phase due to the symmetry \rf{phasetransf}. 
 
The equation \rf {eqofmot} admits an anomalous zero curvature representation (Lax-Zakharov-Shabat equation) with the connection given by 
\br
A_x&=& -i\, T_3^1+ {\bar \gamma}\,\psib\,T_{+}^0+\gamma\, \psi\, T_{-}^0,
\lab{lax}\\
A_t&=& i\, T_3^2+i\,\frac{\delta \, V}{\delta\, \mid\psi\mid^2}\, T_3^0-
\({\bar \gamma}\,\psib\, T_{+}^1+\gamma\,\psi\, T_{-}^1\) 
-i \({\bar \gamma}\,\partial_x\, \psib\, T_{+}^0-\gamma\,\partial_x\psi\, T_{-}^0\),
\nonumber
\er
where the generators $T_i^n$, $i=3,+,-$, and $n$ integer, satisfy the so-called $SL(2)$ loop algebra commutation relations 
\br
\sbr{T_3^m}{T_{\pm}^n}=\pm T_{\pm}^{m+n} \; ; \qquad\qquad\qquad 
\sbr{T_{+}^m}{T_{-}^n}=2\, T_3^{m+n},
\lab{loopalg}
\er
which can be realized in terms of the finite $SL(2)$ algebra generators as $T_i^n\equiv \lambda^n\, T_i$, with $\lambda$ an arbitrary complex parameter. The curvature of the connection \rf{lax} is given by
\br
\partial_t A_x-\partial_x A_t + \sbr{A_x}{A_t}&=& X\, T_3^0+i\,{\bar \gamma}\,\left[-i\,\pa_t \psib+\pa_x^2\psib-\psib \,\frac{\delta \, V}{\delta\, \mid\psi\mid^2}\right]\, T_{+}^0\nonumber\\
&-&i\,\gamma\,\left[i\,\pa_t \psi+\pa_x^2\psi-\psi \,\frac{\delta \, V}{\delta\, \mid\psi\mid^2}\right]\, T_{-}^0
\lab{curvature}
\er
with
\be
X\equiv -i\,\partial_x\(\frac{\delta\, V}{\delta  \mid\psi\mid^2}
-2\, \gamma\,{\bar \gamma}\,\mid\psi\mid^2\)
\lab{xdef}
\ee
In consequence, when the equations of motion \rf {eqofmot} are imposed, the terms, on the r.h.s. of \rf{curvature}, proportional to $T_{+}^0$ and $T_{-}^0$ vanish. Note also that, by taking 
\be
\eta\equiv \gamma\,{\bar \gamma},
\lab{gammadef}
\ee
the anomaly $X$, given in \rf{xdef}, vanishes for the NLS potential \rf{nlspot}. In fact, this vanishing of the  curvature \rf{curvature}  for the NLS equation  makes this
 classical field theory integrable.

 In this paper we will  consider the generalisation of this theory  ({\it i.e.} the deformations of the NLS potential) which make the resultant theory 
nonintegrable,  {\it i.e.} those for which the anomaly \rf{xdef} does not vanish. However, as we will show, the corresponding theories exhibit  properties very similar to the integrable ones, like the solitons preserving their shapes after the scattering {\it etc}. In addition, we will show, using the algebraic technioques borrowed from integrable field theories, that the anomalous Lax-Zakharov-Shabat equation \rf{curvature} leads to an infinite number of quasi-conservation laws. And, we will find that, under some special circumstances, the corresponding charges are conserved asymptotically in the scattering of  soliton type solutions of these (non-integrable) theories. 

 In order, to employ the algebraic techniques mentioned above it is more convenient to work with a new basis of the $SL(2)$ loop algebra and a new parameterization of the fields.  In our work we will use the modulus  $R$ of $\psi^2$ and its phase $\varphi$, defined as
\be
\psi=\sqrt{R}\, e^{i\frac{\vp}{2}}. 
\lab{rphidef}
\ee

In addition, we will parameterize the complex parameters $\gamma$ and ${\bar \gamma}$, appearing in the connection \rf{lax}, as 
\be
\gamma= i\,\sqrt{\mid \eta\mid}\,e^{i\phi},\qquad\qquad 
{\bar \gamma}= -i\,\sigma\, \sqrt{\mid \eta\mid}\,e^{-i\phi},\qquad\qquad
\gamma\, {\bar \gamma}=\eta, \qquad \qquad \sigma={\rm sign} \,\eta.
\lab{gammapar}
\ee
The new basis of the $SL(2)$ loop algebraic is then defined as  
\be
b_n=T_3^n, \qquad \quad 
F_1^n=\frac{1}{2}\(\sigma \, T_{+}^n-T_{-}^n\),\qquad\quad 
F_2^n=\frac{1}{2}\(\sigma \, T_{+}^n+T_{-}^n\),
\ee
which satisfy
\br
\sbr{b_m}{b_n}=0\; ; \quad\;
\sbr{b_n}{F_1^m}= F_2^{n+m}\; ; \quad\;
\sbr{b_n}{F_2^m}= F_1^{n+m}\; ; \quad\;
\sbr{F_1^n}{F_2^m}= \sigma\, b_{n+m}.
\lab{newbasis}
\er

As usual we perform the gauge transformation 
\be
A_{\mu} \rightarrow  {\tilde A}_{\mu}\equiv {\tilde g}\, A_{\mu}\, {\tilde g}^{-1}+\partial_{\mu}{\tilde g}\, {\tilde g}^{-1}\; ; 
\qquad\qquad {\rm with} \qquad\qquad {\tilde g}=e^{i\(\frac{\vp}{2}+\phi\)\,b_0}
\ee
and find that the connection \rf{lax} has now become
\br
{\tilde A}_x&=& -i\, b_1+ \frac{i}{2}\pa_x\vp \,b_0
-2\,i\,\sqrt{\mid \eta\mid}\,\sqrt{R}\,F_1^0,
\lab{atildedef}\\
{\tilde A}_t&=& i\, b_2+ \frac{i}{2}\pa_t\vp \,b_0+i\,\frac{\delta \, V}{\delta\, R}\, b_0
+2\,i\,\sqrt{\mid \eta\mid}\,\sqrt{R}\,F_1^1
\nonumber\\
&+&\sqrt{\mid \eta\mid}\,\sqrt{R}\, \left[-\frac{\partial_x\, R}{R}\, F_2^0 +i\,\pa_x\vp \, F_1^0\right].
\nonumber
\er

For the fields which satisfy the equations of motion \rf{eqofmot}  the curvature becomes 
\be
{\tilde F}_{tx}=\partial_t {\tilde A}_x-\partial_x {\tilde A}_t + \sbr{{\tilde A}_x}{{\tilde A}_t}= X\, b_0 \; ; 
\qquad\quad {\rm with} \qquad\quad X\equiv -i\,\partial_x\(\frac{\delta\, V}{\delta  R}
-2\, \eta \,R\).
\lab{xdef2}
\ee

To go further we carry out the usual abelianization technique of the integrable field theories \cite{drinfeld,oliveabelian,afgzcharges,simple}; {\it i.e.} we  perform a further gauge transformation
\be
{\tilde A}_{\mu}\rightarrow a_{\mu}=g\,{\tilde A}_{\mu}\,g^{-1}+\pa_{\mu}g\,g^{-1}
\lab{gaugetransfatilde}
\ee
with
\be
g=e^{\sum_{n=1}^{\infty} {\cal F}^{(-n)}} \; ; \qquad \qquad {\rm where} \qquad\qquad 
{\cal F}^{(-n)}\equiv \zeta_1^{(-n)}\, F_1^{-n}+\zeta_2^{(-n)}\, F_2^{-n}.
\lab{abelg}
\ee
The parameters $\zeta_i^{(-n)}$ are chosen, as we will explain below, so that the $a_x$ component of the transformed connection lies in the infinite abelian subalgebra spanned  by the generators $b_n$.

 An important role in our construction is played by the grading operator $d$ defined as
\be
d\equiv \lambda\,\frac{d\, }{d\lambda}\;, \qquad\qquad \qquad
\sbr{d}{b_n}= n\, b_n,\qquad\qquad \qquad \sbr{d}{F_i^n}=n\, F_i^n.
\lab{gradingdef}
\ee 

The ${\tilde A}_x$ component of the connection \rf{atildedef} has generators of grade $0$ and $1$. Since the group element \rf{abelg} is an exponentiation of negative grade  generators, the $a_x$ component of the transformed connection has generators of grades ranging from $1$ to $-\infty$. Splitting the transformed potential \rf{gaugetransfatilde} into its eigen-subspaces under the grading operator 
\rf{gradingdef}, {\it i.e.} $a_x=\sum_{n=1}^{\infty} a_x^{(n)}$, we find that
\br
a_x^{(1)}&=&-i\,b_1,
\nonumber\\
a_x^{(0)}&=&i\,\sbr{b_1}{{\cal F}^{(-1)}}+ {\tilde A}_{x}^{(0)}
\nonumber\\
a_x^{(-1)}&=&i\,\sbr{b_1}{{\cal F}^{(-2)}} +\sbr{{\cal F}^{(-1)}}{{\tilde A}_{x}^{(0)}}
-\frac{i}{2!}\,\sbr{{\cal F}^{(-1)}}{\sbr{{\cal F}^{(-1)}}{b_1}}
\nonumber\\
&+& \pa_x {\cal F}^{(-1)},
\lab{axn}\\
a_x^{(-2)}&=&i\,\sbr{b_1}{{\cal F}^{(-3)}} +\sbr{{\cal F}^{(-2)}}{{\tilde A}_{x}^{(0)}}
-\frac{i}{2!}\,\sbr{{\cal F}^{(-2)}}{\sbr{{\cal F}^{(-1)}}{b_1}}
\nonumber\\
&-&\frac{i}{2!}\,\sbr{{\cal F}^{(-1)}}{\sbr{{\cal F}^{(-2)}}{b_1}}
+\frac{1}{2!}\,\sbr{{\cal F}^{(-1)}}{\sbr{{\cal F}^{(-1)}}{{\tilde A}_{x}^{(0)}}}
\nonumber\\
&-&\frac{i}{3!}\,\sbr{{\cal F}^{(-1)}}{\sbr{{\cal F}^{(-1)}}{\sbr{{\cal F}^{(-1)}}{b_1}}}
\nonumber\\
&+& \pa_x {\cal F}^{(-2)}+\frac{1}{2!}\,\sbr{{\cal F}^{(-1)}}{\pa_x {\cal F}^{(-1)}},
\nonumber\\
&\vdots&
\nonumber
\er
where we have denoted ${\tilde A}_{x}^{(0)}=\frac{i}{2}\pa_x\vp \,b_0
-2\,i\,\sqrt{\mid \eta\mid}\,\sqrt{R}\,F_1^0$ (see \rf{gaugetransfatilde}). 

An important ingredient of this construction is the observation that the generator $b_1$ is a semi simple element  (in fact any $b_n$ is) in the sense that it splits the $SL(2)$ loop algebra ${\cal G}$ into the kernel and image of its adjoint action, {\it  i.e.}
\be
 {\cal G}= {\rm Ker} + {\rm Im}\; ; \qquad\qquad {\rm with} \qquad\qquad \sbr{b_1}{{\rm Ker}}=0\; ; \qquad\qquad {\rm Im}=\sbr{b_1}{{\cal G}}.
 \ee
The Ker and Im subspaces do not have common elements, {\it  i.e.}  any element of ${\cal G}$  commuting with $b_1$ cannot be written as a commutator of $b_1$ with some other element of ${\cal G}$. One notes from \rf{newbasis} that $b_n$ constitute a basis of Ker, and $F_i^n$, $i=1,2$, a basis of Im. In addition, one notes from \rf{axn} that the first time that ${\cal F}^{(-n)}$ appears in the expansion of $a_x$, is in the component $a_x^{-n+1}$ of grade $-n+1$, and it appears in the form $\sbr{b_1}{{\cal F}^{(-n)}}$. Therefore, one can choose the parameters in ${\cal F}^{(-n)}$ so that they cancel the image  component of $a_x^{-n+1}$. This can be done recursively starting at the component of grade $0$ and working downwards. It is then clear that the gauge transformation \rf{abelg} can rotate the $a_x$ component of the connection into the abelian subalgebra generated by the $b_n$'s, {\it  i.e.} 
\be
a_x=-i\, b_1+\sum_{n=0}^{\infty} a_x^{(3,-n)}\, b_{-n}.
\lab{axcomp}
\ee

Note from \rf{atildedef} that ${\tilde A}_x$ depends on the real fields $R$ and $\partial_x\varphi$. Thus, the components $a_x^{(3,n)}$ are polynomials in these fields and their $x$-derivatives, and they do not depend on the potential $V$. In consequence, the $a_x$ component of the connection is the same for any choice of the potential. In the appendix \ref{sec:rotatedpot} we give explicit expressions for the first few components of $a_x$.

On the other hand the ${\tilde A}_t$ component of the connection \rf{atildedef}  depends on the choice of the potential $V$. In fact, for the case of the  NLS potential \rf{nlspot} we note that the gauge transformation \rf{gaugetransfatilde}, with the group element \rf{abelg} fixed as above, does rotate $a_t$ into an abelian subalgebra generated by the $b_n$'s, when the equations of motion \rf{nlseq} are satisfied.   For other choices of potentials $V$ this does  not take place even when  the equations of motion \rf{eqofmot} are imposed. Thus, we find that 
\be
a_t=i\,b_2+\sum_{n=0}^{\infty} \left[ a_t^{(3,-n)}\, b_{-n}+a_t^{(1,-n)}\, F_1^{-n}+a_t^{(2,-n)}\, F_2^{-n}\right]. 
\ee

Next we note that $a_t$ does not have the grade $1$ component due to the fact that  the coefficient of $F_1^0$ in ${\tilde A}_x$, and the coefficient of $F_1^1$ in ${\tilde A}_t$, are the same up to a sign (see \rf{atildedef}).   Under the gauge transformation \rf{gaugetransfatilde} the curvature $F_{tx}$ transforms to  $F_{tx}\rightarrow g\,F_{tx}\,g^{-1}$, and so from \rf{xdef2} we see  that
\be
\partial_t a_x-\partial_x a_t + \sbr{a_t}{a_x}= X\, g\,b_0\,g^{-1}.
\ee
Since $a_x$ lies in the kernel of $b_1$ it follows that $\sbr{a_t}{a_x}$ has components only in the image of $b_1$. Thus, denoting 
\be
g\, b_0\, g^{-1}= \sum_{n=0}^{\infty} \left[ \alpha^{(3,-n)}\, b_{-n} + \alpha^{(1,-n)}\, F_1^{-n} +\alpha^{(2,-n)}\, F_2^{-n}\right]
\lab{rotateb0}
\ee 
we find that
\be
\partial_t a_x^{(3,-n)}-\partial_x a_t^{(3,-n)} = X\,  \alpha^{(3,-n)}\; ; \qquad\qquad\qquad
n=0,1,2,\ldots.
\ee

The explicit expressions for the first few $\alpha^{(i,-n)}$, $i=1,2,3$, are given in  appendix \ref{sec:rotatedpot}. 
Note that if the time component of the connection satisfies the boundary condition $a_t^{(3,-n)}\(x=\infty\)=a_t^{(3,-n)}\(x=-\infty\)$, which is the case in the example we consider, then we have anomalous conservation laws
\be
\frac{d\,Q^{(n)}}{dt}= \beta_n \; ; \qquad\quad {\rm with} \quad\quad
Q^{(n)}=\int_{-\infty}^{\infty}dx\, a_x^{(3,-n)}
\; ; \qquad\quad{\rm where}\qquad  \beta_n=\int_{-\infty}^{\infty}dx\,X\,  \alpha^{(3,-n)}.
\lab{chargedef}
\ee
Of course, in the case of the NLS theory we get an infinite number of conserved quantities since the anomaly $X$, given in \rf{xdef} or \rf{xdef2}, vanishes for the NLS potential \rf{nlspot}. 

We now use a more refined algebraic technique to explore the structure of the anomalies $\beta_n$. The key ingredients are the two $\IZ_2$ transformations, one in the internal space of the loop algebra and the other in space-time. The first $\IZ_2$ transformation is an order $2$ automorphism of the $SL(2)$ loop algebra \rf{newbasis} given by 
\be
\Sigma\(b_n\)=-b_n,\qquad\qquad 
\Sigma\(F_1^n\)=-F_1^n,\qquad\qquad
\Sigma\(F_2^n\)= F_2^n.
\lab{sigmadef}
\ee
The second $\IZ_2$ transformation is a space-time reflection around a given point $\(x_{\Delta},t_{\Delta}\)$, {\it i.e.}
\be
P:\qquad \({\tilde x},{\tilde t}\)\rightarrow \(-{\tilde x},-{\tilde t}\) \qquad\qquad {\rm with} \qquad \quad{\tilde x}= x-x_{\Delta} \qquad \quad{\tilde t}=t-t_{\Delta}.
\lab{paritydef}
\ee

Consider now solutions of the equations of motion \rf{eqofmot} of the theory \rf{lagrangian} such that,
in addition, they satisfy the following property under the parity \rf{paritydef} (see \rf{rphidef})
\be
P:\qquad \qquad R\rightarrow R\; ; \qquad\qquad \qquad 
\varphi \rightarrow -\varphi +{\rm constant}.
\lab{nicerphi}
\ee

Then, the $x$-component of the connection \rf{atildedef} transforms as
\be
\Sigma\({\tilde A}_{x}\)=-{\tilde A}_{x}\qquad\qquad \qquad \qquad
P\({\tilde A}_{x}\)={\tilde A}_{x}
\ee
and so it is odd under the joint action of the two $\IZ_2$ transformations:
\be
\Omega\({\tilde A}_{x}\)=-{\tilde A}_{x},\qquad\qquad \qquad \qquad
\Omega\equiv \Sigma\, P.
\ee

In fact, this property is valid for every individual component of  ${\tilde A}_{x}$. 
 Thus we see that we have  $\Omega\(\sbr{b_1}{{\cal F}^{(-n)}}\)= -\sbr{b_1}{\Omega\({\cal F}^{(-n)}\)}$, and so
\be
\(1+\Omega\)\(\sbr{b_1}{{\cal F}^{(-n)}}\)=\sbr{b_1}{\(1-\Omega\){\cal F}^{(-n)}}
\ee
Since ${\tilde A}_{x}^{(0)}$ is odd under $\Omega$, it follows from the second equation of \rf{axn} that
\be
\(1+\Omega\)a_x^{(0)}= i\,\sbr{b_1}{\(1-\Omega\){\cal F}^{(-1)}}.
\lab{nicerel1}
\ee
The r.h.s. of \rf{nicerel1} is clearly in the image of the adjoint action, and we have chosen the ${\cal F}^{(-n)}$ to rotate $a_x$ into the kernel of that same adjoint action. Therefore, the only possibility for \rf{nicerel1} to hold  is that both sides vanish, {\it i.e.} that 
\be
\(1+\Omega\)a_x^{(0)}=0, \qquad\qquad\qquad \(1-\Omega\){\cal F}^{(-1)}=0
\ee 
and so that ${\cal F}^{(-1)}$ is even under $\Omega$. Using this fact we see from the third equation in \rf{axn} that 
\be
\(1+\Omega\)a_x^{(-1)}= i\,\sbr{b_1}{\(1-\Omega\){\cal F}^{(-2)}}.
\ee

Furthermore, using  same arguments we conclude also that 
\be
\(1+\Omega\)a_x^{(-1)}=0, \qquad\qquad\qquad \(1-\Omega\){\cal F}^{(-2)}=0
\ee
and so that ${\cal F}^{(-2)}$ is even under $\Omega$ as well. Again,  from the fourth equation in \rf{axn} 
we see that $\(1+\Omega\)a_x^{(-2)}= i\,\sbr{b_1}{\(1-\Omega\){\cal F}^{(-3)}}$, and so by the same arguments as before we conclude that 
\be
\(1+\Omega\)a_x^{(-2)}=0, \qquad\qquad\qquad \(1-\Omega\){\cal F}^{(-3)}=0.
\ee

Repeating this reasoning we reach the conclusion that all ${\cal F}^{(-n)}$ are even under $\Omega$. So, the group element $g$, given in \rf{abelg}, is even under $\Omega$
\be
\Omega\(g\)=g.
\lab{eveng}
\ee

To go further we note that since ${\tilde A}_x$ and $\partial_x$ are odd under $\Omega$, and since $g$ is even
\rf{gaugetransfatilde} demonstrates that $a_x$ has to be odd under $\Omega$. One can verify all these claims by inspecting the explicit expressions for the parameters $\zeta_i^{(-n)}$ given in appendix \ref{sec:rotatedpot}. Since the ${\cal F}^{(-n)}$ are even under $\Omega$, and since the generators satisfy \rf{sigmadef}, it follows from \rf{abelg} that $P\(\zeta_1^{(-n)}\)=-\zeta_1^{(-n)}$ and $P\(\zeta_2^{(-n)}\)=\zeta_2^{(-n)}$.

Next we use the Killing form of the $SL(2)$ loop algebra given by
\be
{\rm Tr}\(b_n\,b_m\)=\frac{1}{2}\,\delta_{n+m,0}\; ; \qquad\qquad {\rm Tr}\(b_n\,F_i^m\)=0 \; ; \qquad i=1,2,
\ee
which can be realized by ${\rm Tr}\(\star\)\equiv \frac{1}{2\,\pi\,i}\,\oint \frac{d\lambda}{\lambda} {\rm tr}\(\star\)$, with ${\rm tr}$ being the ordinary finite matrix trace, and $T_i^n=\lambda\, T_i$, $i=3,\pm$. 
In this case we see from \rf{rotateb0} that
\be
\alpha^{(3,-n)}= 2\,{\rm Tr}\(g\,b_0\,g^{-1}\, b_{n}\)=2\,{\rm Tr}\(\Sigma\(g\)\,b_0\,\Sigma\(g^{-1}\)\, b_{n}\),
\ee
where in the last equality we have used the fact that the Killing form is invariant under $\Sigma$, and that all the $b_n$'s are odd under it. Thus, using \rf{eveng} we have that 
\be
P\(\alpha^{(3,-n)}\)=2\,{\rm Tr}\(\Omega\(g\)\,b_0\,\Omega\(g^{-1}\)\, b_{n}\)
=2\,{\rm Tr}\(g\,b_0\,g^{-1}\, b_{n}\)= \alpha^{(3,n)}
\ee
and so we see that all the $\alpha^{(3,-n)}$'s are even under $P$. 
Note that $X$, given in \rf{xdef2}, is an $x$-derivative of a functional of $R$. Since we have assumed that $R$ is even under $P$, we see from \rf{nicerphi} that $X$ is odd, {\it i.e.} that  $P\(X\)=-X$ and so that 
\be
\int_{-{\tilde t}_0}^{{\tilde t}_0} dt\,\int_{-{\tilde x}_0}^{{\tilde x}_0} dx\, X\, \alpha^{(3,-n)}=0,
\lab{mm}
\ee
where ${\tilde t}_0$ and ${\tilde x}_0$ are given fixed values of the shifted time ${\tilde t}$ and space coordinate ${\tilde x}$ respectively, introduced in \rf{paritydef}. Therefore,  by taking ${\tilde x}_0\rightarrow \infty$, we conclude that the non-conserved charges \rf{chargedef} satisfy the following mirror time-symmetry around the point: $t_{\Delta}$.
\be
Q^{(n)}\(t={\tilde t}_0 +t_{\Delta}\)=Q^{(n)}\(t=-{\tilde t}_0 +t_{\Delta}\).
\lab{mirrorcharge}
\ee

In consequence, even though the charges $Q^{(n)}$ vary in time, they are symmetric w.r.t. to $t=t_{\Delta}$. Note that we have derived this property for any potential $V$ which depends only on the modulus of $\psi$. The only assumption we have made is that we are considering fields $\psi$ which satisfy \rf{nicerphi}.

In the next sections we will show that such solutions are very plausible and that, in fact, the one and two-soliton solutions of the theories \rf{lagrangian} can always be chosen to satisfy \rf{nicerphi}. This fact has far reaching consequences for the properties of the theories \rf{lagrangian}. For instance, by taking ${\tilde t}_0\rightarrow \infty$ one concludes that the scattering of two-soliton solutions presents an infinite number of charges which are asymptotically conserved. Since the $S$-matrix relies only on asymptotic states, it is quite plausible that the theories \rf{lagrangian} share a lot of interesting properties
with integrable theories (but which have been believed to be only true for integrable field theories).

\subsection{Another way of understanding it}
\label{sec:simplerargument}

The properties leading to charges satisfying \rf{mirrorcharge} can be realized, in fact, in a much wider context. Indeed, consider a field theory in a space-time of $(d+1)$ dimensions with fields labelled by $\vp_a$, $a=1,2,\ldots n$. These fields can be scalars, vectors, spinors, etc, and the indices $a$ just label their components.  Consider a fixed point $x_{\Delta}^{\mu}$ in space-time, and a reflection $P$ around it, {\it i.e.}
\be
P:\qquad\quad {\tilde x}^{\mu}\rightarrow -\,{\tilde x}^{\mu} \qquad\quad {\rm with} \qquad\quad 
{\tilde x}^{\mu}=x^{\mu}-x_{\Delta}^{\mu}\qquad\quad \mu=0,1,2\ldots d.
\lab{generalparity}
\ee 
Suppose that a such field theory possesses a classical solution $\vp_a^s$ such that the fields evaluated on it are eigenvectors of $P$ up to constants, {\it i.e.} that
\be
P\(\vp_a^s\)=\ve_a\, \vp_a^s + c_a, \qquad\qquad\qquad \ve_a=\pm 1\; ; \qquad c_a={\rm const.}
\lab{eigengeneralparity}
\ee
Consider now a functional of the fields and of their derivatives $F=F\(\vp_a, \pa_{\mu}\vp_a, \pa_{\mu}\pa_{\nu}\vp_a,\ldots \)$, that is even under $P$ when evaluated on a particular solution, {\it i.e.}
\be
P\left[F\(\vp_a^s, \pa_{\mu}\vp_a^s, \pa_{\mu}\pa_{\nu}\vp_a^s,\ldots \)\right]= F\(\vp_a^s, \pa_{\mu}\vp_a^s, \pa_{\mu}\pa_{\nu}\vp_a^s,\ldots \).
\ee

Next, look at a rectangular spatial volume ${\cal V}$ bounded by hyperplanes crossing the axes of the space coordinates at the points $\pm {\tilde x}^i_0$, $i=1,2,\ldots d$, corresponding to fixed values of the shifted space coordinates ${\tilde x}^i$ introduced in \rf{generalparity}, {\it i.e.} such that the point $x^i_{\Delta}$ lies in the very center of ${\cal V}$. The integral of this functional over ${\cal V}$ 
\be
Q=\int_{{\cal V}} d^dx \, F
\lab{generalchargedef}
\ee
satisfies
\be
\frac{d\,Q}{d\,x^0}= \int_{{\cal V}} d^dx \, \frac{d\,F}{d\,x^0}=\int_{{\cal V}} d^dx \, \left[ \frac{\delta\,F}{\delta \vp_a}\,\pa_0\vp_a+\frac{\delta\,F}{\delta \pa_{\mu}\vp_a}\,\pa_0\pa_{\mu}\vp_a
+\frac{\delta\,F}{\delta \pa_{\mu}\pa_{\nu}\vp_a}\,\pa_0\pa_{\mu}\pa_{\nu}\vp_a
+\ldots \right].
\lab{dqdt}
\ee
When evaluated on the solution $\vp_a^s$ each term in the integrand in \rf{dqdt} is odd under $P$. The reasons for this are simple:  any derivative of the form $\partial_0\partial_{\mu_1}\ldots \partial_{\mu_m} \vp_a$, when evaluated on $\vp_a^s$, has an eigenvalue of $P$ equal to $\varepsilon_a\,\(-1\)^{m+1}$. Since  $F$ evaluated on $\vp_a^s$  is even under $P$, it follows that any derivative of the form $\frac{\delta\,F}{\delta \pa_{\mu_1}\ldots \pa_{\mu_m}\vp_a}$ has an eigenvalue of $P$ equal to  $\varepsilon_a\,\(-1\)^{m}$, when evaluated on $\vp_a^s$. Therefore, when evaluated on $\vp_a^s$  each term of the integrand on the r.h.s. of \rf{dqdt} is odd under $P$. Consequently, one finds that 
\br
Q^s\({\tilde x}^0\)&-&Q^s\(-{\tilde x}^0\)= \int_{-{\tilde x}^0}^{{\tilde x}^0} dx^0\,\frac{d\,Q^s}{d\,x^0}
\lab{generalmirror}\\
&=& \int_{-{\tilde x}^0}^{{\tilde x}^0}dx^0\,\int_{{\cal V}} d^d x \, \left[ \frac{\delta\,F^s}{\delta \vp_a^s}\,\pa_0\vp_a^s+\frac{\delta\,F^s}{\delta \pa_{\mu}\vp_a^s}\,\pa_0\pa_{\mu}\vp_a^s
+\frac{\delta\,F^s}{\delta \pa_{\mu}\pa_{\nu}\vp_a^s}\,\pa_0\pa_{\mu}\pa_{\nu}\vp_a^s
+\ldots \right] = 0,
\nonumber
\er
where the superscript $s$ denotes that $Q$ is evaluated  on the solution $\vp_a^s$, and ${\tilde x}^0$ is a given fixed value of the shifted time introduced in 
\rf{generalparity}.

Summarizing our results: if one has a solution of the theory such that all the fields evaluated on  this solution are eigenstates of $P$, {\it i.e.} they satisfy \rf{eigengeneralparity}, then any even functional of these fields and their derivatives leads to charges that satisfy a mirror time-symmetry like \rf{generalmirror}. 

In the case studied in this paper we have shown that the $x$-component of the connection, $a_x$, is odd under the transformation $\Omega=\Sigma \, P$, {\it i.e.} $\(1+\Omega\)a_x=0$. Since $a_x$ lies in the abelian subalgebra generated by the $b_n$'s (see \rf{axcomp}), which are odd under $\Sigma$ (see \rf{sigmadef}), it follows that the charge densities $a_x^{(3,-n)}$ are even under $P$. Therefore, the charges $Q^{(n)}$ introduced in \rf{chargedef} are in the class of charges \rf{generalchargedef} discussed  in this subsection. So, the assumption of the existence of a solution satisfying \rf{nicerphi} has much deeper consequences. It implies  not only that the charges \rf{chargedef} satisfy the mirror time-symmetry \rf{mirrorcharge}, but also that any charge built out of a density that is even under $P$ when evaluated on this solution, also satisfies  \rf{mirrorcharge}. The fact that a solution satisfies \rf{nicerphi} implies that its past and future w.r.t. to the point in time $t_{\Delta}$, are strongly linked and, in consequence, so are many of its properties. Indeed, the mirror time-symmetry \rf{mirrorcharge} is a direct consequence of such a link between the past and the future. The non-linear phenomena behind the quasi-integrability properties we are discussing are certainly driven by the parity property \rf{nicerphi}. However, we still have to understand the basic physical processes guarantee that a given solution satisfies \rf{nicerphi}. This is one of the great challenges for our techniques  to understand. In the next section, we argue that for the theories \rf{lagrangian} for which  the potential $V$ is a deformation of the NLS potential \rf{nlspot}, the solutions satisfying  \rf{nicerphi} are favoured  by the dynamics if the corresponding undeformed solution of the integrable NLS theory also satisfies \rf{nicerphi}.

\section{Dynamics versus parity}
\label{sec:deformnls}
\setcounter{equation}{0} 

In terms of fields $R$ and $\vp$ introduced in \rf{rphidef}, the equations of motion \rf{eqofmot} become
\br
\pa_t R&=&-\pa_x\(R\, \pa_x\vp\),\nonumber\\
-R^2\,\pa_t \vp&=&-R\,\pa_x^2 R+\frac{R^2}{2}\,\(\pa_x\vp\)^2+\frac{1}{2}\(\pa_xR\)^2 +2\, R^2\, \frac{\partial\,V}{\partial R}.
\lab{rvpeqs}
\er

Let us analyze what type of solutions these equations admit if we assume that the fields of these solutions are eigenstates of the of parity transformation $P$ introduced in \rf{paritydef}. We split the fields as
\be
R= R^{(+)}+  R^{(-)}\; ; \qquad\qquad\qquad \vp=\vp^{(+)}+\vp^{(-)},
\ee
where 
\be
P\(R^{(\pm)}\)=\pm R^{(\pm)}, \qquad\qquad \qquad
P\(\vp^{(\pm)}\)=\pm \vp^{(\pm)} + {\rm constant.}
\ee

Let us now assume that we have a solution for which $R^{(+)}=\vp^{(-)}=0$. Then, the l.h.s. of the first equation in \rf{rvpeqs} is even under $P$, and its r.h.s. is odd. Thus, $\pa_t R^{(-)}=0$ and $\pa_x\(R^{(-)}\, \pa_x\vp^{(+)}\)=0$. In addition, the second equation in \rf{rvpeqs} implies that $\partial_t\vp^{(+)}=-2\,\left[\frac{\partial\,V}{\partial R}\right]^{(-)}$.

Note also that if we have a solution for which $R^{(-)}=\vp^{(-)}=0$ we get very similar results, namely that $\pa_t R^{(+)}=0$,  $\pa_x\(R^{(+)}\, \pa_x\vp^{(+)}\)=0$, and that $\partial_t\vp^{(+)}=-2\,\left[\frac{\partial\,V}{\partial R}\right]^{(-)}$.

In a similar way, if we assume that our solution satisfies $R^{(+)}=\vp^{(+)}=0$, then the second equation in \rf{rvpeqs} implies that $\left[\frac{\partial\,V}{\partial R}\right]^{(-)}=0$. This condition, however, excludes potentials that are even functions of $R$, like the integrable NLS potential  \rf{nlspot}. 
Thus, we would not expect interesting non-trivial solutions, like a two-soliton solution, with one of these three classes of cases in which the fields are eigenstates of $P$. 

The only remaining case is the one we assumed in \rf{nicerphi}, namely, that $R^{(-)}=\vp^{(+)}=0$. 
One can easily check that the equations \rf{rvpeqs} do not impose any restrictions on the solutions of this type. Indeed, $\frac{\partial\,V}{\partial R}$ is always even under $P$ for any $V$, if $R$ is even under $P$.

Consequently, we would expect most of the interesting non-trivial results for solutions of the theories \rf{lagrangian}, for which the fields evaluated on them are eigenstates of $P$,  to fall into the class \rf{nicerphi}. Of course, there can also exist classes of non-trivial solutions for which the parity components are mixed
and the above arguments do not apply. However, this does not mean that the results 
of these arguments are necessarily incorrect. Sometimes they may still hold even though one has to work harder to prove them. In the next section we present a detailed analysis of the case in which the potential $V$ is a deformation of the NLS potential  \rf{nlspot}. Our analysis shows that the mixed solutions can always be ``gauged away'', order by order, in the perturbation expansion around the NLS theory.

\subsection{Deformations of the NLS theory}

We now consider the theories \rf{lagrangian} for which the potential $V$ is a deformation of the NLS potential \rf{nlspot}. The deformation is introduced through a parameter $\ve$ such that for $\ve=0$, $V$ corresponds to \rf{nlspot}. We will not consider here the deformations for which the potential depends upon the phase of $\psi$. Examples of such potentials are
\be
V^{(1)}= \eta \, R^{2+\ve}\; ; \qquad\qquad 
V^{(2)}= \eta\, R^2+\ve\, R^3\; ;\qquad\qquad
V^{(3)}= \eta\, R^2\, e^{-\ve\, R},
\ee
where $R=\mid \psi\mid^2$ was introduced in  \rf{rphidef}. 

We start our analysis by  expanding the solutions of the corresponding equations 
of motion in powers of $\ve$ around the solution of the integrable NLS theory as
\be
R=R_0+\ve\,R_1+\ve^2\,R_2+\ldots, \qquad\qquad \quad
\vp=\vp_0+\ve\,\vp_1+\ve^2\,\vp_2+\ldots
\ee

Of course, the deformed potential $V$ has the expansion
\be
V= V\mid_{\ve=0}+ \ve\, \left[\frac{\partial V}{\partial \ve}\mid_{\ve=0}+\frac{\partial V}{\partial R}\mid_{\ve=0}\, R_1\right] + O\(\ve^2\)
\ee
and its gradient has the expansion
\br
\frac{\partial\,V}{\partial R}&=& \frac{\partial\,V}{\partial R}\mid_{\ve=0} + 
\ve\,\left[\frac{\partial^2\,V}{\pa\ve\,\partial R}\mid_{\ve=0}  +\frac{\partial^2\,V}{\partial R^2}\mid_{\ve=0} \, R_1  \right]\\
&+&\frac{\ve^2}{2}\,\left[ \frac{\partial^3\,V}{\pa\ve^2\,\partial R}\mid_{\ve=0}  +2\,\frac{\partial^3\,V}{\pa\ve\partial R^2}\mid_{\ve=0} \, R_1+  \frac{\partial^3\,V}{\partial R^3}\mid_{\ve=0} \, R_1^2 +2\,\frac{\partial^2\,V}{\partial R^2}\mid_{\ve=0} \, R_2 \right] +\ldots
\nonumber
 \er

We also expand the equations of motion \rf{rvpeqs} into powers of $\ve$ and at the same time we split the equations, and so the fields, into their even and odd parts under a given space-time parity $P$ of the type  \rf{paritydef}. At this stage the value of the point $\(x_{\Delta},t_{\Delta}\)$ around which we perform the reflection is not yet important. We just use the fact that the operation $P$ satisfies $P^2=\one$, and so it has eigenvalues $\pm1$. Next we introduce the following notation for the eigen-components of the fields: 
\be
\star^{(\pm)}\equiv \frac{1}{2}\,\(1\pm P\)\, \star.
\ee

Then the zero order part of the equations of motion \rf{rvpeqs} splits under $P$  as
\br
\pa_t R_0^{(-)}&=&-\pa_x\(R_0^{(+)}\, \pa_x\vp_0^{(+)}+R_0^{(-)}\, \pa_x\vp_0^{(-)}\), 
\lab{splitzero1}\\
\pa_t R_0^{(+)}&=&-\pa_x\(R_0^{(+)}\, \pa_x\vp_0^{(-)}+R_0^{(-)}\, \pa_x\vp_0^{(+)}\) 
\lab{splitzero2}
\er
and
\br
&-&\(\(R_0^{(+)}\)^2+\(R_0^{(-)}\)^2\)\,\pa_t \vp_0^{(-)} -2\,R_0^{(+)}\,R_0^{(-)}\,\pa_t \vp_0^{(+)}=-R_0^{(+)}\,\pa_x^2 R_0^{(+)}\, -R_0^{(-)}\,\pa_x^2 R_0^{(-)}
\nonumber\\
&+&\frac{1}{2}\,\(\(R_0^{(+)}\)^2+\(R_0^{(-)}\)^2\)\,\(\(\pa_x\vp_0^{(+)}\)^2+\(\pa_x\vp_0^{(-)}\)^2\)+2 \,R_0^{(+)}\,R_0^{(-)}\,\pa_x\vp_0^{(+)}\,\pa_x\vp_0^{(-)}
\nonumber\\
&+&\frac{1}{2}\(\(\pa_x R_0^{(+)}\)^2+\(\pa_x R_0^{(-)}\)^2\) 
\lab{splitzero3}\\
&+&2\, \(\(R_0^{(+)}\)^2+\(R_0^{(-)}\)^2\)\, \left[\frac{\partial\,V}{\partial R}\mid_{\ve=0}\right]^{(+)} 
+4\, R_0^{(+)}\,R_0^{(-)}\,\left[\frac{\partial\,V}{\partial R}\mid_{\ve=0}\right]^{(-)} 
\nonumber
\er
and
\br
&-&\(\(R_0^{(+)}\)^2+\(R_0^{(-)}\)^2\)\,\pa_t \vp_0^{(+)} -2\,R_0^{(+)}\,R_0^{(-)}\,\pa_t \vp_0^{(-)}=-R_0^{(+)}\,\pa_x^2 R_0^{(-)}\, -R_0^{(-)}\,\pa_x^2 R_0^{(+)}
\nonumber\\
&+&\(\(R_0^{(+)}\)^2+\(R_0^{(-)}\)^2\)\,\pa_x\vp_0^{(+)}\,\pa_x\vp_0^{(-)}
+R_0^{(+)}\,R_0^{(-)}\,\(\(\pa_x\vp_0^{(+)}\)^2+\(\pa_x\vp_0^{(-)}\)^2\)
\nonumber\\
&+&\pa_x R_0^{(+)}\,\pa_x R_0^{(-)}
\lab{splitzero4}\\
&+&2\, \(\(R_0^{(+)}\)^2+\(R_0^{(-)}\)^2\)\, \left[\frac{\partial\,V}{\partial R}\mid_{\ve=0}\right]^{(-)} 
+4\, R_0^{(+)}\,R_0^{(-)}\,\left[\frac{\partial\,V}{\partial R}\mid_{\ve=0}\right]^{(+)}. 
\nonumber
\er

As we have shown in section \ref{sec:model}, and in particular in sub-section \ref{sec:simplerargument}, the mirror time-symmetry property of the charges,  given in \rf{mirrorcharge}, is valid for solutions for which the components of the fields with different eigenvalues of $P$ are not mixed. Since, we have two fields $R$ and $\vp$ we have four possibilities for non-mixing solutions. In our analysis we shall assume that the potentials satisfy the property 
\be
\frac{\partial\,V}{\partial R}\mid_{\ve=0}\; \sim R_0.
\ee
If one considers solutions for which $R_0^{(+)}=0$ and $\pa\vp_0^{(+)}=0$ (with $\partial$ standing for time and space derivatives), then the zero order equations of motion \rf{splitzero1}-\rf{splitzero4} impliy that $R_0^{(-)}=0$. In addition, if one assumes $R_0^{(+)}=0$ and $\pa\vp_0^{(-)}=0$ then  \rf{splitzero1}-\rf{splitzero4} imply $\pa_t R_0^{(-)}=0$. Finally, if one assumes $R_0^{(-)}=0$ and $\pa\vp_0^{(-)}=0$ then one finds that $\pa_t R_0^{(+)}=0$ and $\pa_t\vp_0^{(+)}=0$. Therefore, in none of those three cases one should expect to get interesting solutions, specially two-soliton solutions. Therefore, we shall restrict our attention to the class of solutions for which $R_0^{(-)}=0$ and $\pa_{t,x}\vp_0^{(+)}=0$, {\it i.e.} those that satisfy
\be
P:\qquad\qquad R_0 \rightarrow R_0 \qquad\qquad \vp_0 \rightarrow -\vp_0 + {\rm const.}
\lab{nicezeroordersol}
\ee

Note that with $R_0$ even under $P$ it follows that all derivatives of the form $\frac{\partial^{n+m}\,V}{\pa\ve^n\,\partial R^m}\mid_{\ve=0}$ are even under  $P$. 
Now, assuming \rf{nicezeroordersol} one gets that the first order part of the equations of motion \rf{rvpeqs} split under $P$  as
\br
\pa_t R_1^{(-)}&=&-\pa_x\(R_0^{(+)}\, \pa_x\vp_1^{(+)} +R_1^{(-)}\, \pa_x\vp_0^{(-)}\),
\lab{splione1}\\
\pa_t R_1^{(+)}&=&-\pa_x\(R_0^{(+)}\, \pa_x\vp_1^{(-)} +R_1^{(+)}\, \pa_x\vp_0^{(-)}\),
\lab{splione2}\\
-\(R_0^{(+)}\)^2\,\pa_t \vp_1^{(-)}&=& 2 R_0^{(+)} R_1^{(+)}\,\pa_t \vp_0^{(-)}
-R_0^{(+)}\,\pa_x^2 R_1^{(+)}-R_1^{(+)}\,\pa_x^2 R_0^{(+)}
\lab{splione3}\\
&+&R_0^{(+)} R_1^{(+)}\,\(\pa_x\vp_0^{(-)}\)^2+\(R_0^{(+)}\)^2\,\pa_x\vp_0^{(-)}\, \pa_x\vp_1^{(-)}
+\pa_x R_0^{(+)}\, \pa_x R_1^{(+)} 
\nonumber\\
&+&4\, R_0^{(+)}\,R_1^{(+)}\, \frac{\partial\,V}{\partial R}\mid_{\ve=0}
+2\, \(R_0^{(+)}\)^2\, \left[\frac{\partial^2\,V}{\pa\ve\,\partial R}\mid_{\ve=0}  +\frac{\partial^2\,V}{\partial R^2}\mid_{\ve=0} \, R_1^{(+)}  \right],
\nonumber\\
-\(R_0^{(+)}\)^2\,\pa_t \vp_1^{(+)}&=& 2 R_0^{(+)} R_1^{(-)}\,\pa_t \vp_0^{(-)}
-R_0^{(+)}\,\pa_x^2 R_1^{(-)}-R_1^{(-)}\,\pa_x^2 R_0^{(+)}
\lab{splione4}\\
&+&R_0^{(+)} R_1^{(-)}\,\(\pa_x\vp_0^{(-)}\)^2+\(R_0^{(+)}\)^2\,\pa_x\vp_0^{(-)}\, \pa_x\vp_1^{(+)}
+\pa_x R_0^{(+)}\, \pa_x R_1^{(-)} 
\nonumber\\
&+&4\, R_0^{(+)}\,R_1^{(-)}\, \frac{\partial\,V}{\partial R}\mid_{\ve=0}
+2\, \(R_0^{(+)}\)^2\, \frac{\partial^2\,V}{\partial R^2}\mid_{\ve=0} \, R_1^{(-)}.  
\nonumber
\er
Once the zero order solutions for $R_0^{(+)}$ and $\vp_0^{(-)}$ have been found, we put them into \rf{splione1}-\rf{splione4} and get four  coupled partial differential equations with non-constant coefficients which are linear in the first order fields $R_1^{(\pm)}$ and $\vp_1^{(\pm)}$. 

There are two important facts about \rf{splione1}-\rf{splione4}. First they couple $R_1^{(+)}$ only to $\vp_1^{(-)}$   and $R_1^{(-)}$ only to $\vp_1^{(+)}$, {\it i.e.} the pair of equations \rf{splione1} and \rf{splione4} is decoupled from the pair formed by \rf{splione2} and \rf{splione3}. Secondly, the pair of equations \rf{splione1} and \rf{splione4} is homogeneous in the first order fields, but 
the equation \rf{splione3} is non-homogeneous due to the term $2\, \(R_0^{(+)}\)^2\, \frac{\partial^2\,V}{\pa\ve\,\partial R}\mid_{\ve=0}$, which does not involve the first order fields. Therefore, there are no solutions for which  $R_1^{(+)}=0$  and $\vp_1^{(-)}={\rm constant}$. On the other hand we can have solutions for which $R_1^{(-)}=0$  and $\vp_1^{(+)}={\rm constant}$. In addition, if $R_1$ and $\vp_1$, are solutions with a non-definite parity,  then 
$R_1-R_1^{(-)}$ and $\vp_1-\vp_1^{(+)}$ are also   solutions but now with a definite parity. So, we can always choose the first order solutions to satisfy
\be
P:\qquad\qquad R_1 \rightarrow R_1 \qquad\qquad \quad \vp_1 \rightarrow -\vp_1 + {\rm const.}
\lab{nicefirstordersol}
\ee

If we now take the zero and first order solutions satisfying \rf{nicezeroordersol} and \rf{nicefirstordersol}, respectively, then the second order part of the equations of motion \rf{rvpeqs} splits under $P$  as
\br
\pa_t R_2^{(-)}&=&-\pa_x\(R_2^{(-)}\, \pa_x\vp_0^{(-)}\)-\pa_x\(R_0^{(+)}\, \pa_x\vp_2^{(+)}\),
\lab{splitwo1}\\
\pa_t R_2^{(+)}&=&-\pa_x\(R_2^{(+)}\, \pa_x\vp_0^{(-)}\)-\pa_x\(R_0^{(+)}\, \pa_x\vp_2^{(-)}\)
-\pa_x\(R_1^{(+)}\, \pa_x\vp_1^{(-)}\)
\lab{splitwo2}
\er
and
\br
&-&\(R_0^{(+)}\)^2\,\pa_t \vp_2^{(-)} -2R_0^{(+)}\,R_1^{(+)}\,\pa_t \vp_1^{(-)} 
-\(\(R_1^{(+)}\)^2+2\,R_0^{(+)}\,R_2^{(+)}\)\,\pa_t \vp_0^{(-)}
\nonumber\\
&=&-R_0^{(+)}\,\pa_x^2 R_2^{(+)} -R_2^{(+)}\,\pa_x^2 R_0^{(+)} -R_1^{(+)}\,\pa_x^2 R_1^{(+)}
\nonumber\\
&+&\(R_0^{(+)}\)^2\,\pa_x\vp_0^{(-)} \, \pa_x\vp_2^{(-)}
+\frac{1}{2}\, \(R_0^{(+)}\)^2\,\(\pa_x\vp_1^{(-)}\)^2
+2\,R_0^{(+)}\,R_1^{(+)}\,\pa_x\vp_0^{(-)} \, \pa_x\vp_1^{(-)}
\nonumber\\
&+& R_0^{(+)}\, R_2^{(+)}\,\(\pa_x\vp_0^{(-)}\)^2
+\frac{1}{2}\, \(R_1^{(+)}\)^2\,\(\pa_x\vp_0^{(-)}\)^2
\lab{splitwo3}\\
&+&\frac{1}{2}\(\pa_xR_1^{(+)}\)^2 + \pa_xR_0^{(+)}\,\pa_xR_2^{(+)}
+\left[2\, \(R_1^{(+)}\)^2+ 4\, R_0^{(+)}\,R_2^{(+)}\right]\, \frac{\partial\,V}{\partial R}\mid_{\ve=0}
\nonumber\\
&+& 4\, R_0^{(+)}\,R_1^{(+)}\,\left[\frac{\partial^2\,V}{\pa\ve\,\partial R}\mid_{\ve=0}  +\frac{\partial^2\,V}{\partial R^2}\mid_{\ve=0} \, R_1^{(+)}  \right]
\nonumber\\
&+& \(R_0^{(+)}\)^2\,\left[ \frac{\partial^3\,V}{\pa\ve^2\,\partial R}\mid_{\ve=0}  +2\,\frac{\partial^3\,V}{\pa\ve\partial R^2}\mid_{\ve=0} \, R_1^{(+)}+  \frac{\partial^3\,V}{\partial R^3}\mid_{\ve=0} \, \(R_1^{(+)}\)^2 +2\,\frac{\partial^2\,V}{\partial R^2}\mid_{\ve=0} \, R_2^{(+)} \right] 
\nonumber
\er
and
\br
&-&\(R_0^{(+)}\)^2\,\pa_t \vp_2^{(+)}  -2\,R_0^{(+)}\,R_2^{(-)}\,\pa_t \vp_0^{(-)}=
-R_0^{(+)}\,\pa_x^2 R_2^{(-)} -R_2^{(-)}\,\pa_x^2 R_0^{(+)} 
\nonumber\\
&+&\(R_0^{(+)}\)^2\,\pa_x\vp_0^{(-)} \, \pa_x\vp_2^{(+)}
+R_0^{(+)}\, R_2^{(-)}\,\(\pa_x\vp_0^{(-)}\)^2
\lab{splitwo4}\\
&+&\pa_xR_0^{(+)}\,\pa_xR_2^{(-)}
+ 4\, R_0^{(+)}\,R_2^{(-)}\, \frac{\partial\,V}{\partial R}\mid_{\ve=0}
\nonumber\\
&+& 2\,\frac{\partial^2\,V}{\partial R^2}\mid_{\ve=0} \, R_2^{(-)}.
\nonumber
\er

Again we have a structure very similar to that discussed in the case of the equations \rf{splione1}-\rf{splione4}. Indeed, having found the solutions for the zero and first order fields, we put them into \rf{splitwo1}-\rf{splitwo4} and get four coupled partial differential equations with non-constant coefficients which are linear in $R_2^{(\pm)}$ and $\vp_2^{(\pm)}$. In addition, the pair of equations \rf{splitwo1} and \rf{splitwo4} is decoupled from the pair \rf{splitwo2} and \rf{splitwo3}, {\it i.e.} $R_2^{(+)}$ couples only to $\vp_2^{(-)}$  and $R_2^{(-)}$ also only to $\vp_2^{(+)}$. Again, the pair of equations \rf{splitwo1} and \rf{splitwo4} is homogeneous in the second order fields and the pair \rf{splitwo2} and \rf{splitwo3} is non-homogeneous. Thus, as before, $R_2^{(-)}=0$ and $\vp_2^{(+)}={\rm constant}$ is a solution, but $R_2^{(+)}=0$ and $\vp_2^{(-)}={\rm constant}$, cannot be a solution. In addition, if $R_2$ and $\vp_2$, are solutions, with a non-definite parity, then 
$R_2-R_2^{(-)}$ and $\vp_2-\vp_2^{(+)}$ are also   solutions but now with a definite parity. So, we can always choose the second order solutions to satisfy
\be
P:\qquad\qquad R_2 \rightarrow R_2 \qquad\qquad \quad \vp_2 \rightarrow -\vp_2 + {\rm const.}
\lab{nicesecondordersol}
\ee

We can repeat this process, and even though we have not proved this here, this structure repeats
itself at every order of perturbation in $\ve$. Therefore, the fields $R^{(-)}$ and $\vp^{(+)}$ can always be ``gauged away'' since they satisfy homogeneous equations (order by order), but the fields $R^{(+)}$ and $\vp^{(-)}$ are robust in the sense that they always have  to be present in the solution. Thus, we have shown that the solutions satisfying \rf{nicerphi} are favoured by the dynamics when the potential $V$ in \rf{lagrangian} is a deformation of the integral  NLS potential \rf{nlspot}.  We point out however, that the property of being even or odd under $P$  is not something that can be encoded into the initial boundary conditions at a given initial time $t_0$. The properties under $P$ involve a link between the past and the future of the solution and so, perhaps, cannot be understood using the usual techniques (specially numerical) of investigating the coupling of the normal modes as the systems evolves in time.  We are perhaps facing a new and intriguing  non-liner phenomenon. In the next section, we go further in our analysis and study the properties under $P$ of the exact one and two-soliton solutions of the integral NLS theory.

\section{The NLS solitons and their parity properties}
\label{sec:nlssoliton}
\setcounter{equation}{0}

We will now analyze the one and two soliton solutions of the integrable NLS theory \rf{nlseq} under the parity transformation  \rf{paritydef}. The solutions are constructed by the Hirota's method described in the appendix \ref{sec:hirota}.

\subsection{The one-soliton solutions}

 In terms of the fields $R$ and $\vp$ introduced in \rf{rphidef}  the one-bright-soliton solution \rf{brightsol} is given by
\be
R_0^{{\rm bright}}= 
\frac{\rho^2}{\mid \eta\mid}\, 
\frac{1}{\cosh^2\left[\rho\,\(x-v\,t-x_0\)\right]}\; ;\qquad\qquad\qquad
\vp_0^{{\rm bright}}=2\,\left[\(\rho^2-\frac{v^2}{4}\)\,t+ \frac{v}{2}\, x\right]
\lab{brightsolrvp}
\ee
Analogously, the one-dark-soliton solution \rf{darkonesol} is given by 
\be
R_0^{{\rm dark}}= 
 \frac{\rho^2}{\eta}\,\tanh^2\left[\rho\(x-v\,t-x_0\)\right]\; ; \qquad\qquad\qquad
 \vp_0^{{\rm dark}}=2\,\left[\frac{v}{2}\,x-\(2\,\rho^2+\frac{v^2}{4}\)\,t\right]
 \lab{darkonesolrvp}
 \ee
Then it is clear that the relevant parity transformation, in each case,  is
\be
P:\qquad \qquad{\tilde x}\rightarrow -{\tilde x} \qquad\qquad \qquad t\rightarrow -t \qquad\quad {\rm with}\qquad\quad {\tilde x}=x-x_0
\lab{onesolparity}
\ee
Therefore one has that
\be
P:\qquad \quad R_0^{{\rm bright/dark}}\rightarrow R_0^{{\rm bright/dark}}
\; ; \qquad\qquad\quad \vp_0^{{\rm bright/dark}}\rightarrow - \vp_0^{{\rm bright/dark}}+2\,v\,x_0
\ee
which is agreement with \rf{nicezeroordersol} and \rf{nicerphi}. 

If one chooses the potential in \rf{lagrangian} as 
\be
V= \frac{2}{2+\ve}\, \eta \, R^{2+\ve}
\lab{pert}
\ee
then the theory has a one-soliton solution given by
\be
R= \left[\frac{2+\ve}{2}\,
\frac{\rho^2}{\mid \eta\mid}\, 
\frac{1}{\cosh^2\left[\(1+\ve\)\,\rho\,\(x-v\,t-x_0\)\right]}\right]^{\frac{1}{1+\ve}}\; ;\qquad
\vp=2\,\left[\(\rho^2-\frac{v^2}{4}\)\,t+ \frac{v}{2}\, x\right]
\lab{brightsoldeform}
\ee
which is a deformation of the one-bright-soliton \rf{brightsolrvp}. Notice that under the parity \rf{onesolparity} it transforms as
\be
P:\qquad \quad R\rightarrow R
\; ; \qquad\qquad\quad \vp\rightarrow - \vp+2\,v\,x_0
\lab{nicedeformedonesolparity}
\ee
Since \rf{brightsoldeform} is an exact solution of the deformed NLS theory this observation  supports our claims of section \ref{sec:deformnls}, based on the perturbative series in $\ve$, that solutions satisfying the property \rf{nicedeformedonesolparity} are favoured by the dynamics.

\subsection{The two-soliton solutions}

The two-bright-soliton solution of the NLS model can been obtained using the Hirota method. The details are given in the appendix \ref{sec:hirota}.  Its expression is given in \rf{twobrightsoliton},  which can be rewritten as  
\br
\psi_0=\frac{2}{\sqrt{\mid \eta\mid}} \;\frac{{\cal N}}{{\cal D}}, 
   \lab{twobrightsoliton2}
\er
where the overall phase $i\,e^{-i\,\phi}$, has been absorbed using the symmetry \rf{phasetransf}, and where we have defined
\be
{\cal D}=2\, e^{z_{+}}\left[\cosh z_{+}+ e^{-\Delta}\,\cosh z_{-} - 16 \frac{\mid \rho_1\mid\,\mid \rho_2\mid}{\Lambda_{-}}\,\cos\(\Omega_1-\Omega_2-2\,\delta_{+}\)\right]
\lab{calddef}
\ee
and
\br
{\cal N}&=&e^{z_{+}}\,e^{-\frac{\Delta}{2}}\,e^{-i\,\frac{\(\Omega_1+\Omega_2\)}{2}}
 \,e^{-i\,\delta_{-}}\left[
 e^{-\frac{z_{+}}{2}}e^{i\,\delta_{-}}\(\mid \rho_1\mid \,e^{-i\,\frac{\(\Omega_1-\Omega_2\)}{2}}\,e^{\frac{z_{-}}{2}} 
 + \mid \rho_2\mid\,e^{i\,\frac{\(\Omega_1-\Omega_2\)}{2}}\,e^{-\frac{z_{-}}{2}}\)\right.\nonumber\\
& +&\left. e^{\frac{z_{+}}{2}}e^{-i\,\delta_{-}}\left(
 \mid \rho_2\mid\, e^{i\,\frac{\(\Omega_1-\Omega_2\)}{2}}\, \, e^{\frac{z_{-}}{2}}\, e^{-i\,2\delta_{+}}
 +\mid \rho_1\mid\, e^{-i\,\frac{\(\Omega_1-\Omega_2\)}{2}}\, \,e^{-\frac{z_{-}}{2}}\, e^{i\,2\delta_{+}}\right)\right].
 \lab{calndef}
 \er

In this expression we use $\Delta$ defined by
\be
e^{\Delta}=\frac{\Lambda_{-}}{\Lambda_{+}}
=\frac{\(v_1-v_2\)^2+4\(\rho_1-\rho_2\)^2}{\(v_1-v_2\)^2+4\(\rho_1+\rho_2\)^2}
\lab{Deltadef}
\ee 
and the coordinates
\be
z_{+}\equiv X_1+X_2+\Delta \qquad \qquad \qquad z_{-}\equiv X_1-X_2
\lab{zidef}
\ee
with 
\be
X_i= \rho_i\,\(x-v_i\,t-x_i^{(0)}\)
\qquad\qquad 
\Omega_i =  \(\frac{v_i^2}{4}-\rho_i^2\)\,t- \frac{v_i}{2}\, x +\theta_i+\zeta_i
\qquad\qquad i=1,2
\lab{omegaxdef}
\ee
 where
\be
\delta_{\pm}={\rm ArcTan}\left[\frac{2\,\(\rho_1\pm \rho_2\)}{\(v_1-v_2\)}\right].
\lab{deltapmdef}
\ee

The quantities $\Omega_i$ are linear in $x$ and $t$, and so in the new coordinates $z_{\pm}$. We can therefore, separate the homogeneous dependence on $z_{\pm}$ by writing
\br
\frac{\Omega_1-\Omega_2}{2}-\delta_{+}&=& \Omega_{-} + c,
\\
\frac{\Omega_1+\Omega_2}{2}+\delta_{-}&=&\Omega_{+}+d,
\lab{cddef}
\er
where $\Omega_{\pm}$ are homogeneous in $z_{\pm}$, {\it i.e.} $\Omega_{\pm}=\beta_{\pm}^+\,z_{+}+\beta_{\pm}^-\,z_{-}$, with $\beta_{\pm}^{\pm}$ being some constants depending on $v_i$ 
and $\rho_i$, $i=1,2$. Note that the constants $\zeta_i$ appearing in the expression of $\Omega_i$ in \rf{omegaxdef} are the phases of $z_1$ and $w_1$ given in \rf{zetaidef}, and so depend on $v_i$ and $\rho_i$, $i=1,2$. However, the constants $\theta_i$ also appearing in \rf{omegaxdef} are the phases of $a_{+}$ and $b_{+}$ given in \rf{thetadef}, and so are independent of $v_i$ and $\rho_i$. The constants $c$ and $d$ introduced in \rf{cddef} depend on $v_i$, $\rho_i$ and $x_i^{(0)}$ ($i=1,2$) but are linear in $\theta_1-\theta_2$ and $\theta_1+\theta_2$, respectively, and so can be traded for $\theta_i$, $i=1,2$, and be considered as  constants independent of $v_i$, $\rho_i$ and $x_i^{(0)}$. Thus, the two-soliton solution \rf{twobrightsoliton2} depends only on $8$ free parameters;  namely,  $v_i$, $\rho_i$, $x_i^{(0)}$ ($i=1,2$) $c$ and $d$, and can be written as 
\br
\psi_0=\frac{e^{-\frac{\Delta}{2}}}{\sqrt{\mid \eta\mid}} \, e^{-i\, \Omega_{+}}
\;\frac{{\hat {\cal N}}}{{\hat {\cal D}}} 
   \lab{twobrightsoliton3}
\er
where the overall phase $e^{-i\,d}$, has been absorbed using the symmetry \rf{phasetransf}, and where we have introduced 
\be
{\hat {\cal D}}= \cosh z_{+}+ e^{-\Delta}\,\cosh z_{-} - 16 \frac{\mid \rho_1\mid\,\mid \rho_2\mid}{\Lambda_{-}}\,\cos\left[2\, \(\Omega_{-}+c\)\right]
\lab{hatcalddef}
\ee
and
\br
{\hat {\cal N}}
&=& 
 e^{-\frac{z_{+}}{2}}e^{i\,\delta_{-}}\(\mid \rho_1\mid \,e^{-i\,\(\Omega_{-}+ c+\delta_{+}\)}\,e^{\frac{z_{-}}{2}} 
 + \mid \rho_2\mid\,e^{i\,\(\Omega_{-}+ c+\delta_{+}\)}\,e^{-\frac{z_{-}}{2}}\)\nonumber\\
& +& e^{\frac{z_{+}}{2}}e^{-i\,\delta_{-}}\left(
\mid \rho_1\mid\, e^{-i\,\(\Omega_{-}+ c-\delta_{+}\)}\, \,e^{-\frac{z_{-}}{2}}
+ \mid \rho_2\mid\, e^{i\,\(\Omega_{-}+ c-\delta_{+}\)}\, \, e^{\frac{z_{-}}{2}}
 \right).
 \lab{hatcalndef}
\er

We are now in a position to consider the parity transformation \rf{paritydef} relevant for the two-bright-soliton solution, {\it i.e.} 
\be
P: \qquad \qquad \(z_{+},z_{-}\)\rightarrow \(-z_{+},-z_{-}\)
\lab{paritydeftwosol}
\ee
which can be written  in terms of $x$ and $t$ as  
\be
P:\qquad \({\tilde x},{\tilde t}\)\rightarrow \(-{\tilde x},-{\tilde t}\) \qquad\qquad {\rm with} \qquad \quad{\tilde x}= x-x_{\Delta} \qquad \quad{\tilde t}=t-t_{\Delta}
\lab{paritydeftwosol2}
\ee
and
\br
x_{\Delta}&=& 
\frac{\Delta (\rho_1\, v_1 - \rho_2\, v_2) + 2\, \rho_1\, \rho_2 (v_2\, x_1^{(0)} - v_1\, x_2^{(0)})}{(2 \rho_1 \, \rho_2 (v_2 - v_1))}
\nonumber\\
t_{\Delta}&=&
\frac{\Delta (\rho_1 - \rho_2) + 2 \rho_1\, \rho_2 (x_1^{(0)} - x_2^{(0)})}{(2 \rho_1 \, \rho_2 (v_2 - v_1))}.
\lab{tdeltazeroorder}
\er

Note that under this parity transformation $\Omega_{\pm}$ are odd  since they are linear and homogeneous in $z_{\pm}$. Therefore, if 
\be
c= n\,\frac{\pi}{2},\qquad\qquad\qquad n \in \IZ
\lab{nicecchoice}
\ee
 the term $\cos\left[2\, \(\Omega_{-}+c\)\right]$, in  ${\hat {\cal D}}$, given in \rf{hatcalddef}, is invariant under the parity $P$. Consequently, ${\hat {\cal D}}$ is even under $P$
\be
P\( {\hat {\cal D}}\)={\hat {\cal D}}.
\ee
In addition, when $c$ satisfies \rf{nicecchoice}, as one can check,  
\be
P\({\hat {\cal N}}\)= \(-1\)^n\, {\hat {\cal N}}^*.
\ee
Thus, the two-bright-soliton solution \rf{twobrightsoliton3} satisfies
\be
P\(\psi_0\)= \(-1\)^n\,\psi_0^*
\ee
In terms of the fields $R$ and $\vp$ introduced in \rf{rphidef}, {\it i.e.} for $\psi_0=\sqrt{R_0}\, e^{i\frac{\vp_0}{2}}$, we find
 that
\be
P: \qquad\qquad R_0\rightarrow R_0\; ; \qquad \qquad\qquad 
\vp_0 \rightarrow -\vp_0 + 2\,\pi\,n
\lab{r0phi0transformunderp}
\ee
which is what we have assumed in \rf{nicezeroordersol}.

\section{Numerical support}
\label{sec:nsupport}
\setcounter{equation}{0}

In this section we present some numerical results which support the claims  we have made in the preceding sections. 

Our results concern the NLS model and its deformation discussed in the last section {\it i.e.} with the potential of the form \rf{pert}. In our numerical studies we used a fixed lattice of 5001 points with time evolution calculated using the 4th order Runge-Kutta method. The lattice step was taken to be  dx=0.01 (and sometimes 0.05 or 0.1) and the time step used was dt=0.00005. We used both fixed and absorbing boundary conditions (to avoid any reflections from the boundaries) but as our field configurations were always very localised in the main section of the lattice the results did not depend on the boundary conditions (we only 
considered the evolution of the solitons when they were still some 
distance away from these boundaries.

\subsection{The NLS model}

Let us first present some of our results for the NLS model ({\it i.e.} for $\epsilon=0$).
The one soliton solution, \rf{brightsol}, for the case of $v=0$, is shown in fig. 1. In this figure we present a plot of $\vert \psi\vert^2$ as a function of $x$.

\begin{figure}
\begin{center}
\includegraphics[width=0.3\textwidth]{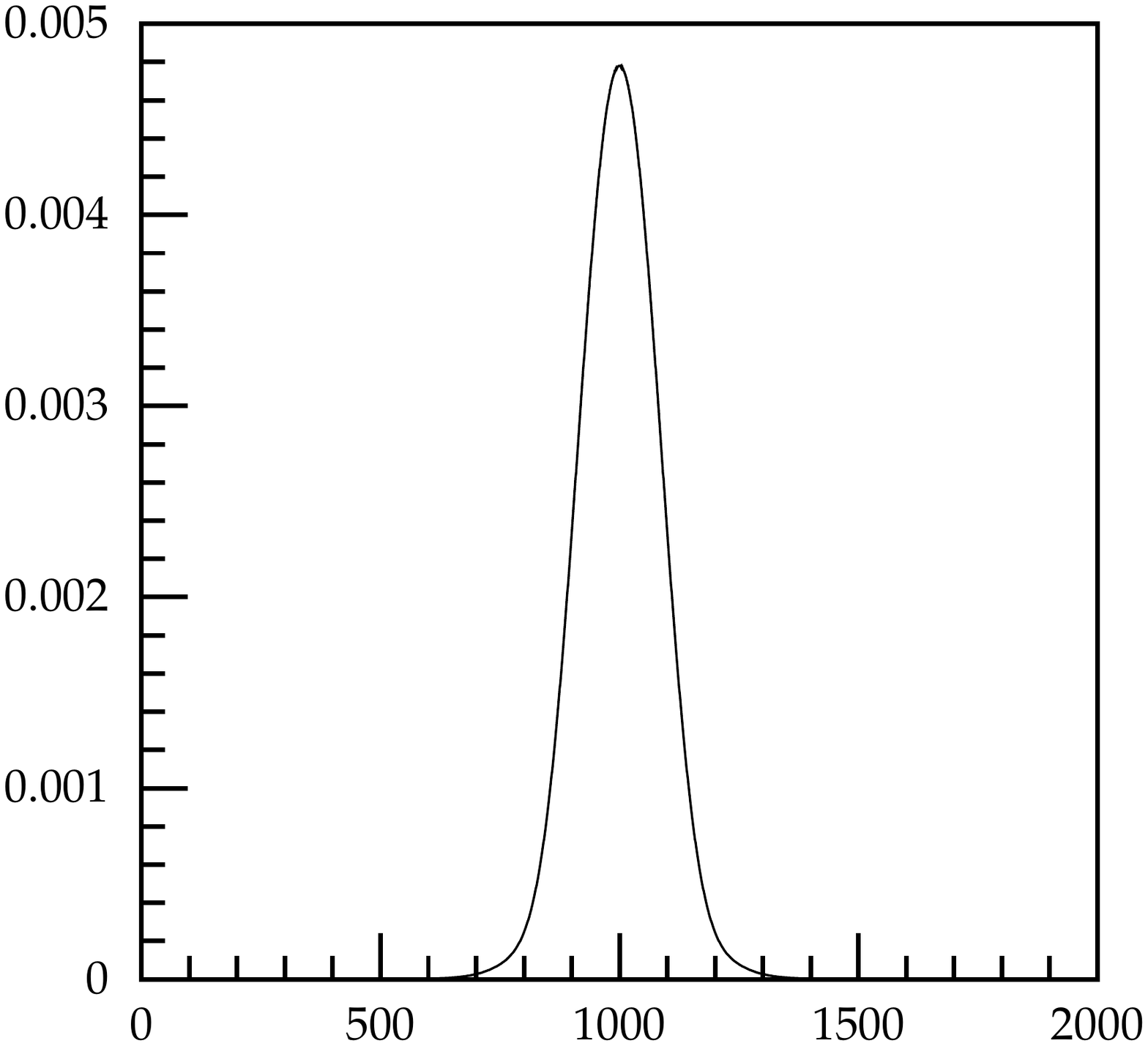}   
\end{center}
\hskip 0.5cm Fig. 1 Plot of $\mid \psi\mid^2$ against $x$ for the one-soliton solution of the unperturbed NLS model.
\end{figure}

Next we have looked at several field configurations involving two solitons ({\it i.e.}) given by \rf{twobrightsoliton3}. In this case we varied the values of the free parameter $c$. As mentioned in the last section, when $c$ is an integer multiple of $\frac{\pi}{2}$ the two-soliton field configuration \rf{twobrightsoliton3} is an eigenfunction of $P$ in the sense of \rf{r0phi0transformunderp}. We have followed the field configuration given by \rf{twobrightsoliton3} and have used this field configuration as an initial condition for a full simulation and the results were virtually 
indistinguishable from each other. This has provided a test of our numerical procedure. In fig 2 (a,b and c) we present plots of the position of one soliton as a function of time for 3 different values of $c$, namely $c=0$, $c=0.7$ and $c=1.4$. The position was determined by looking at the maxima of the energy density and the trajectory of the other soliton was symmetrically placed and to the right of the one that is plotted. We notice a slight dependence on the values of $c$.

\begin{figure}
    \begin{center}
\includegraphics[angle=0,width=0.3\textwidth]{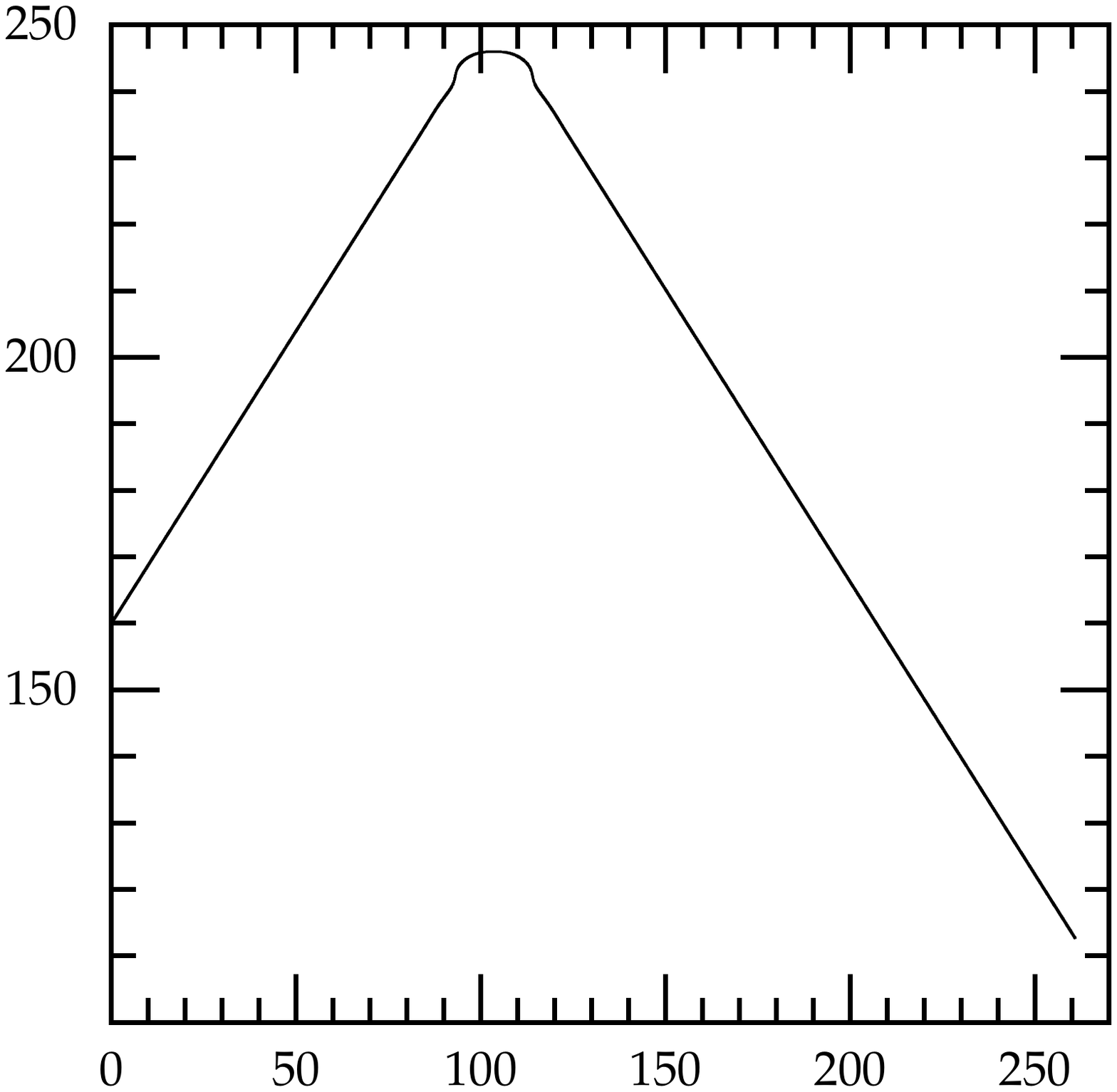}
	\includegraphics[angle=0,width=0.3 \textwidth]{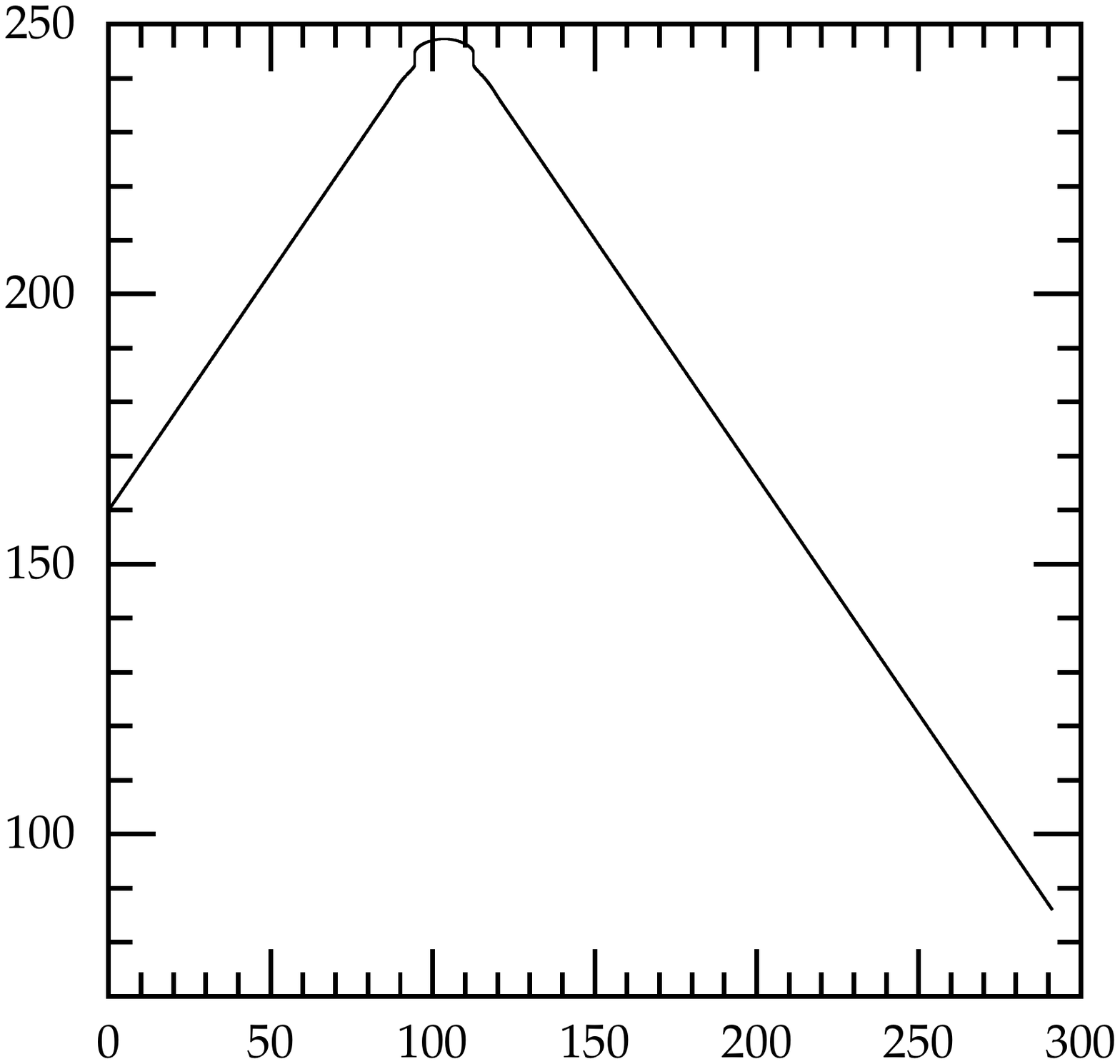}
	\includegraphics[angle=0,width=0.3 \textwidth]{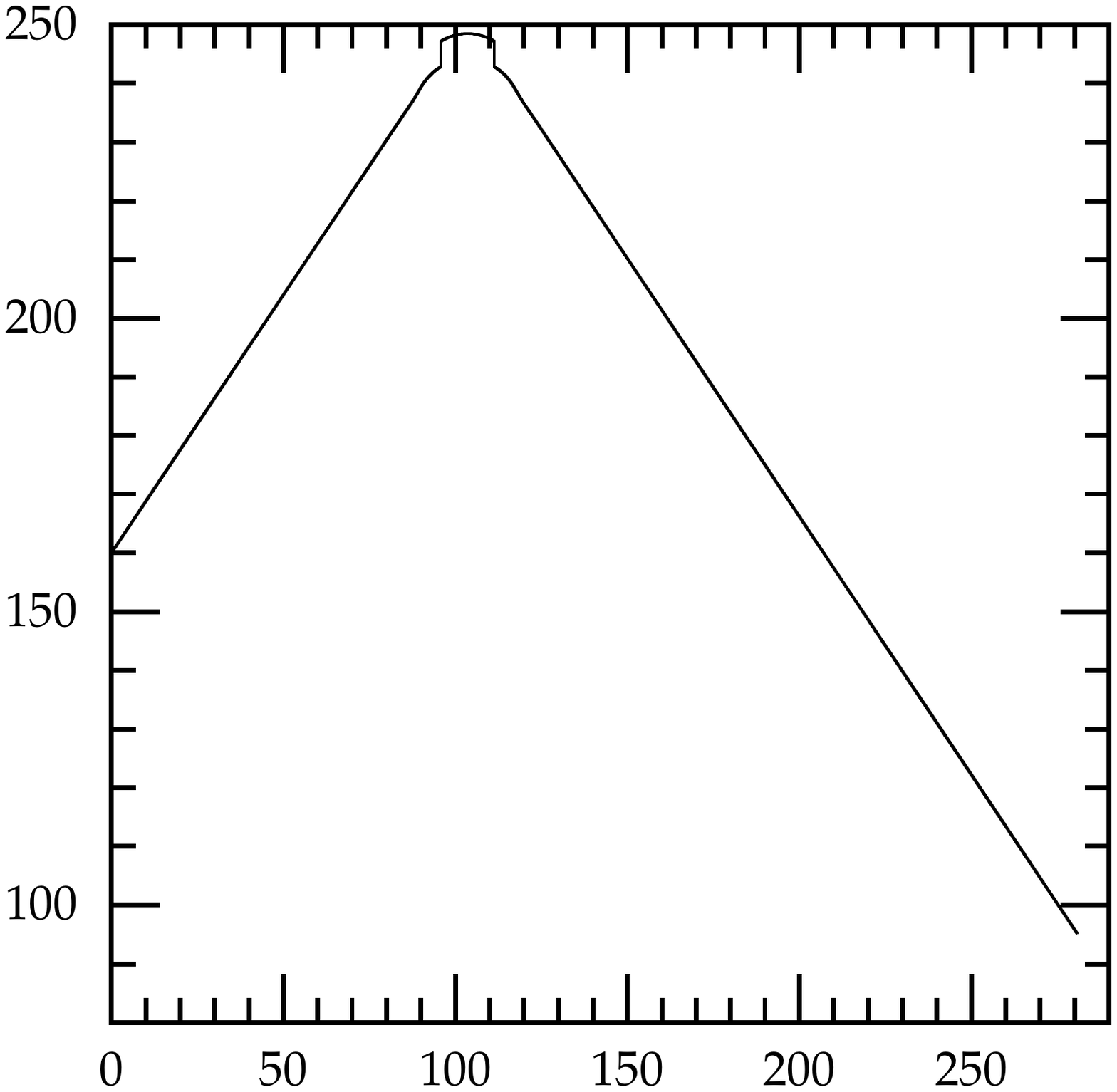}
		\end{center}
{Fig. 2  Trajectories of two Solitons at $v=0.4$  ($\epsilon=0$)}
a) $c=0$, b) $c=0.7$ and c) $c=1.4$
\end{figure}

\begin{figure}
    \begin{center}
\includegraphics[angle=0,width=0.3\textwidth]{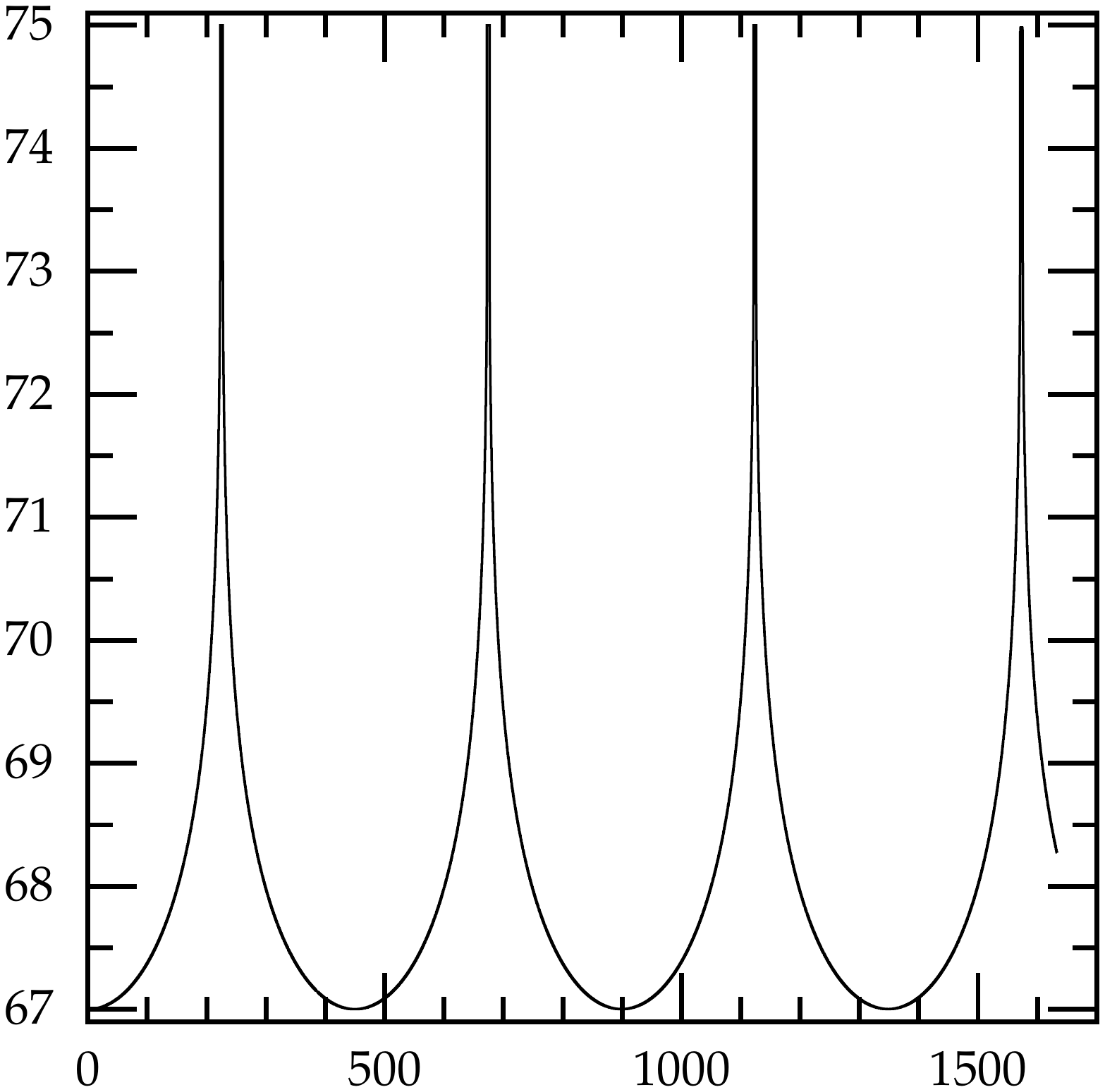}
	\includegraphics[angle=0,width=0.3 \textwidth]{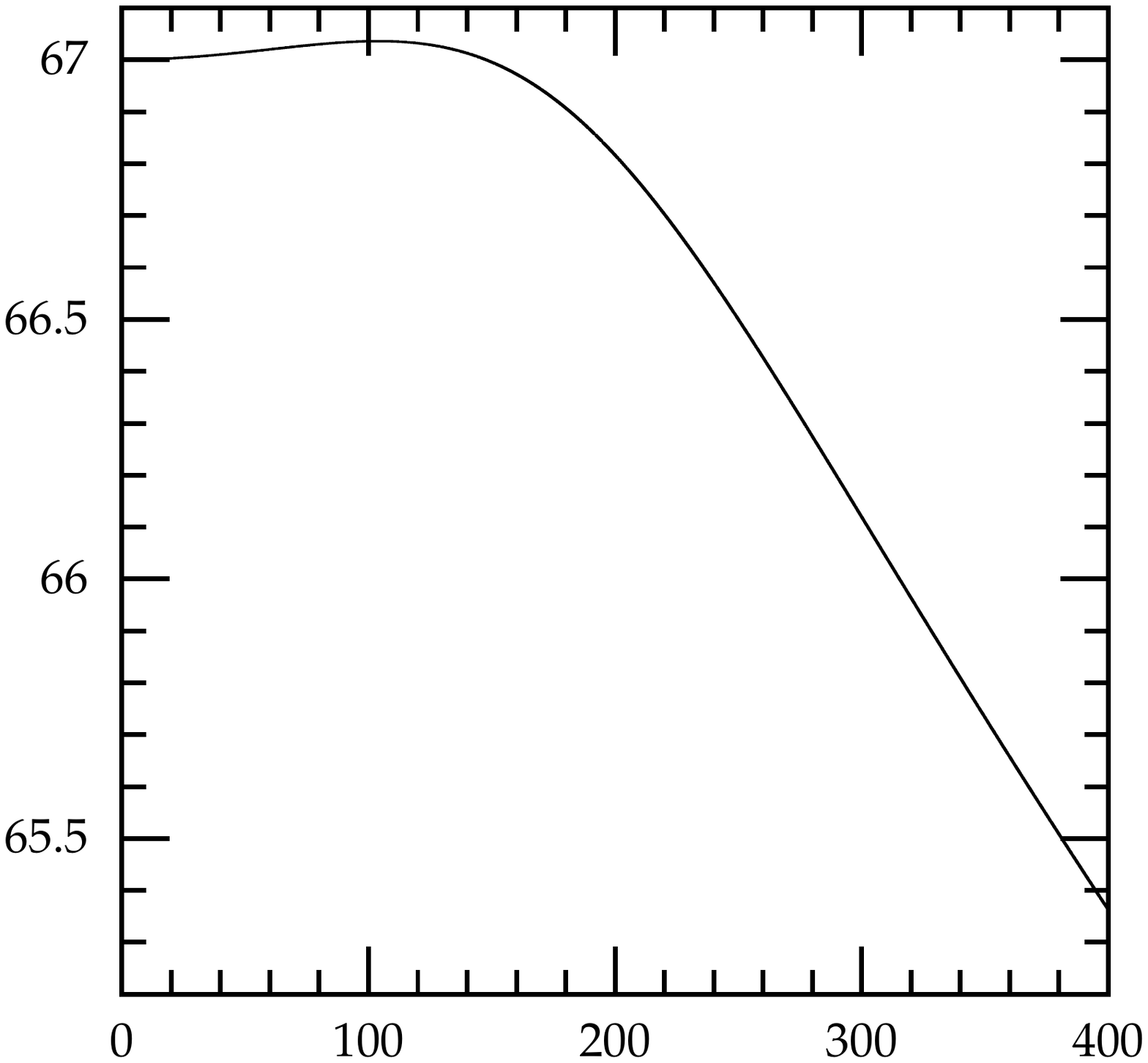}
	\includegraphics[angle=0,width=0.3 \textwidth]{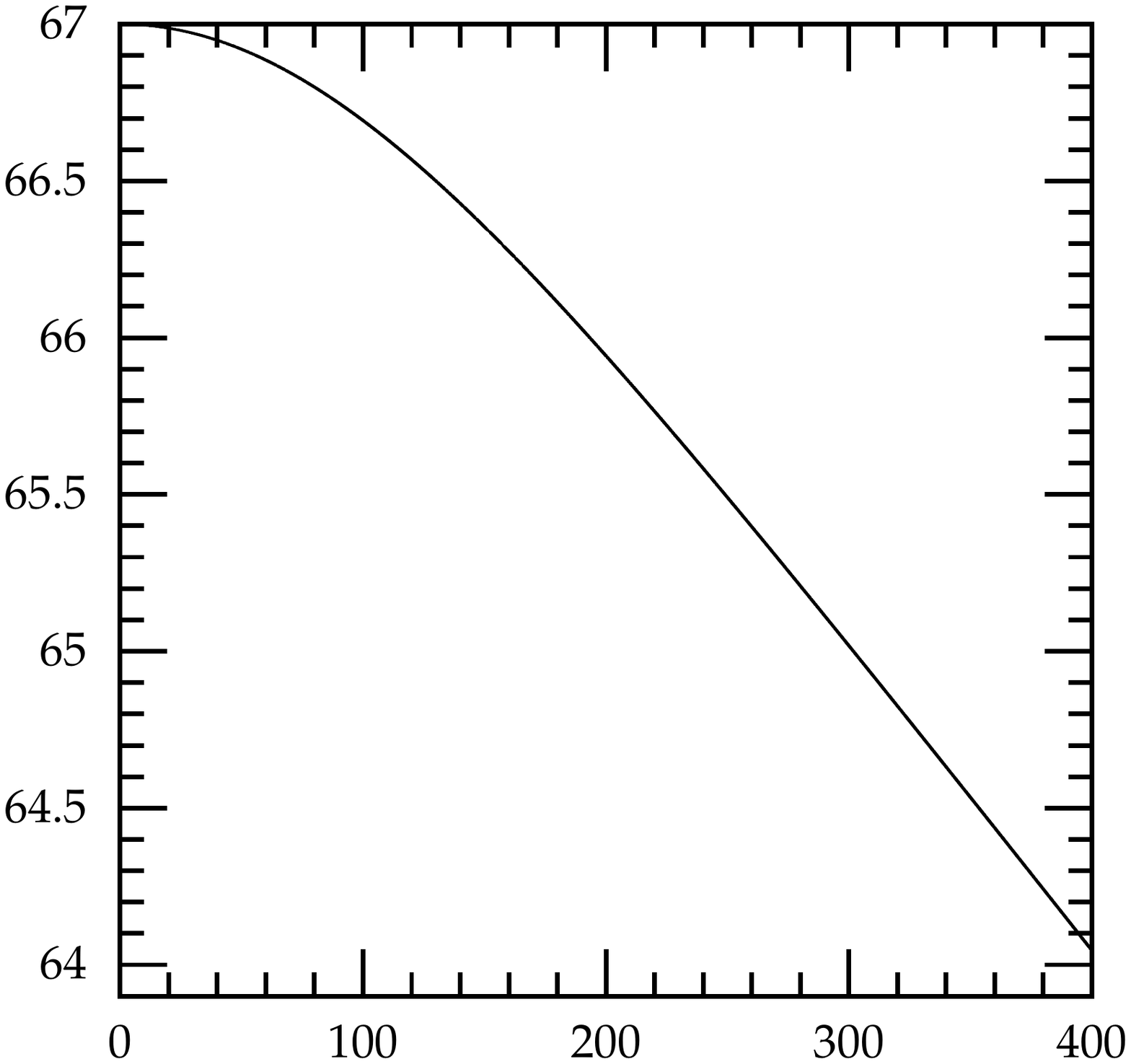}
		\end{center}
{Fig. 3  Trajectories of two solitons at rest ($\epsilon=0$)}
a) $c=0$, b) $c=0.7$ and c) $c=1.4$
\end{figure}

\begin{figure}
    \begin{center}
\includegraphics[angle=0,width=0.2\textwidth]{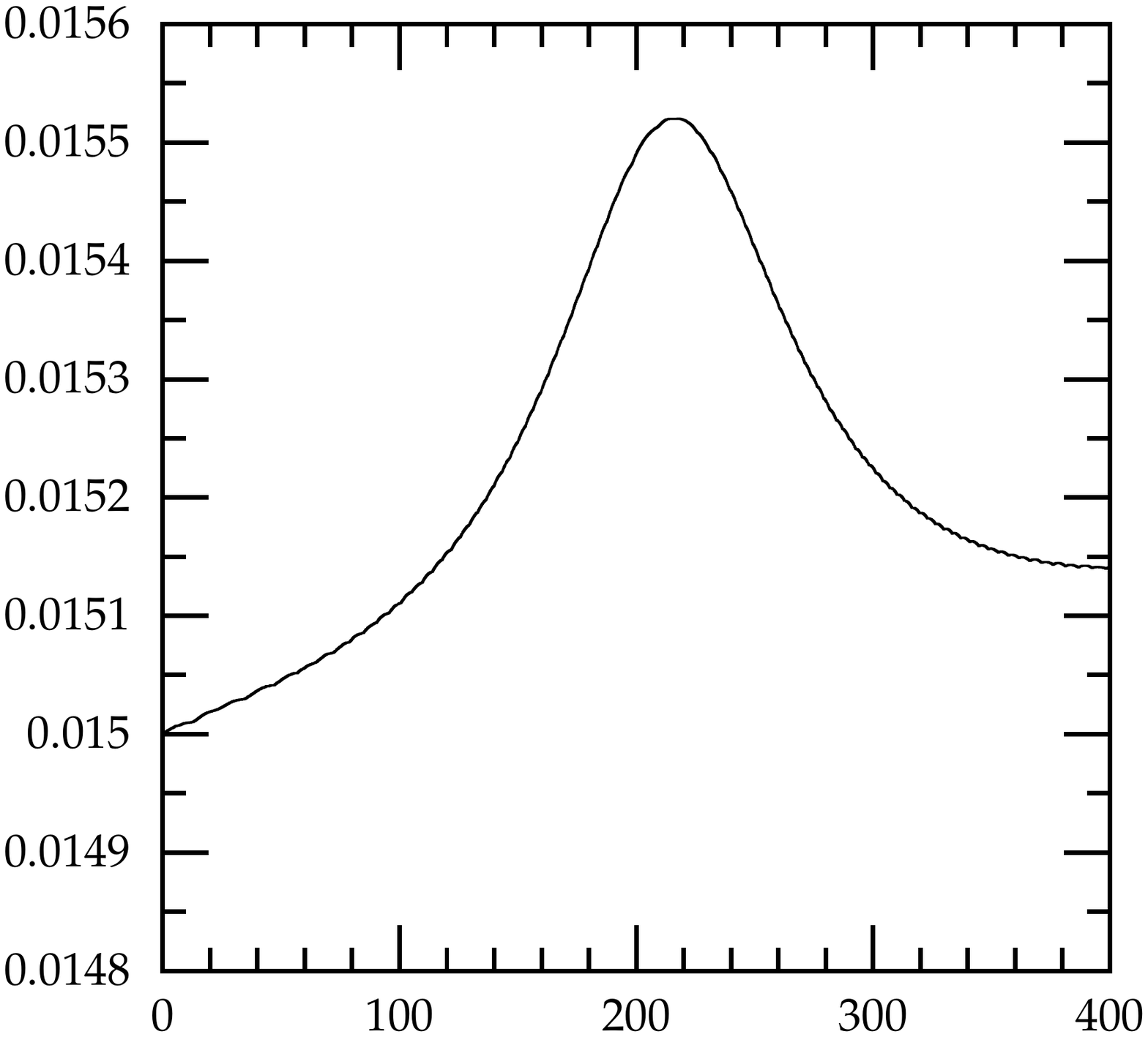}
	\includegraphics[angle=0,width=0.2 \textwidth]{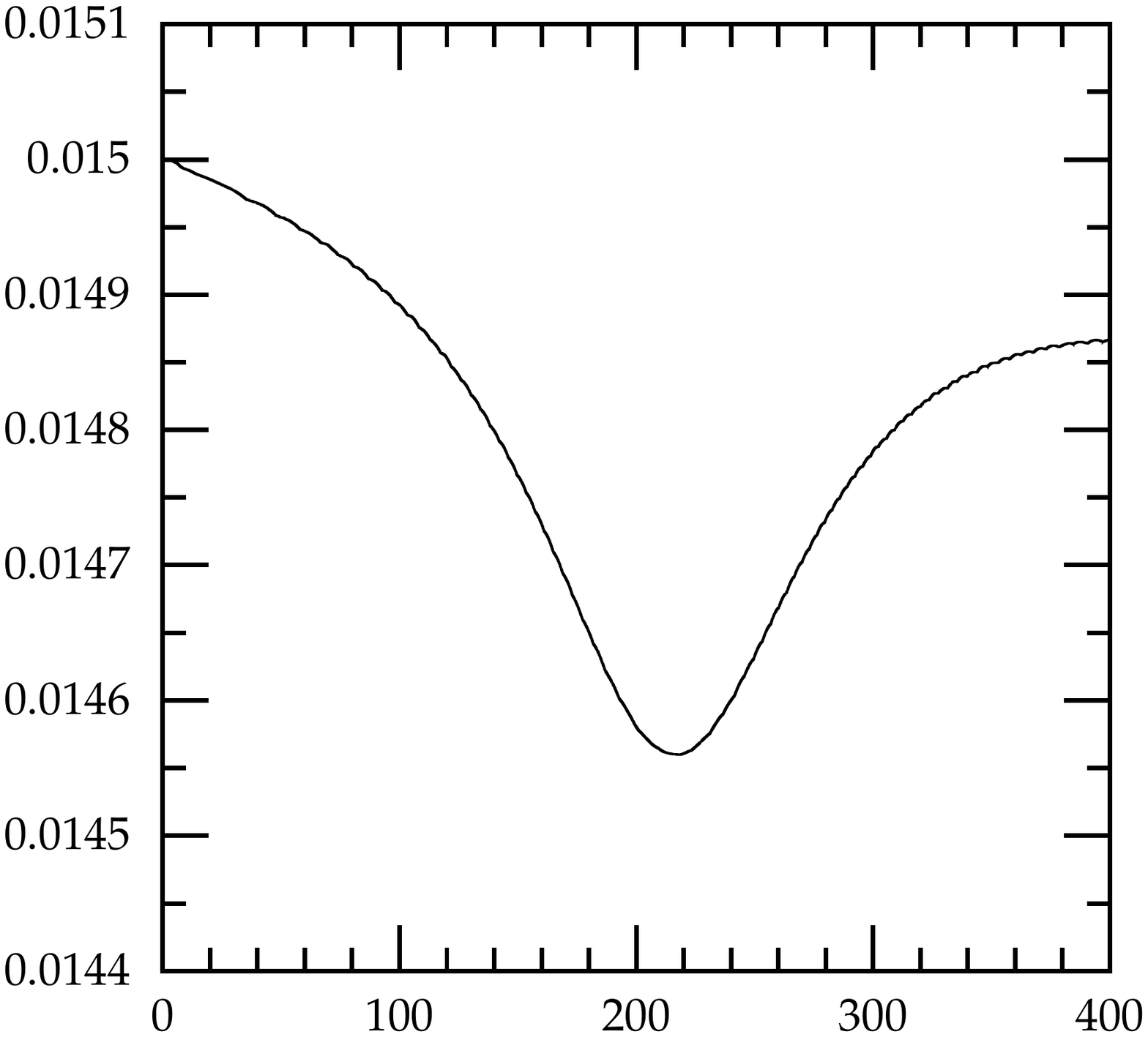}
	\includegraphics[angle=0,width=0.2 \textwidth]{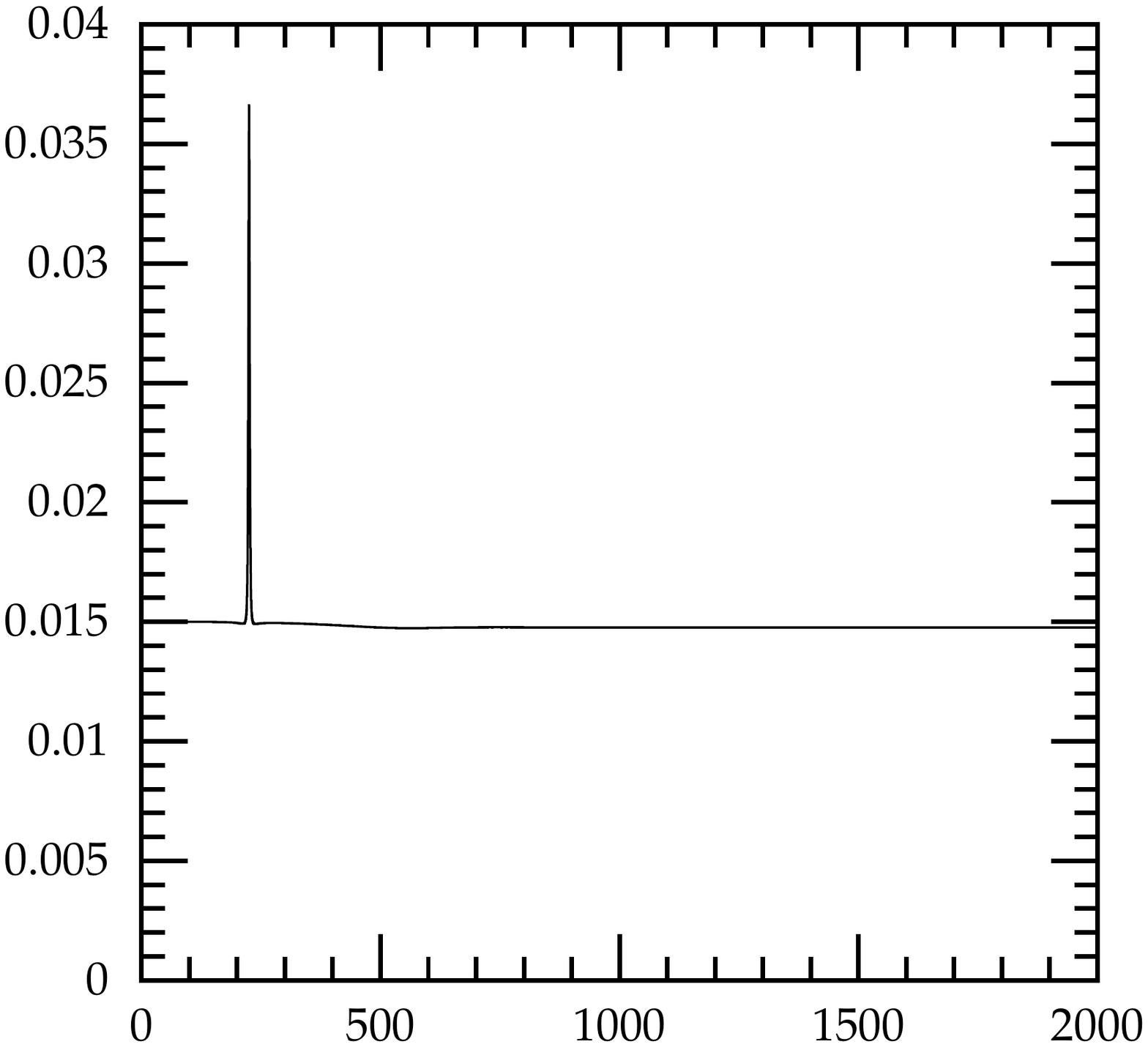}
\includegraphics[angle=0,width=0.2 \textwidth]{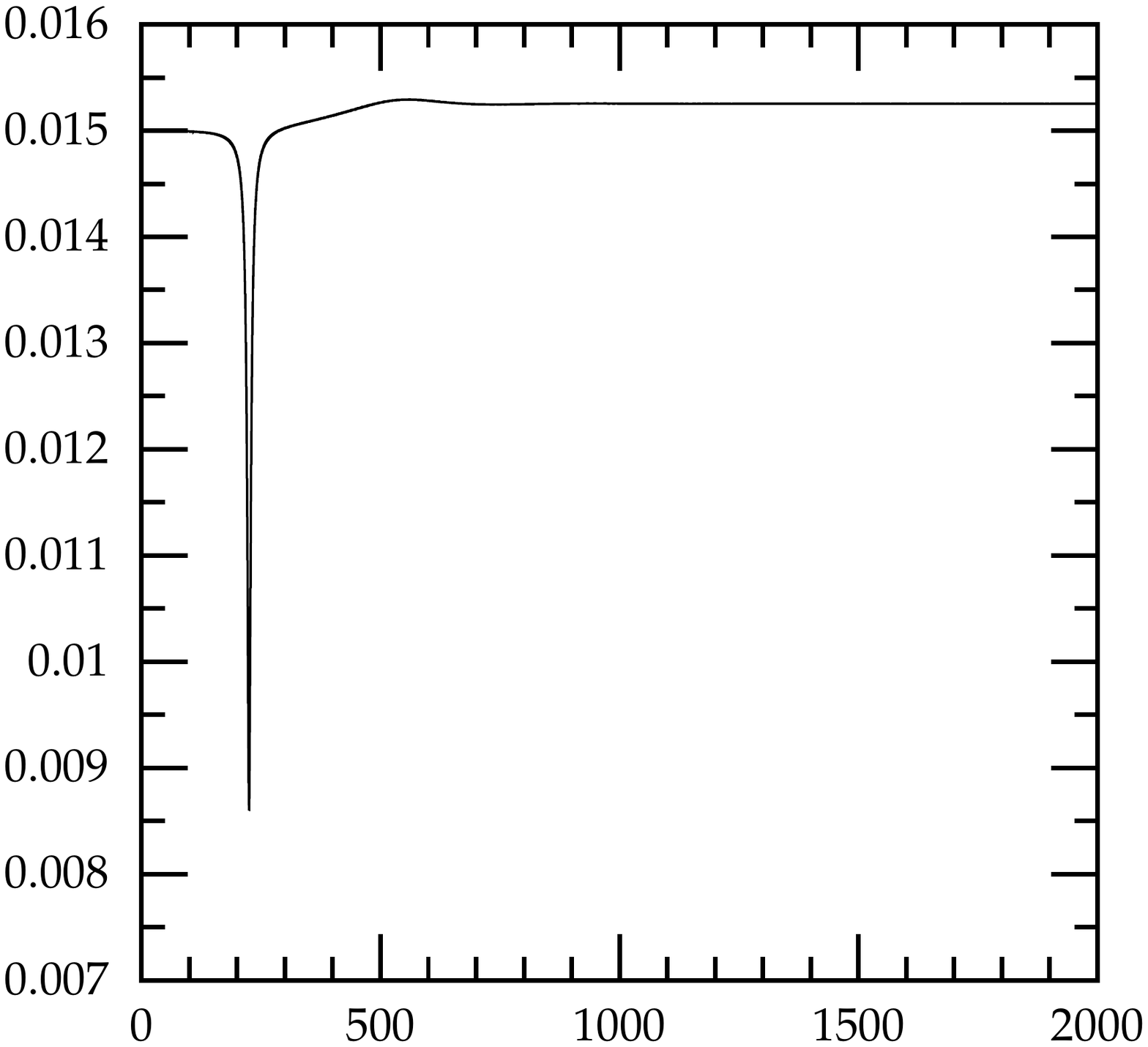}
		\end{center}
{Fig. 4  Heights of the solitons originaly at rest ($\epsilon=0$)}
a) b)  $c=0.3$ and c) and d) $c=0.01$
\end{figure}

The existence of multisoliton solutions does not directly describe the
forces between the solitons. Of course, one can deduce them by analysing in detail the time dependence of their positions {\it etc}. Another way to proceed involves putting two solitons at rest, not too close (not to deform them)  and not too far away (so that they do interact) and see what will happen.

We have performed such a study and in fig 3 we present similar trajectories to those shown in fig 2, for 3 values of the relative phase between them (equivalent to $c$). We see that at $c=0$ the solitons attract, at 
$c=0.7$ the forces are quite complicated resulting in a rather complicated trajectories and for $c=1.4$ they repel.  However, the parameter $c$ has also another role and this
is associated with the heights of the solitons. When $c=0$ both solitons, when they move towards each other, stay of the same size but as they come
towards each other they overlap and some appear to be taller. When $c\ne0$ the situation is more complicated. The nonzero value of $c$ breaks the symmetry and so one soliton 
tends to grow the other to decrease in size. For this to happen they have to interact and so be close enough and so the two effects (both of them growing and one of them growing
and the other one getting smaller) produce a more complicated pattern of their sizes and, in part, is responsible for their repulsion and never being able to come very close to each other. Hence the effect of them overlapping is very small.
In fig 4. we present the time dependence of the heights of the solitons for the cases of $c=0.3$ and $c=0.01$. The first two pictures (from the left) show the time dependence of the heights of the two solitons 
for $c=0.3$, and the other two for $c=0.01$.
The extremum of height seen in plots a) and b) corresponds to the case when the two solitons are at the closest distance from each other. In the plots c) and d) we note that 
after the scattering the values of the heights are slighty different. This may appear strange at first but the two solitons move with marginally different velocities after the scattering; this effect is induced  during the scattering by the nonzero value of $c$.

And what about the conserved charges? Well, the NLS model is integrable so that all anomalies vanish (and so all charges  are conserved).
In the next subsection we look at the same problems for $\epsilon\ne 0$
{\it i.e.} when the model is not integrable.

\subsection{Modified model; {\it i.e.} $\epsilon\ne0$}

Next we have considered the $\epsilon\ne0$ cases. This time we have only one soliton solution \rf{brightsoldeform} which is a simple deformation of the one soliton of the NLS model \rf{brightsol}. In fact, when one plots 
it  for small values of $\epsilon$ it is hard to see any difference.

As for $\epsilon\ne0$ the model is non-integrable and we do not have analytic expressions involving two solitons. Hence we can only use two one solitons some distance apart or use the two-soliton solutions of the NLS model ({\it i.e.} the expression for $\epsilon=0$) and take them as the initial conditions for our numerical simulations.

In fig 5  we present the plots of the trajectories of one soliton
(similar to fig 2) for  $\epsilon=0.06$  for 3 values of $c$. 
\begin{figure}
    \begin{center}
\includegraphics[angle=0,width=0.3\textwidth]{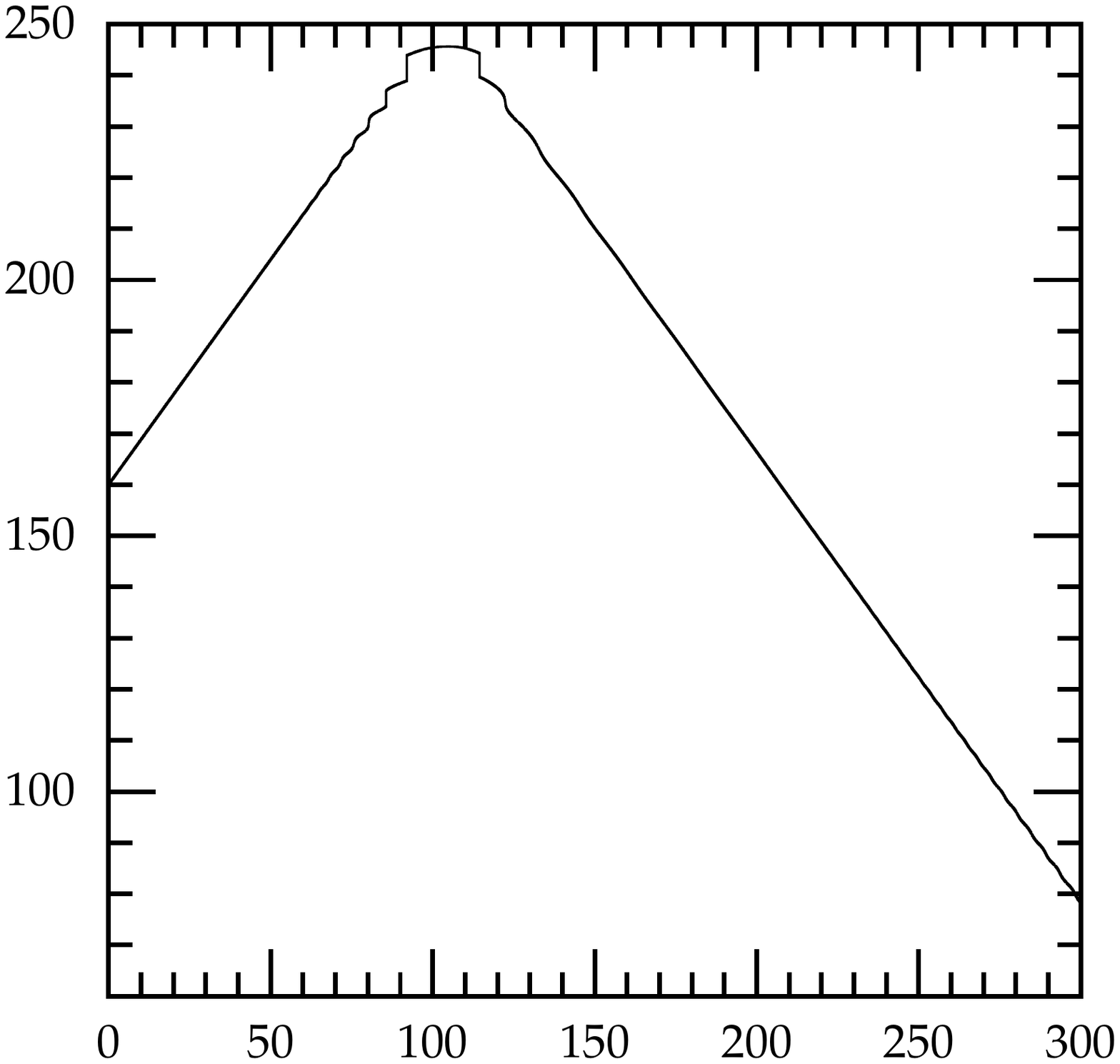}
	\includegraphics[angle=0,width=0.3 \textwidth]{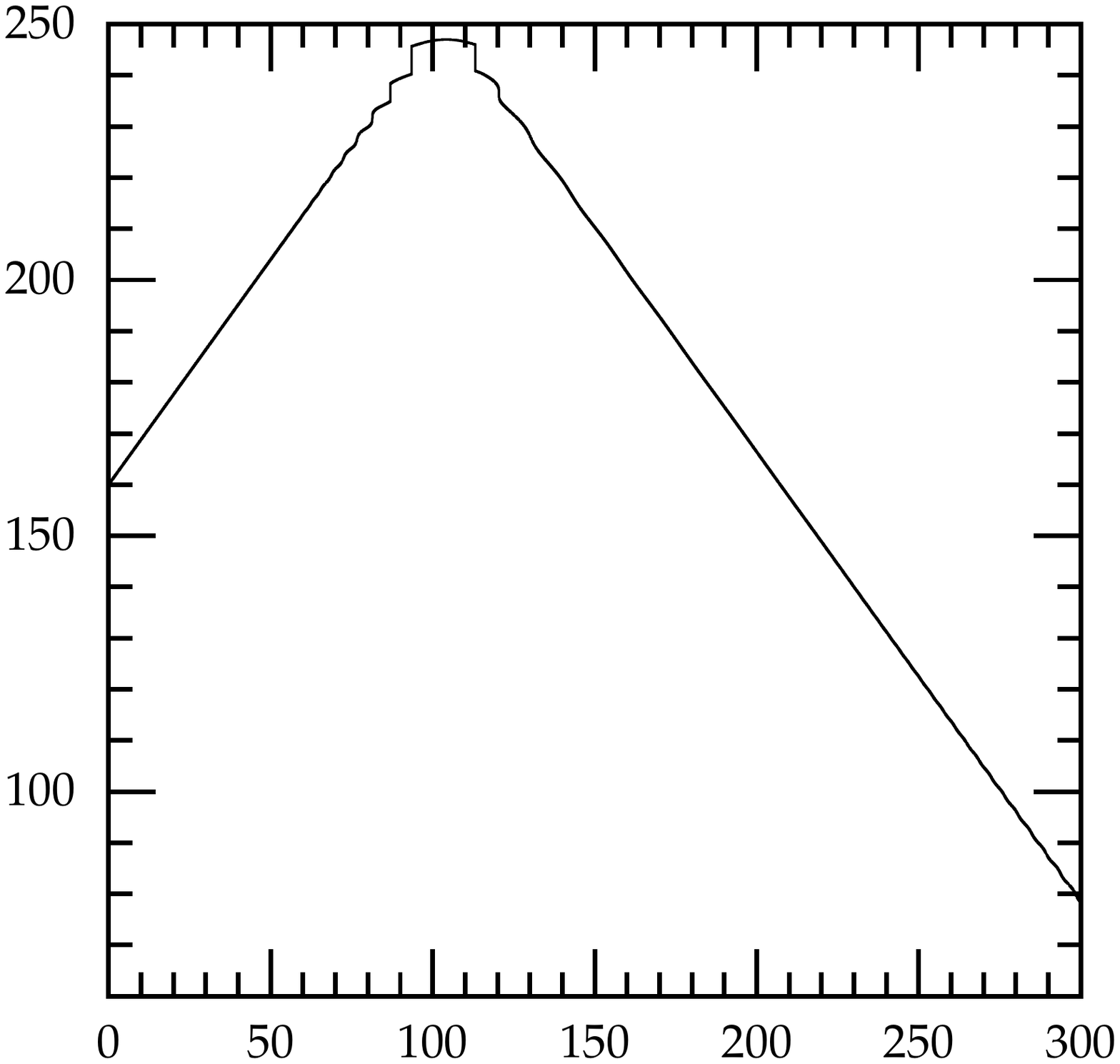}
	\includegraphics[angle=0,width=0.3 \textwidth]{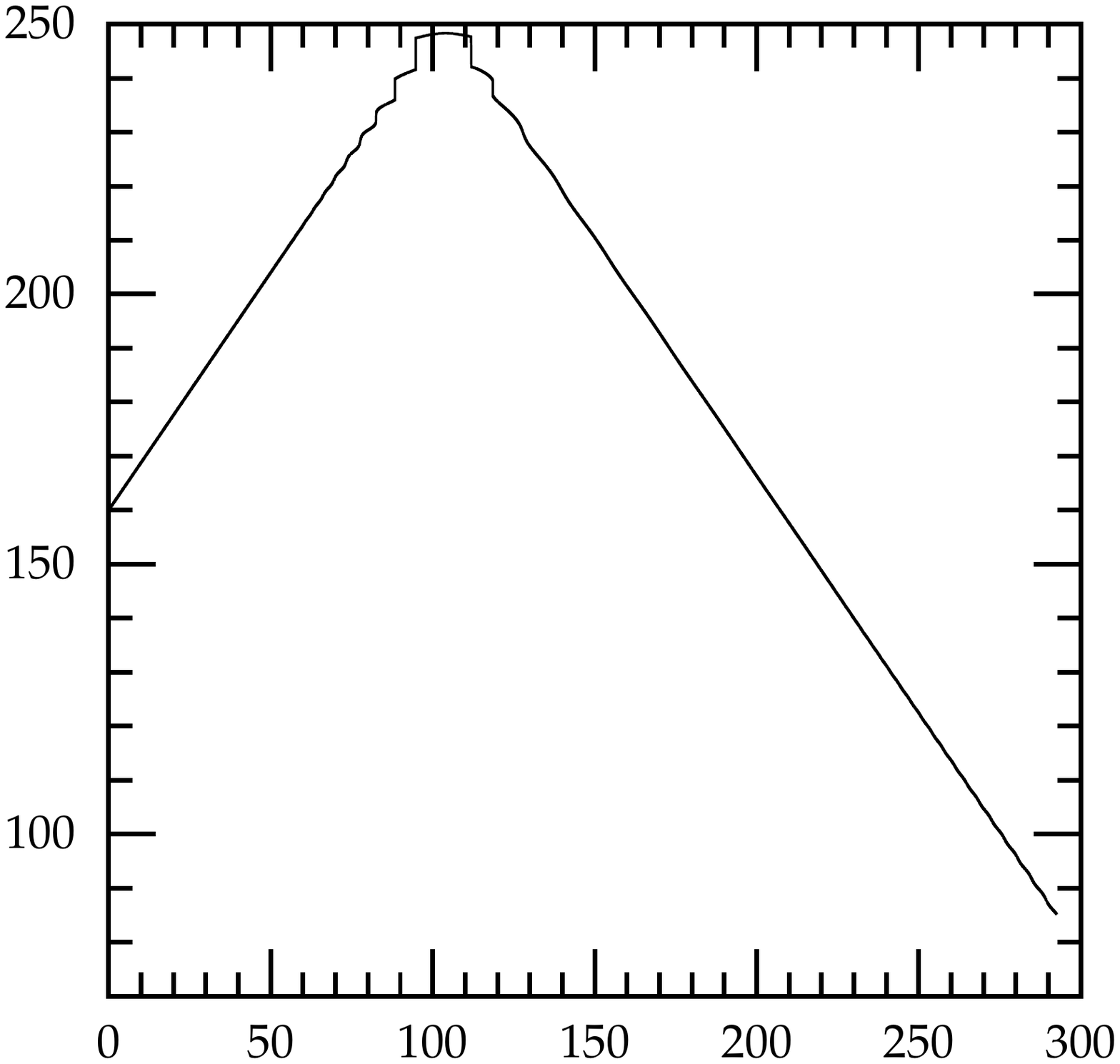}
		\end{center}
{Fig. 5   Trajectories of two solitons at $v=0.4$  ($\epsilon=0.06$)}
a) $c=0$, b) $c=0.7$ and c) $c=1.4$
\end{figure}
Looking at the trajectories and comparing them to those of the NLS model we see very little difference.
The same was observed for other values of $\epsilon$. In fact these trajectories were obtained by starting 
with initial configurations corresponding to the NLS model and then evolving them with $\epsilon\ne0$. We have also
looked at the effects of evolving the initial configurations described by two $\epsilon\ne0$ solitons `sewn' together.
The obtained trajectories were very similar. This is due to the fact that the solitons are well
localised and all the perturbations induced by taking non-exact expressions were very small.

Next we looked at two solitons at rest. In this case we have taken the expressions for two solitons corresponding 
to $\epsilon\ne0$ placed next to each other.
In fig 6 and 7 we present the plots similar to those of fig 3 for $\epsilon=0.06$ and for $\epsilon=-0.06$.

\begin{figure}
    \begin{center}
\includegraphics[angle=0,width=0.2\textwidth]{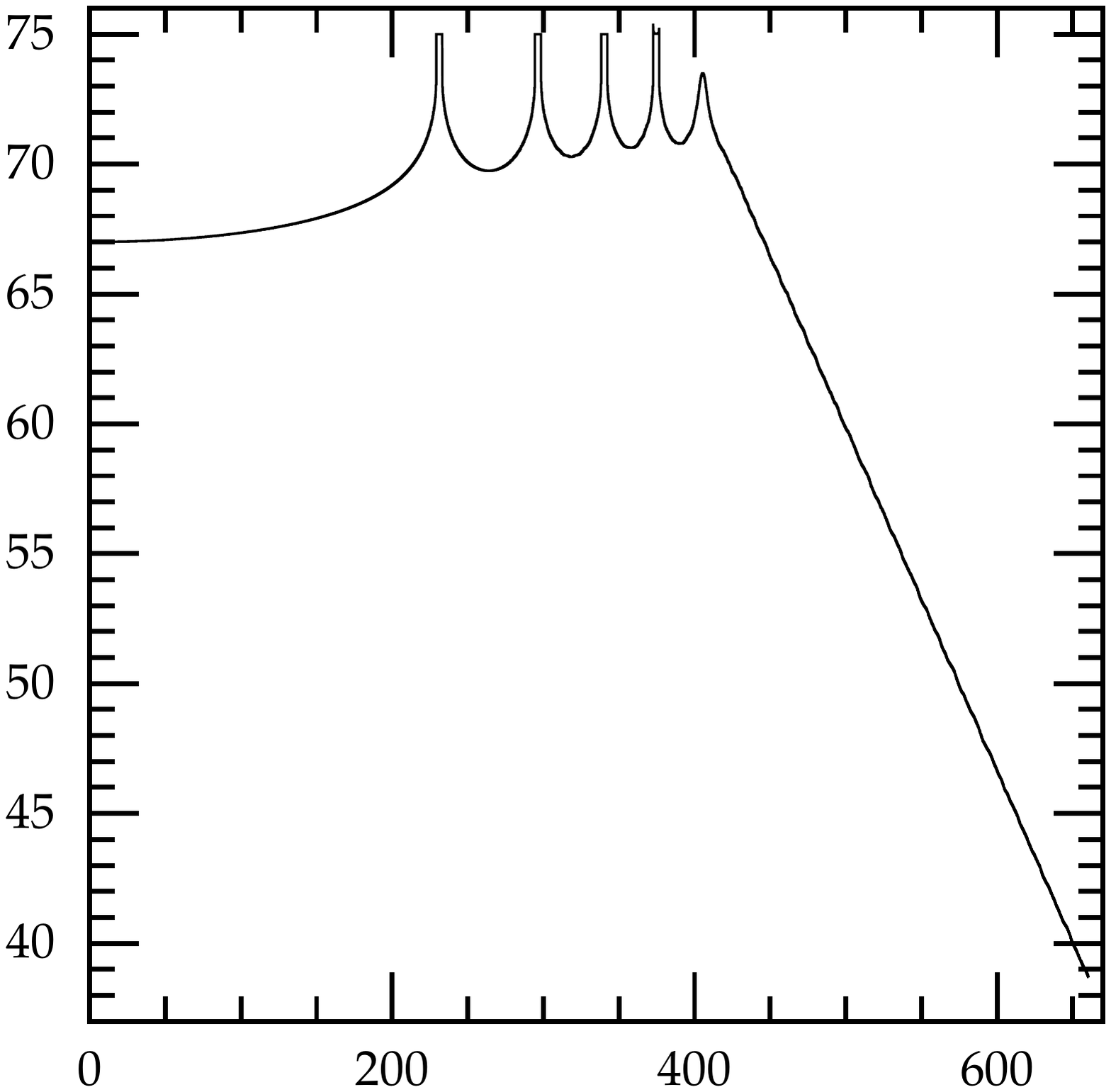}
\includegraphics[angle=0,width=0.2\textwidth]{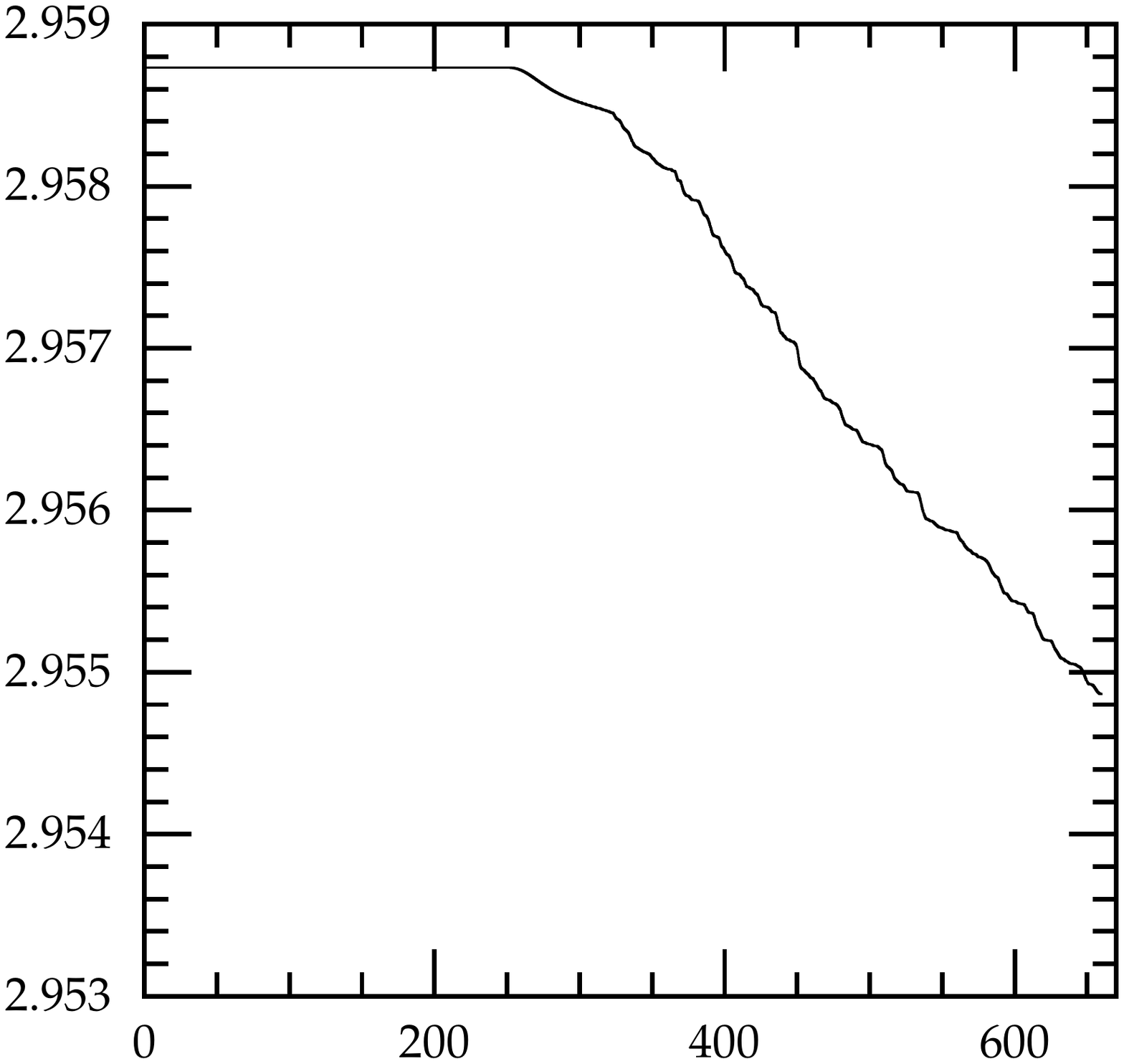}
	\includegraphics[angle=0,width=0.2 \textwidth]{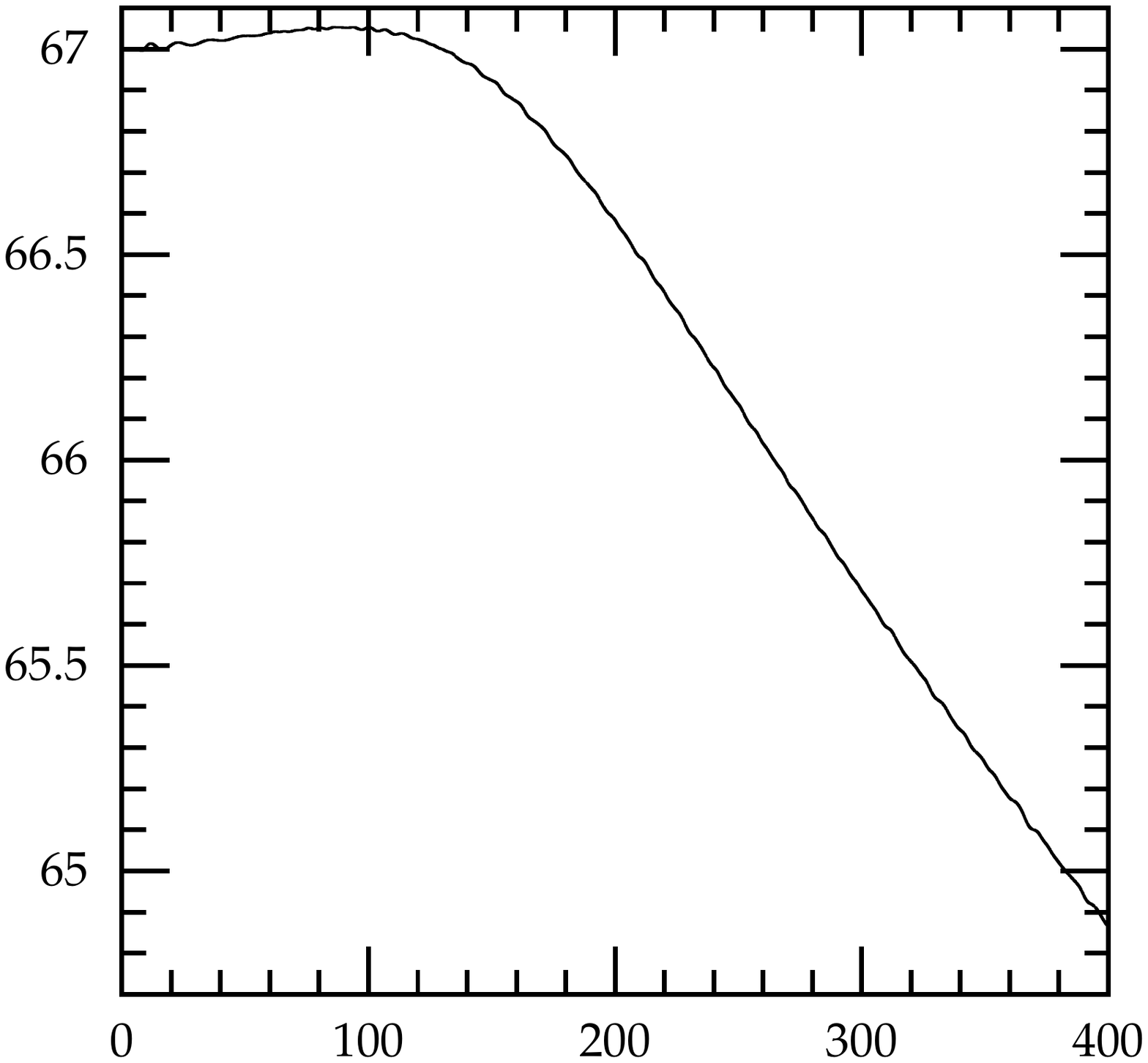}
	\includegraphics[angle=0,width=0.2 \textwidth]{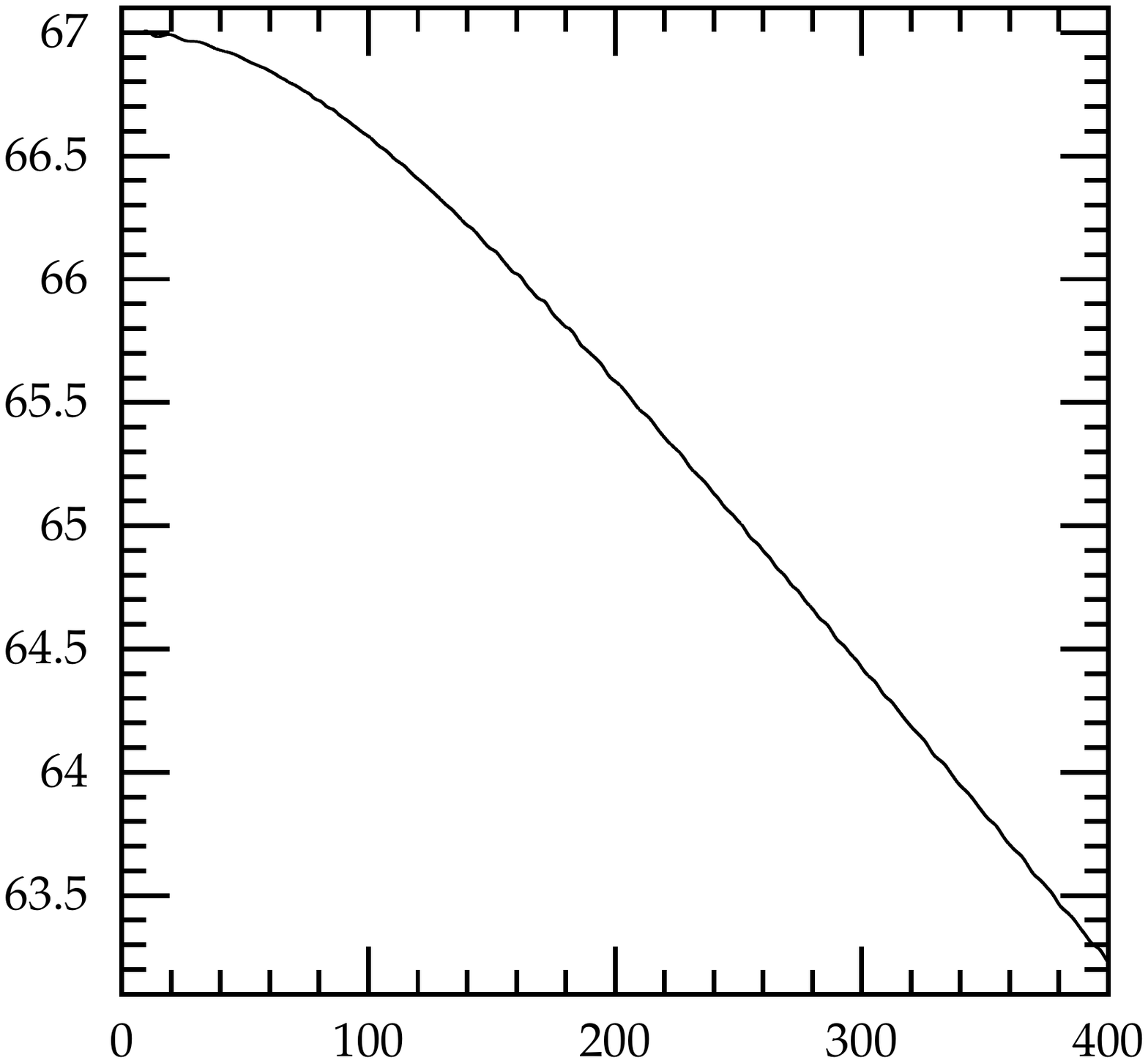}
		\end{center}
{Fig. 6  Trajectories (and the energy) of two solitons at rest ($\epsilon=0.06$)}; Trajectories:
a) $c=0$, c) $c=0.7$ and d) $c=1.4$ and b) the energy for $c=0$.
\end{figure}

\begin{figure}
    \begin{center}
\includegraphics[angle=0,width=0.2\textwidth]{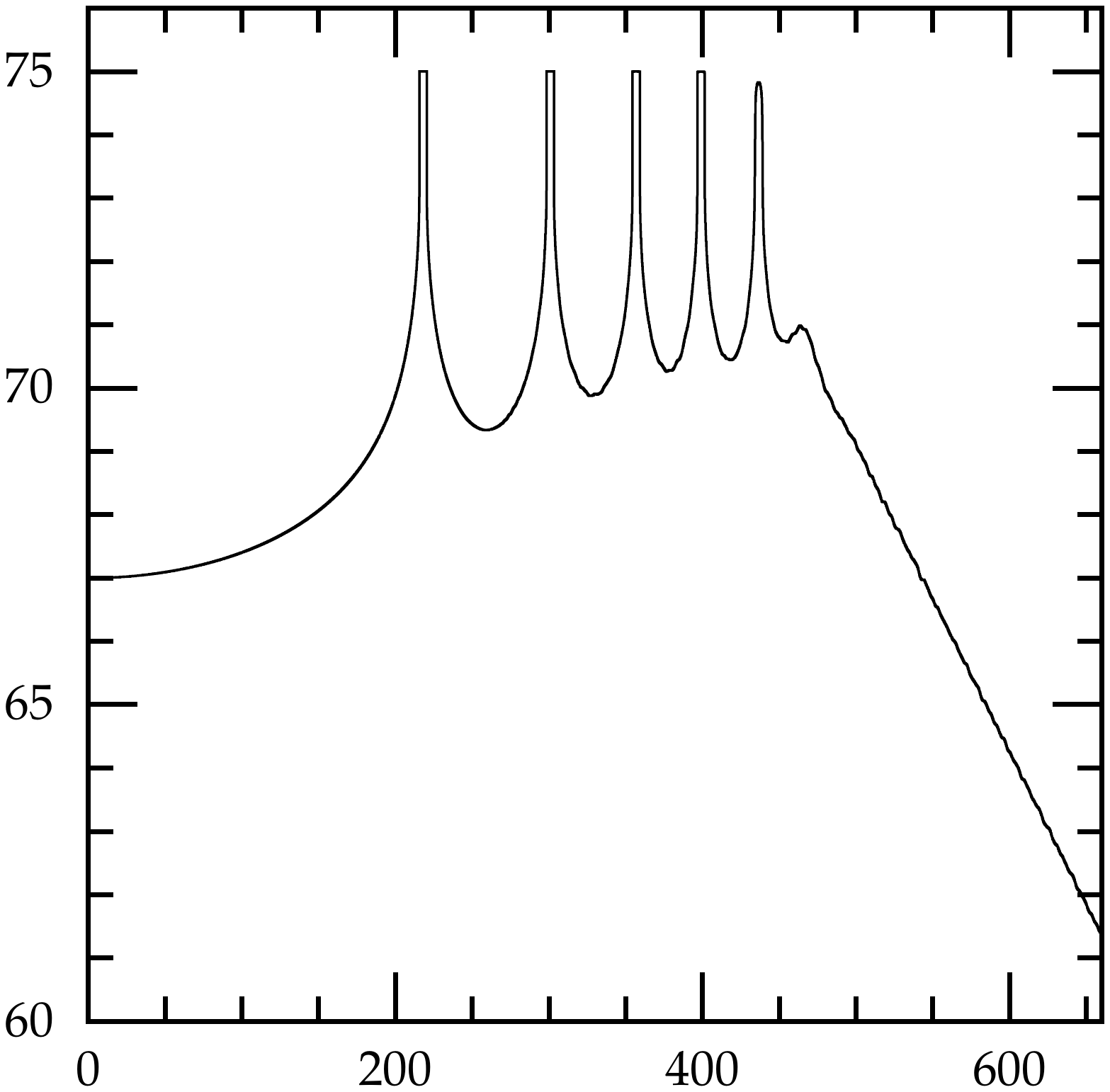}
\includegraphics[angle=0,width=0.2\textwidth]{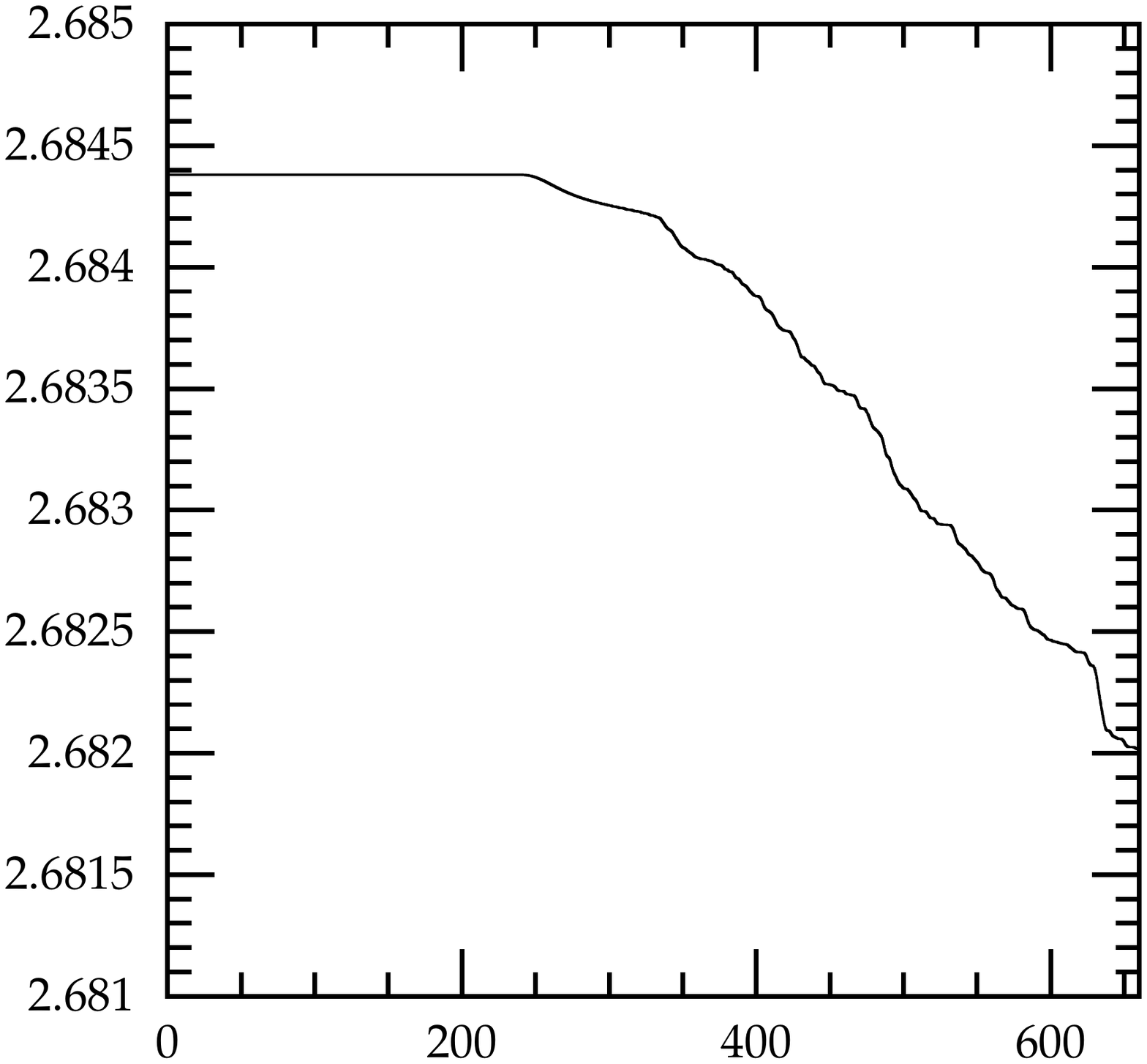}
	\includegraphics[angle=0,width=0.2 \textwidth]{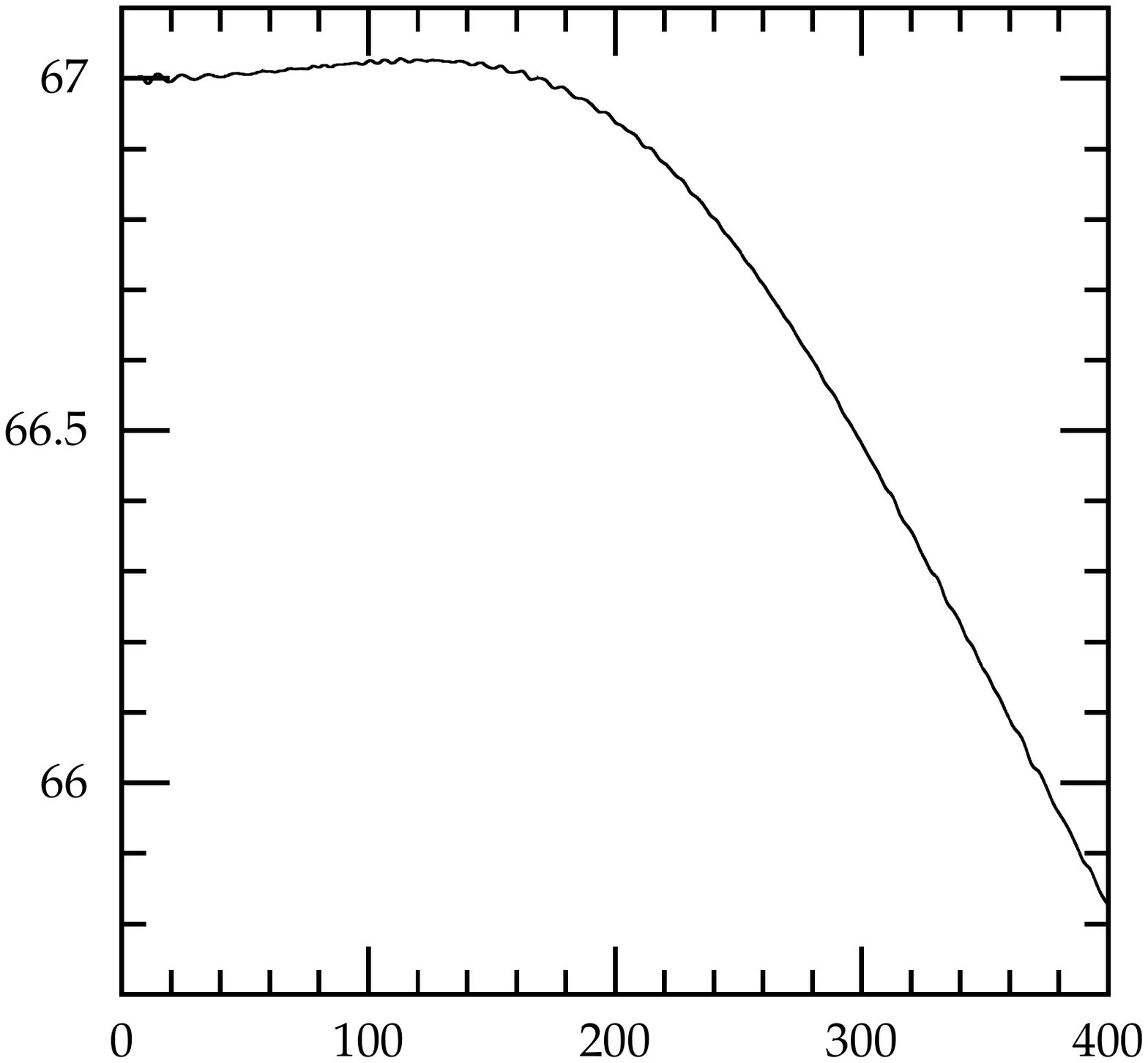}
	\includegraphics[angle=0,width=0.2 \textwidth]{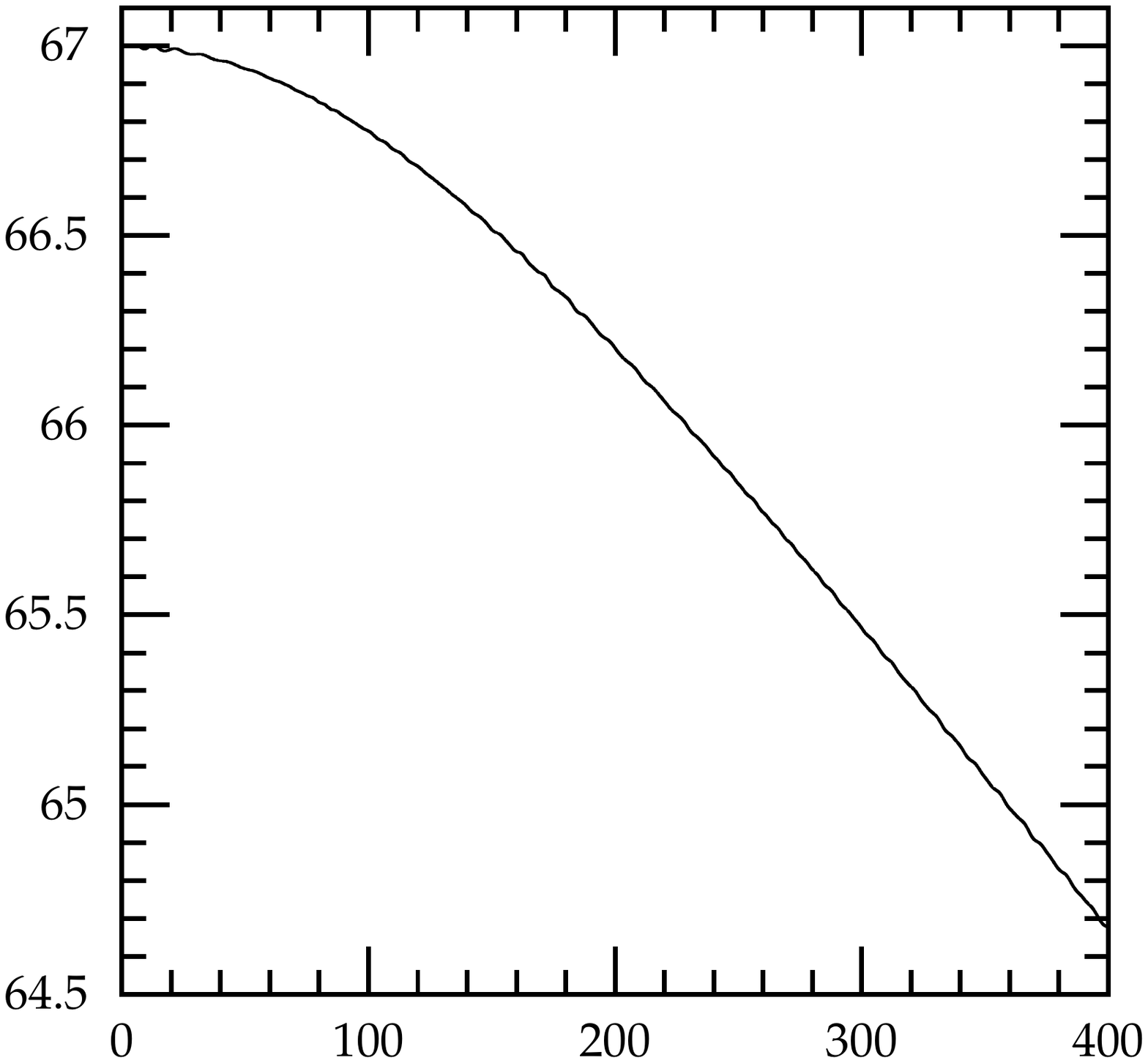}
		\end{center}
{Fig. 7  Trajectories (and the energy) of two solitons at rest ($\epsilon=-0.06$)}; Trajectories:
a) $c=0$, c) $c=0.7$ and d) $c=1.4$ and b) the energy for the case $c=0$.
\end{figure}


 Comparing these plots with those of fig 3 we see only little difference. The dependence on $c$ is very similar although the strength of the attraction (or repulsion) does appear to depend on $\epsilon$.
Clearly the overall
attraction (at least for $c=0$) increases with the increase of $\epsilon$. In addition, we note that for $c=0$, in the 
NLS case, the solitons oscillate around their point of attraction while for $\epsilon\ne0$ the amplitude
of their oscillation decreases (see fig  3a and compare with fig 6a and 7a). This suggests that for $\epsilon\ne0$ 
the solitons radiate a little and so come closer and closer to each other after each oscillation. This is indeed the case
as can be seen from the expressions of the total energy (for $c=0$, the energy is effectively conserved while for 
$c\ne0$ it decreases a little (see figures 6b) and 7b). After a while, however, during these interactions, they gradually change their height and then they split up,
repel and move away from each other. During this last part of the motion they move with slightly different velocities and so their sizes are also 
slightly different. In this their behaviour resembles the $c\ne0$ case; so we note that as $\epsilon\ne0$ the interaction between the solitons gradually induces
their behaviour as if $c$ were not 0.
In fig 8 we plot the heights of the two solitons observed in the scattering in the $\epsilon=0.06$, $c=0$ case.
Fig 8a corresponds to the case of the left hand one, and fig 8b - the right one.
\begin{figure}
    \begin{center}
\includegraphics[angle=0,width=0.2\textwidth]{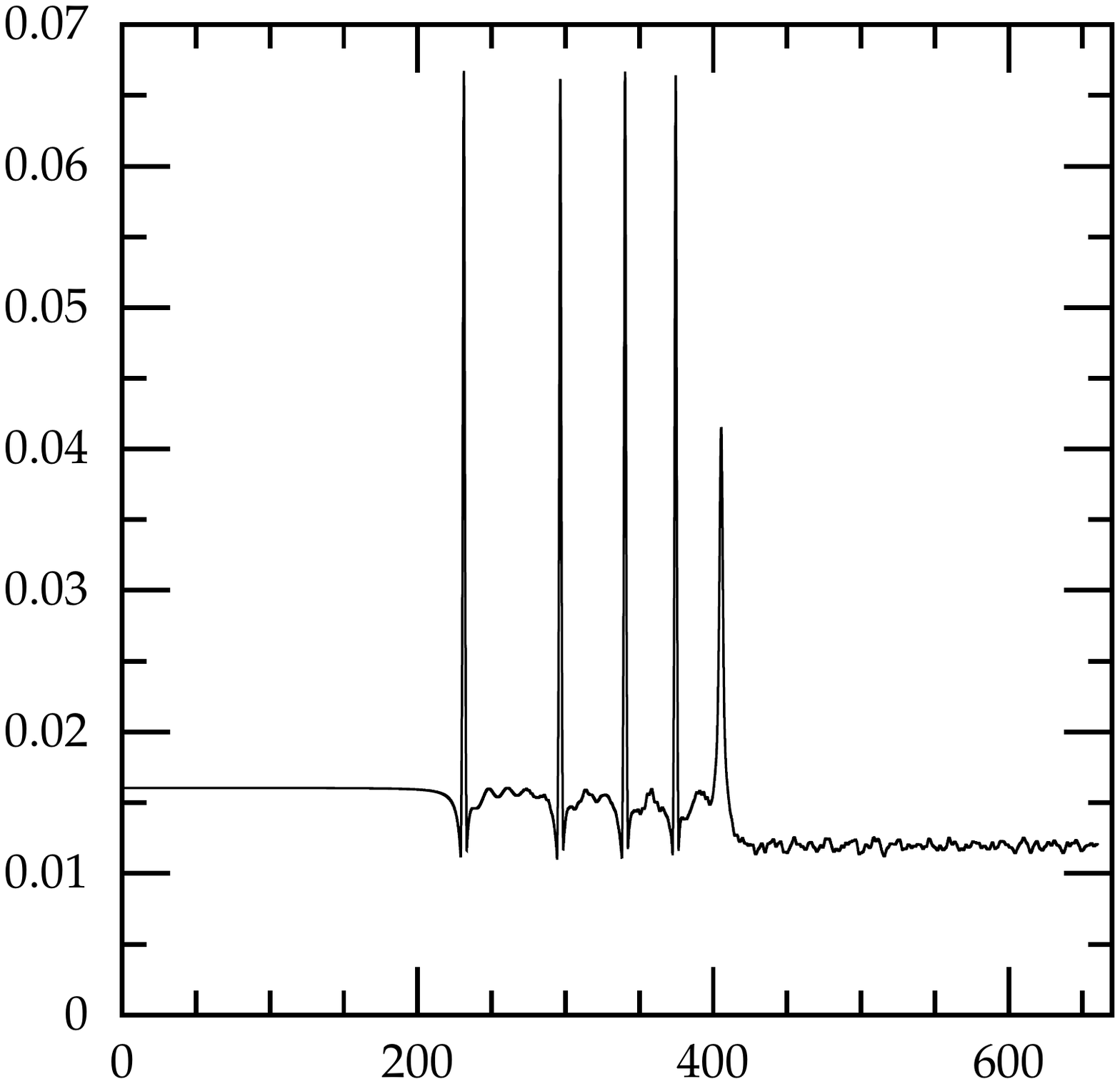}
\includegraphics[angle=0,width=0.2\textwidth]{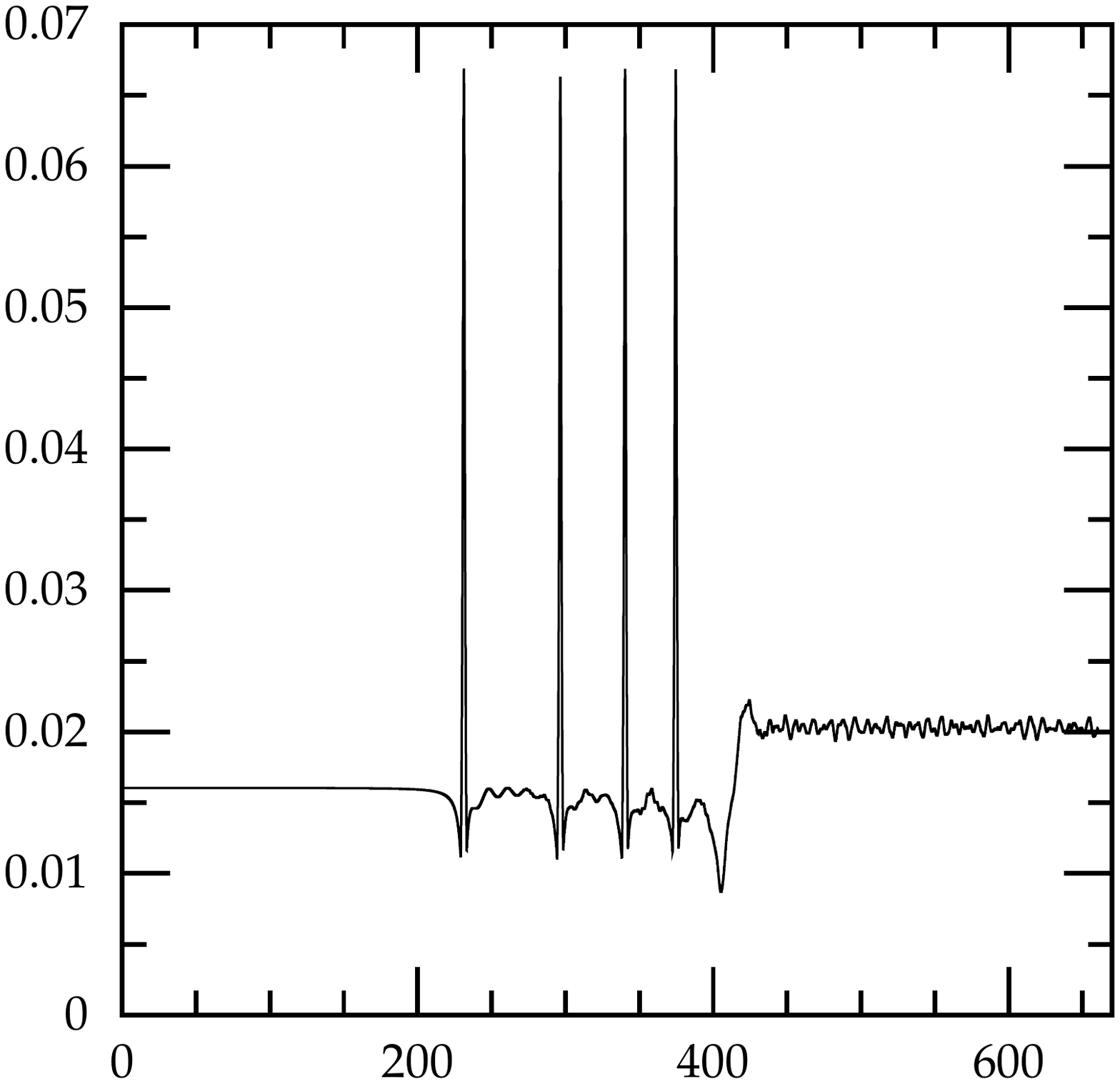}
			\end{center}
{Fig. 8  Heights of the two  solitons observed in ther scattering at rest ($\epsilon=-0.06$ $c=0$)}
a) the left one, b) the right one.
\end{figure}

Furthermore, in the last section we did stress that the cases of 
$c$ given by \rf{nicecchoice} are special for all $\epsilon$'s as then we could use our parity arguments to claim asymptotic conservation of further anomalous conserved quantities \rf{mirrorcharge}.

So we have looked at the first nontrivial anomaly.
To get its form we used the expression of our potential \rf{pert} and so calculated $X$ from the second 
formula in \rf{xdef2}. Then
we put it into the formula for $\alpha^{(3,-4)}$ given in \rf{anomaly3}. In order to avoid using the explicit value of $t_{\Delta}$, which for the zero order solution (expanded in $\ve$) is given in \rf{tdeltazeroorder}, we decided to integrate the resultant expression for $\beta_4$. Therefore, using  \rf {xdef2},  \rf{chargedef} and \rf{anomaly3} we introduce the quantity
\br
\chi^{(4)}\(t\)&\equiv &\int_{-\infty}^t dt^{\prime}\, \beta_4 = 
\int_{-\infty}^t dt^{\prime}\,\int_{-\infty}^{\infty}dx\,X\,  \alpha^{(3,-4)}
\lab{chi4def}\\
&=&-2\,i\,\eta^2\,\int_{-\infty}^t dt^{\prime}\,\int_{-\infty}^{\infty}dx\,
\(R^{\ve}-1\)\left[
6 \,\eta \,R^3
+ \frac{3}{2} \,  \(\partial_x\varphi \)^2 R^2 -2 \, R\,
   \partial_x^2 R +\frac{3}{2}   \(\partial_x R\)^2\right]
   \nonumber
\er
As at large values of $t^{\prime}$ the integrand in \rf{chi4def} vanishes, we can take, in our numerical simulations, the lower end of the  $t^{\prime}$-integral 
to be large in the past but finite. It is the quantity $\chi^{(4)}$ given in \rf{chi4def} whose plots we present next.  

Clearly for $\epsilon=0$ the anomaly vanishes 
so in fig 9 and 10 we present our results for $\epsilon=0.06$ and in fig 11 and 12 those 
for $\epsilon=-0.06$.

 The first figures in each group show the anomaly when the solitons were sent towards each other
at $v=0.4$ and the second ones (10 and 12) those started at rest.
In each case the first figure corresponds to the special value of $c$, {\it i.e.} $c=0$, the others to $c=0.7$ and $c=1.4$.  Note  that the scale on the vertical axis in the figures 
is very different. The anomaly for the cases corresponding to $c=0$ is essentially zero thus supporting our claims of the previous section.
Of course, our results are non-perturbative but they do involve also small corrections due to the numerical errors. In any case the smallness of the corrections suggest to us that our claims 
are correct and the results are stable with respect to small perturbations.  For $c\ne 0$ we do see some important corrections to the anomaly as expected (even though the differences of the trajectories are not very significant).

 \begin{figure}
    \begin{center}
\includegraphics[angle=0,width=0.3\textwidth]{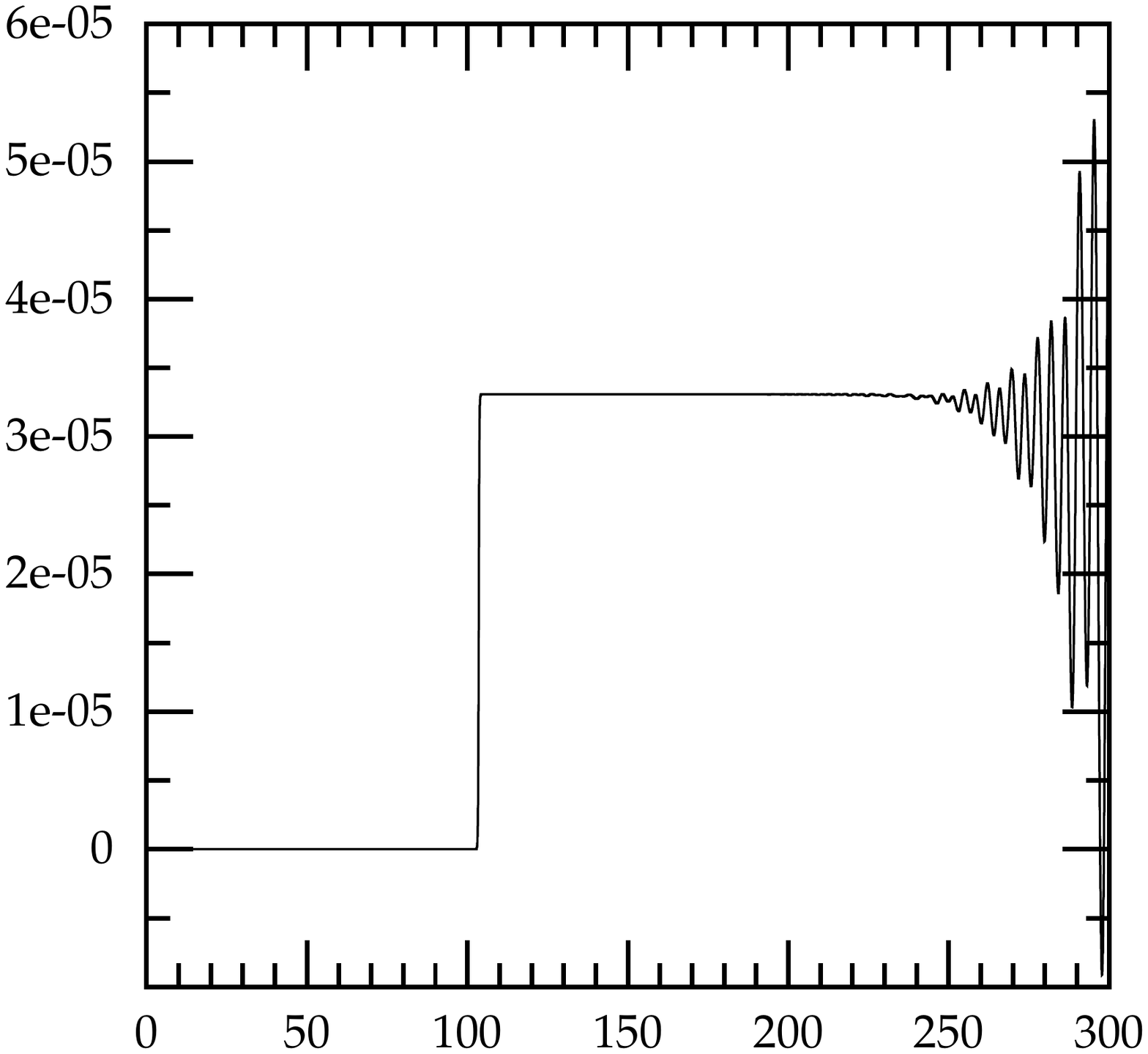}
	\includegraphics[angle=0,width=0.3 \textwidth]{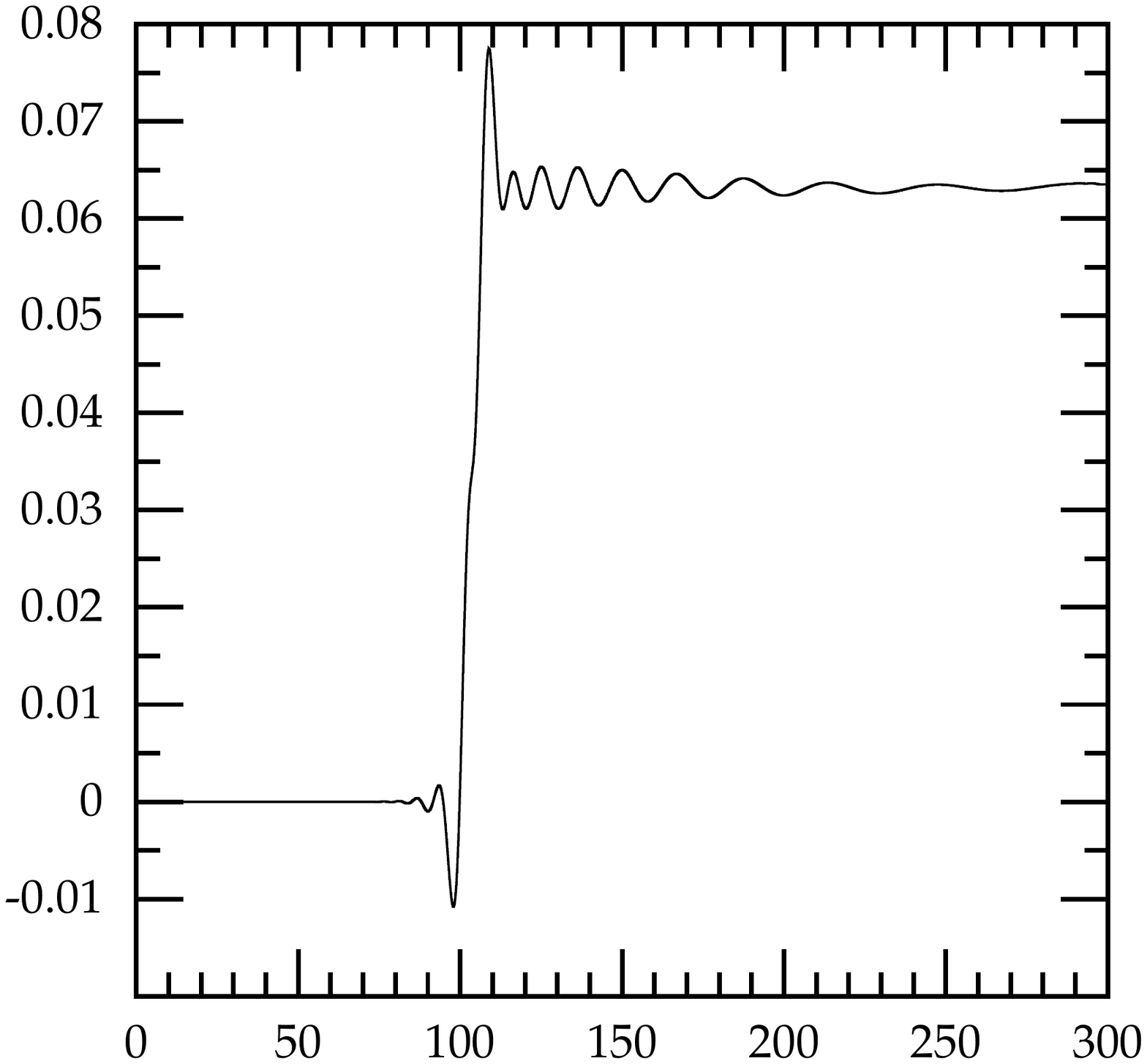}
	\includegraphics[angle=0,width=0.3 \textwidth]{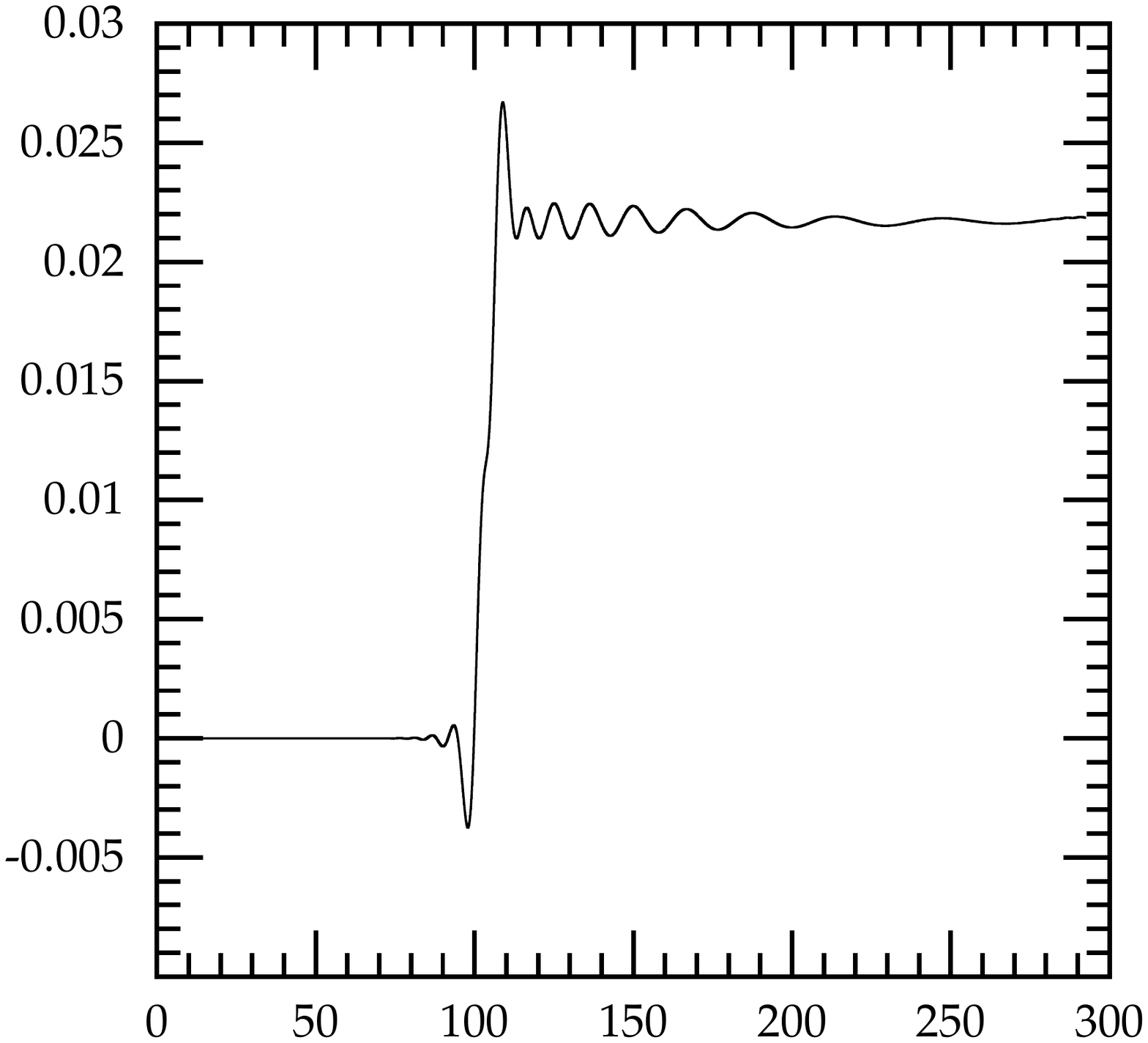}
		\end{center}
{Fig. 9  Time integrated anomaly of two solitons sent at $v=0.4$ ($\epsilon=0.06$)}
a) $c=0$, b) $c=0.7$ and c) $c=1.4$
\end{figure}

\begin{figure}
    \begin{center}
\includegraphics[angle=0,width=0.3\textwidth]{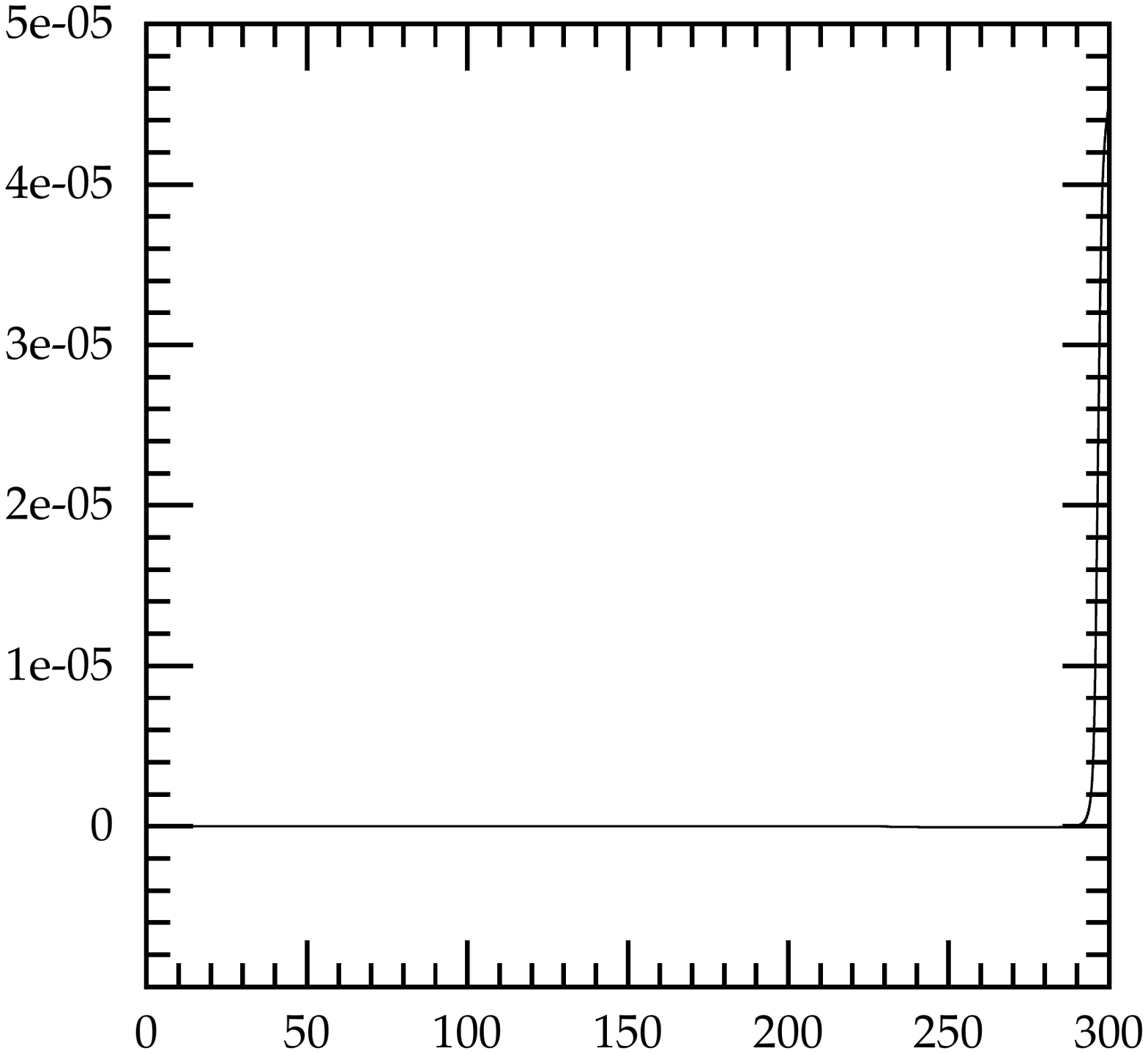}
	\includegraphics[angle=0,width=0.3 \textwidth]{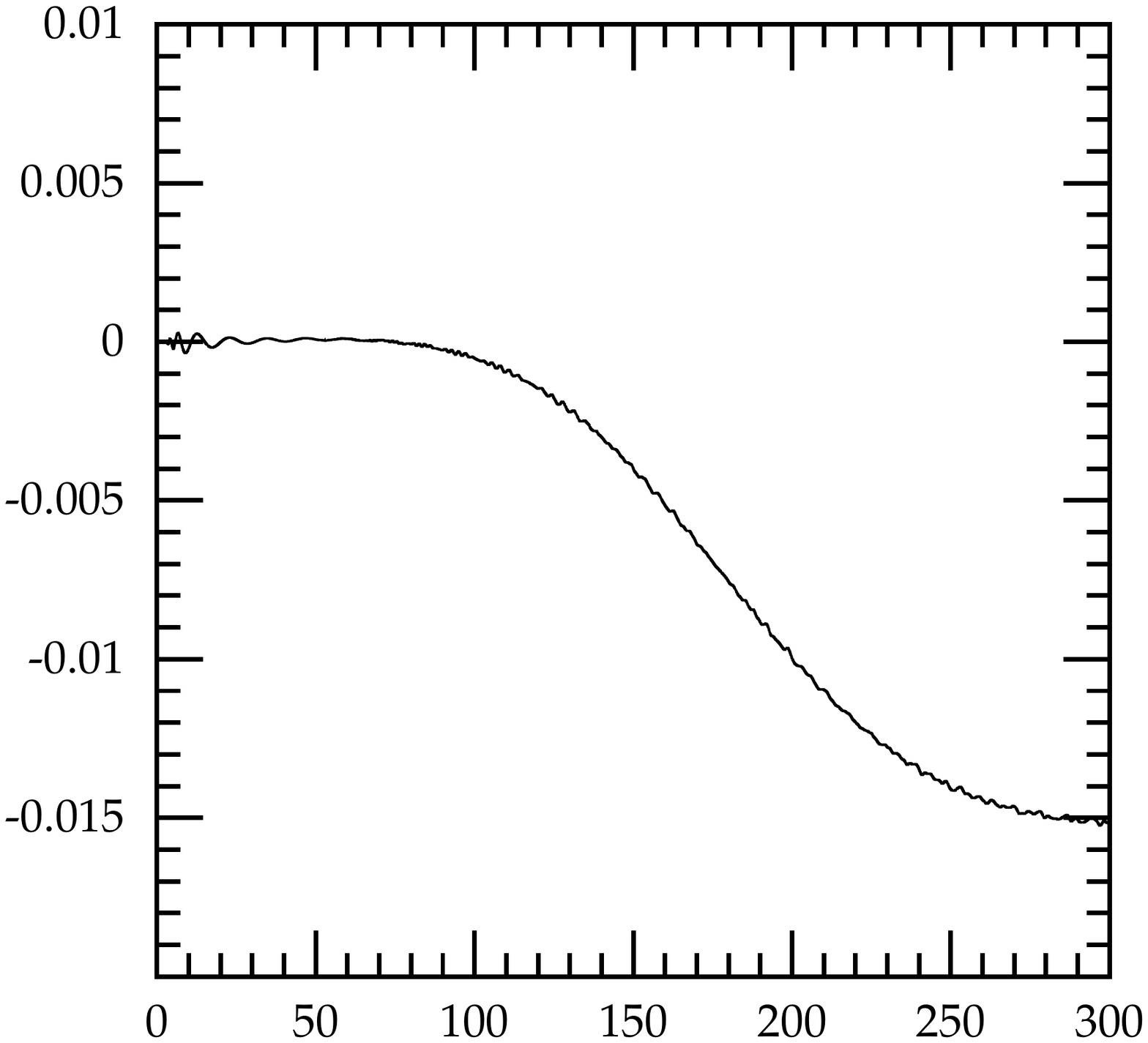}
	\includegraphics[angle=0,width=0.3 \textwidth]{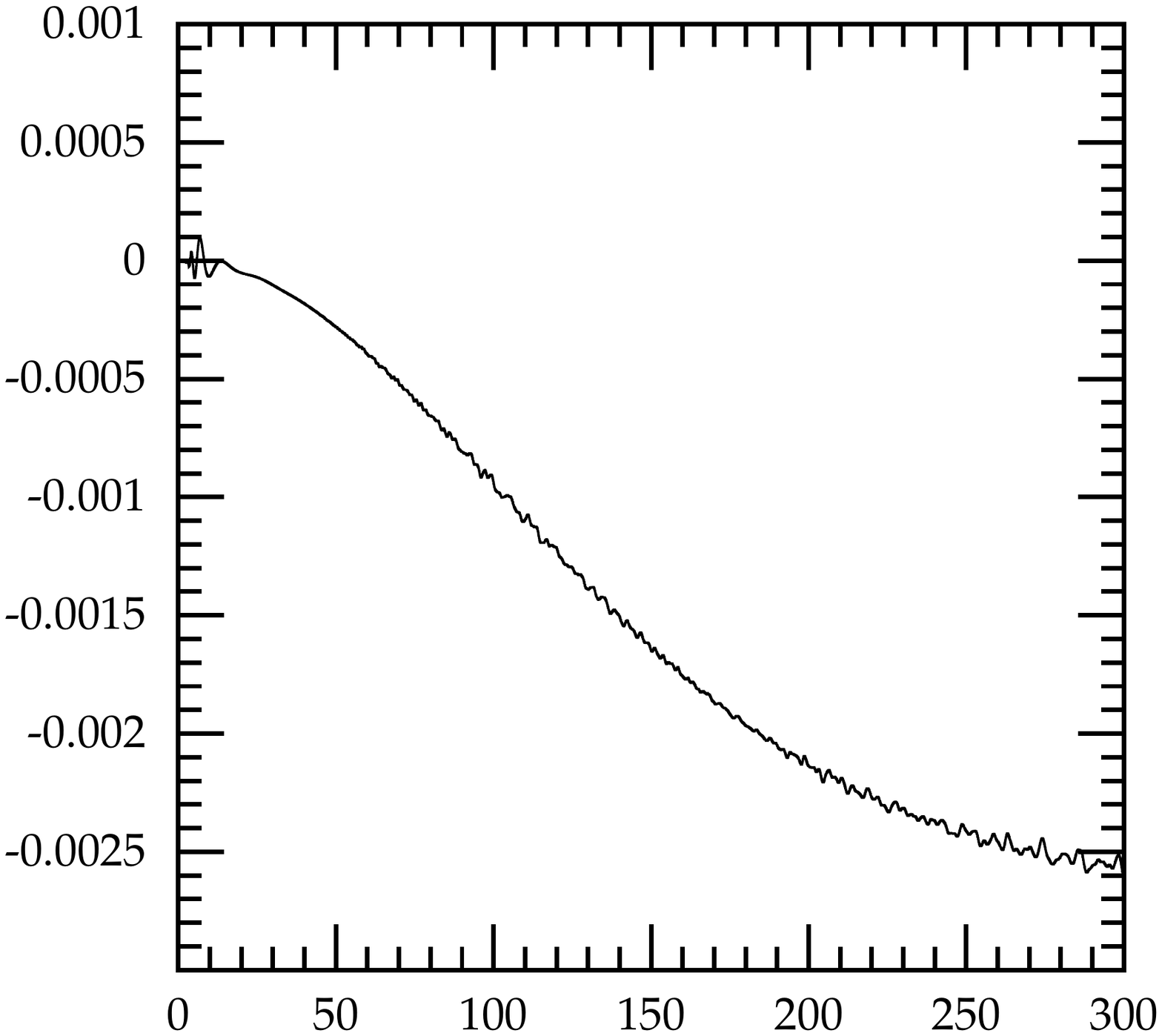}
		\end{center}
{Fig. 10  Time integrated anomaly of two solitons at rest ($\epsilon=0.06$)}
a) $c=0$, b) $c=0.7$ and c) $c=1.4$
\end{figure}

\begin{figure}
    \begin{center}
\includegraphics[angle=0,width=0.3\textwidth]{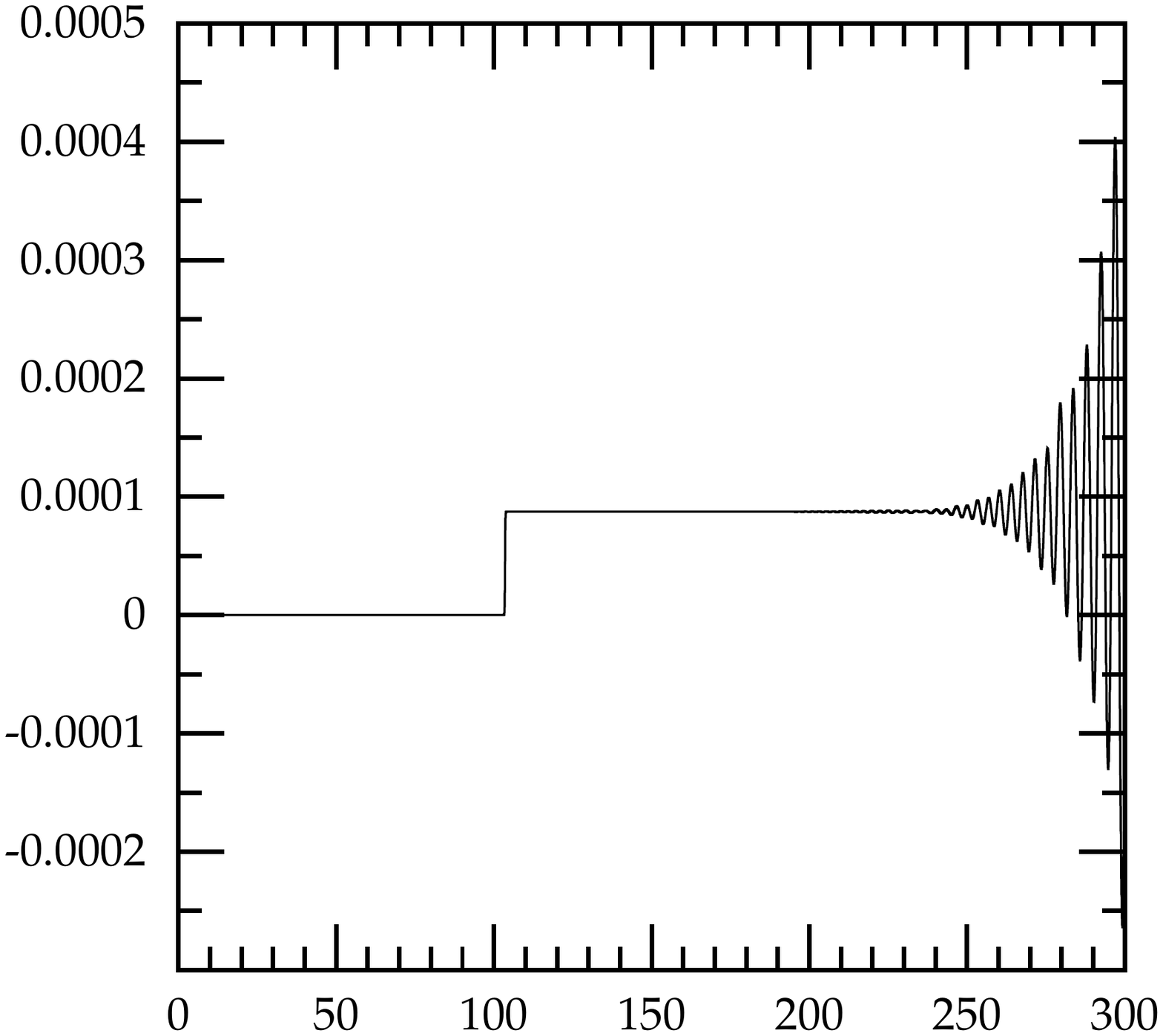}
	\includegraphics[angle=0,width=0.3 \textwidth]{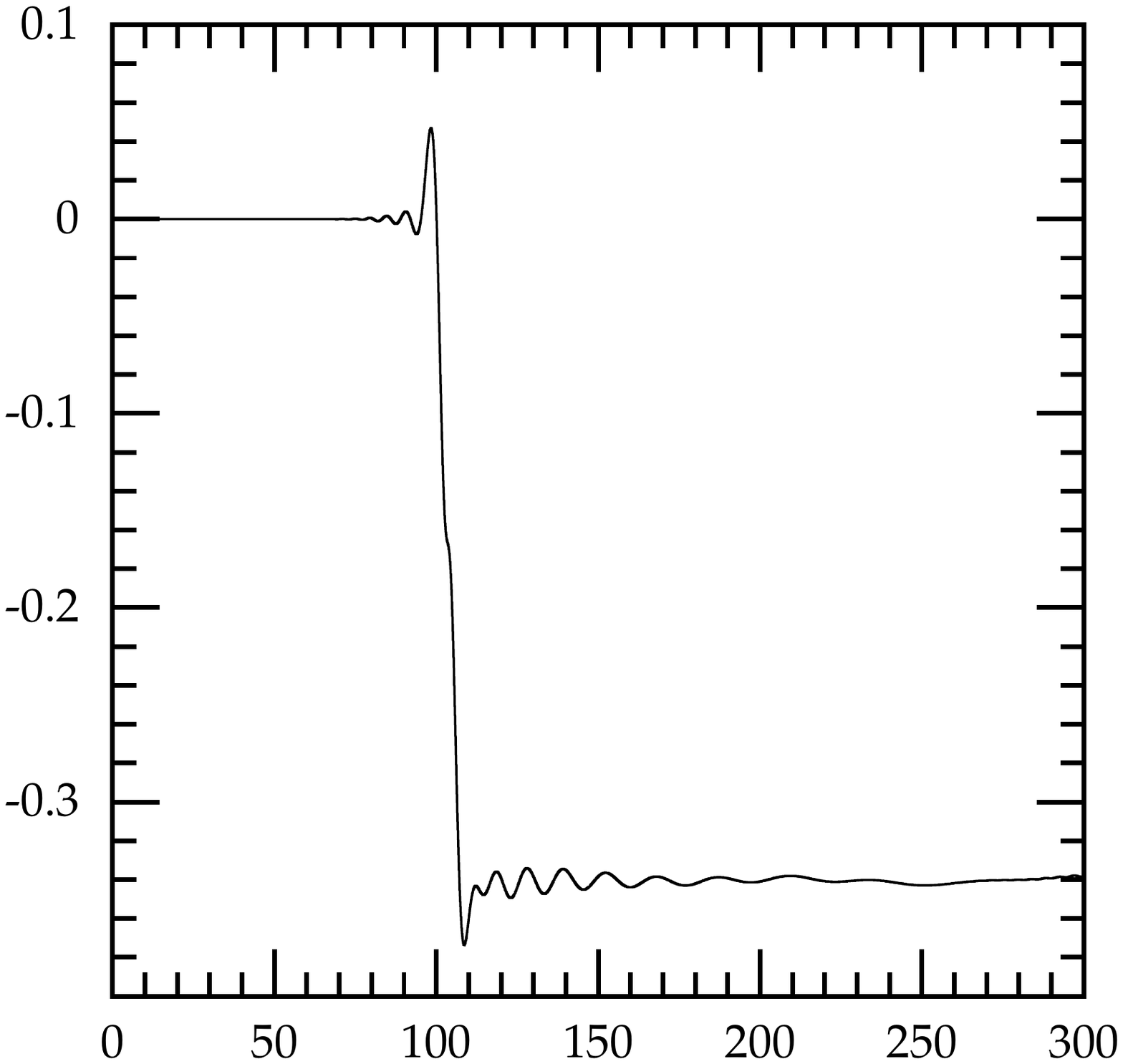}
	\includegraphics[angle=0,width=0.3 \textwidth]{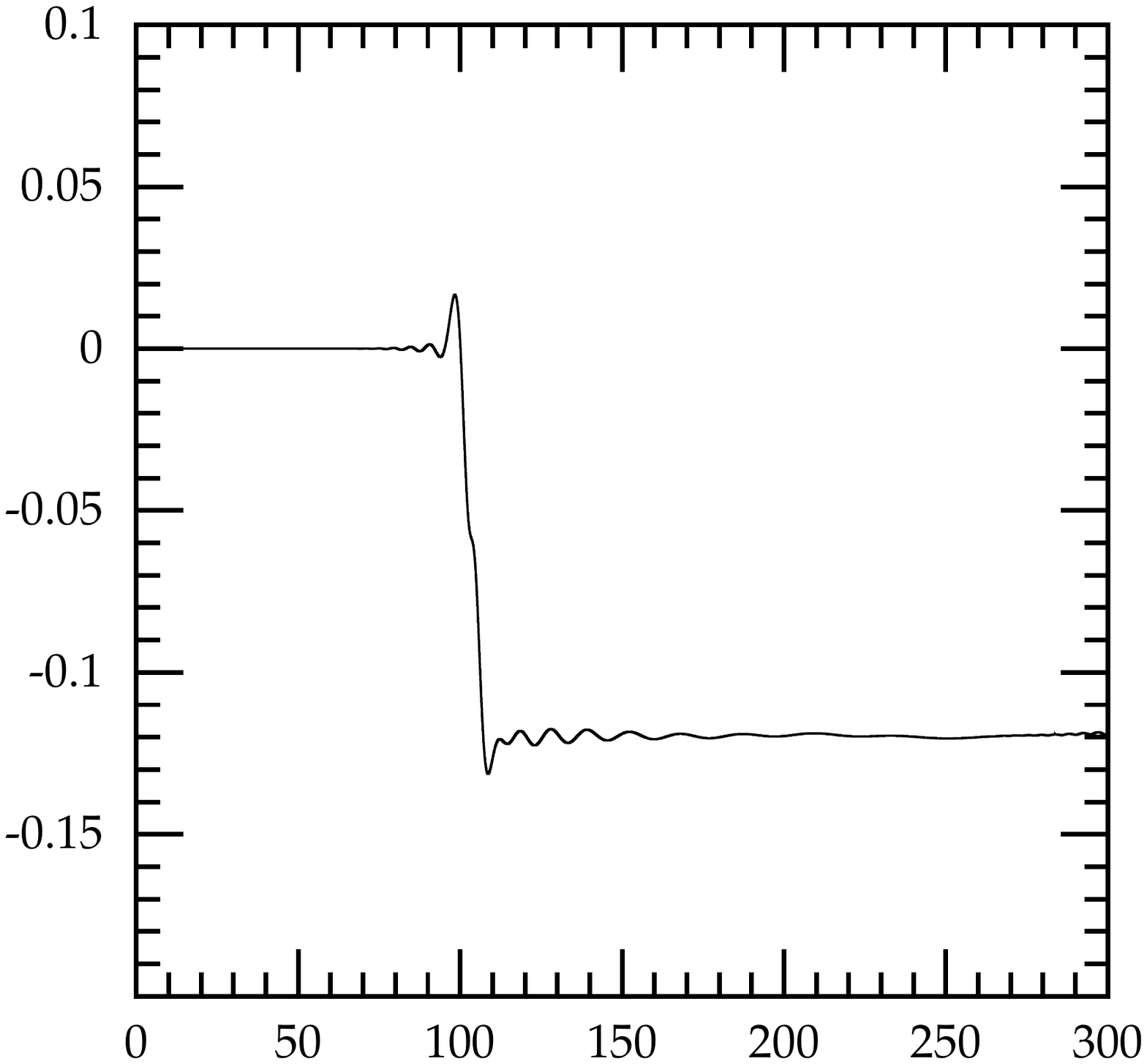}
		\end{center}
{Fig. 11  Time integrated anomaly of two solitons sent at $v=0.4$ ($\epsilon=-0.06$)}
a) $c=0$, b) $c=0.7$ and c) $c=1.4$
\end{figure}

\begin{figure}
    \begin{center}
\includegraphics[angle=0,width=0.3\textwidth]{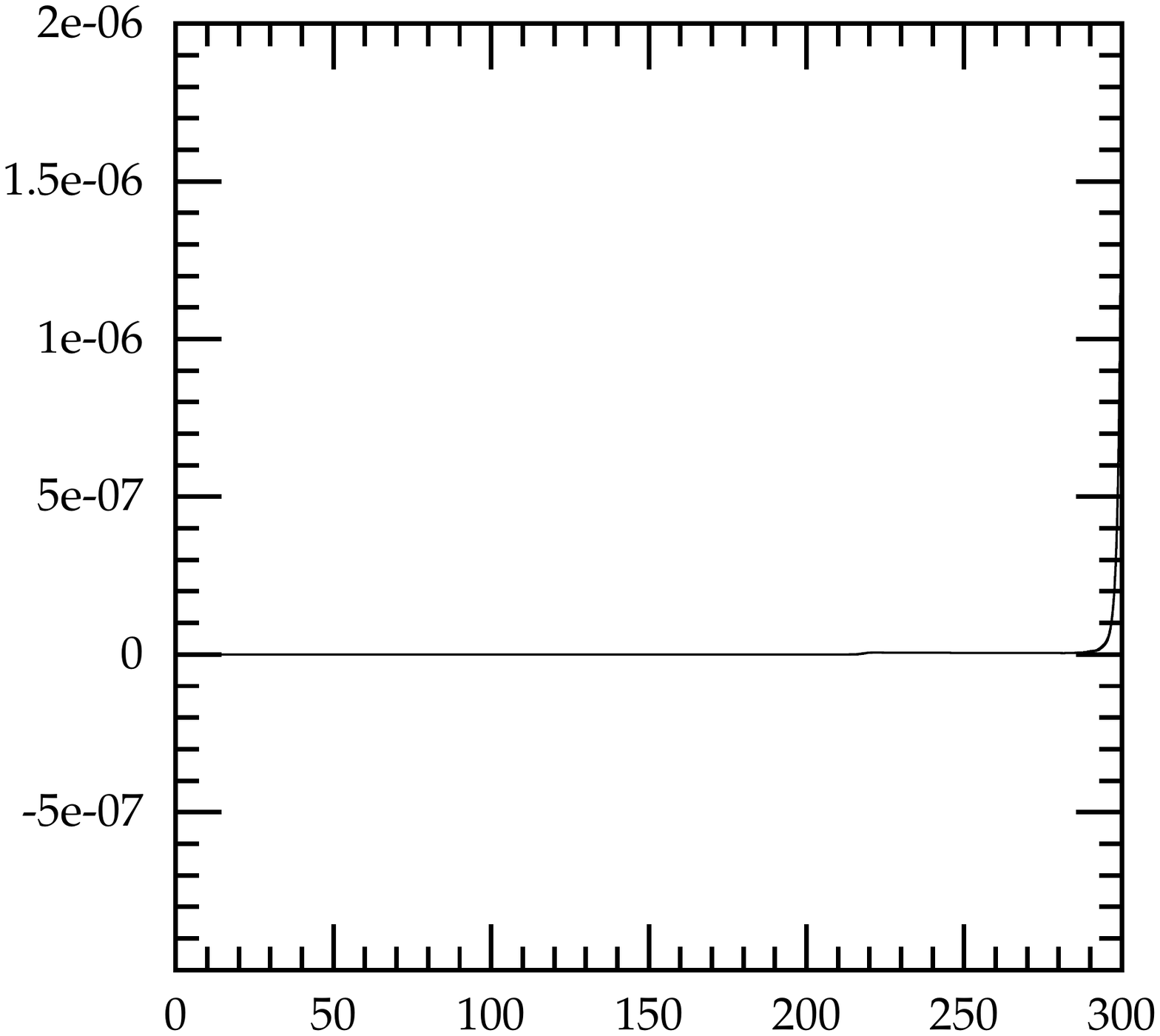}
	\includegraphics[angle=0,width=0.3 \textwidth]{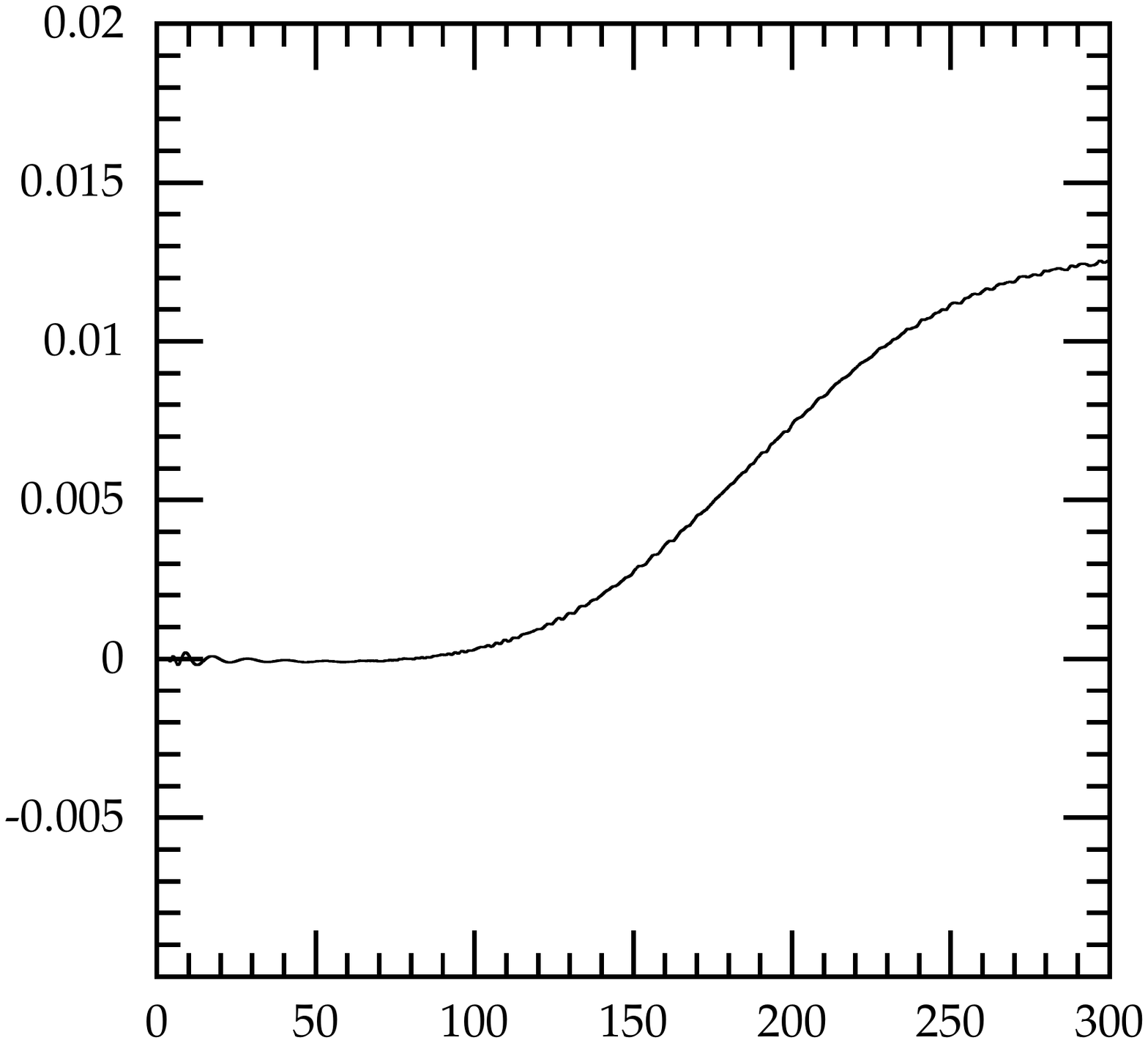}
	\includegraphics[angle=0,width=0.3 \textwidth]{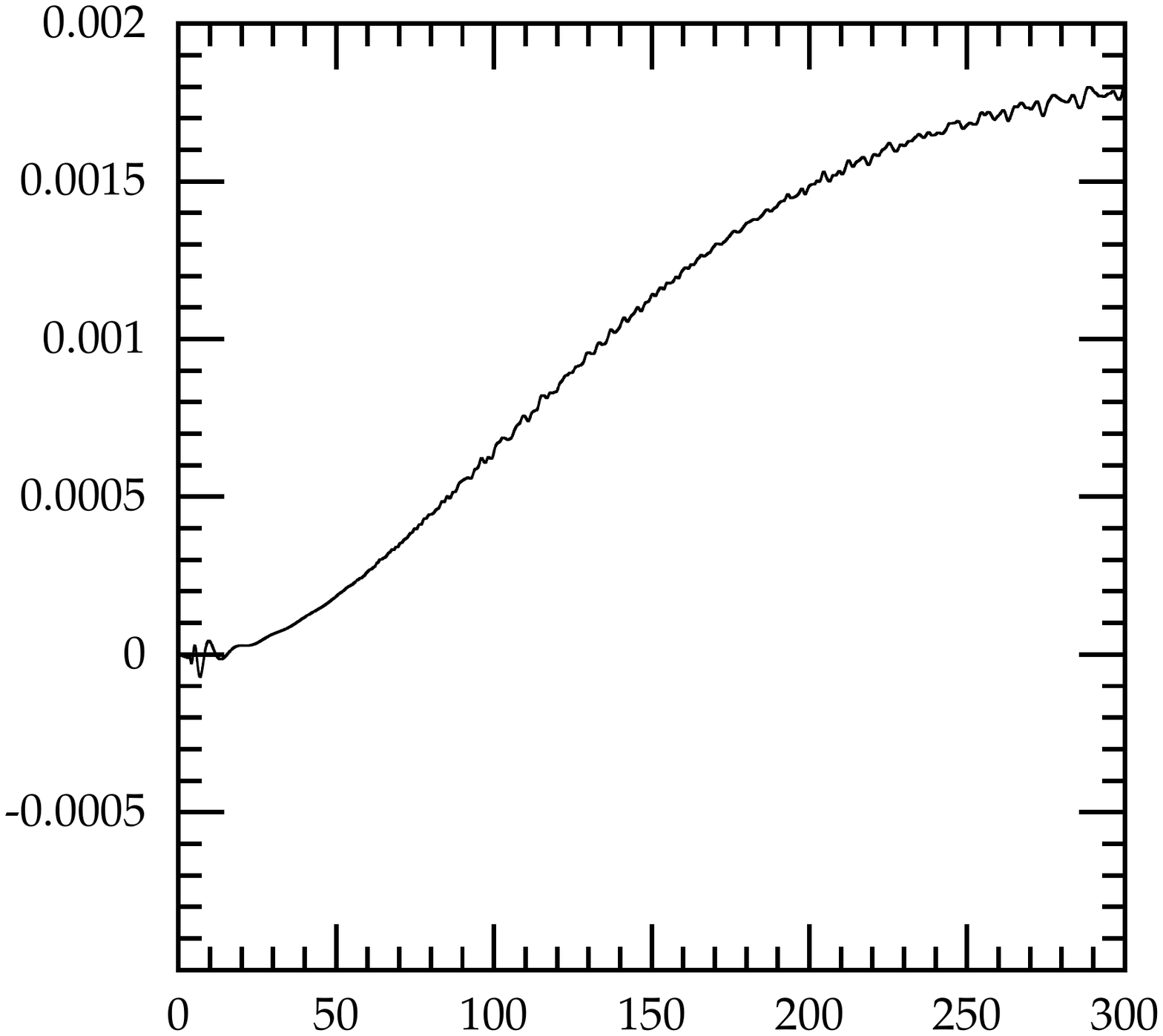}
		\end{center}
{Fig. 12  Time integrated anomaly of two solitons at rest ($\epsilon=-0.06$)}
a) $c=0$, b) $c=0.7$ and c) $c=1.4$
\end{figure}

\section{Conclusions}
\label{sec:conclusions}
\setcounter{equation}{0} 

In this paper we have looked at the concept, recently introduced by  two of us,  of quasi-integrability in the context of the deformations of the NLS model in (1+1) dimensions. The unperturbed model is fully integrable and possesses multisoliton solutions. The perturbations destroy integrability but the perturbative models still possess soliton solutions.

In our work we have looked at the problem of quasi-integrability and in this case related it to the properties of specific field configurations (like those describing multisolitons) under very specific  parity transformations. We have shown that when one considers the perturbed models which are not integrable, the models do not possess an infinite number of conserved charges (like the integrable ones do). However, when we restrict our attention to specific field configurations, sometimes we can say more. Namely, when the field configurations satisfy some very specific parity conditions (which are often physical in nature)
the extra charges, though not conserved, do satisfy some interesting conditions (given in \rf{mirrorcharge}). These conditions do restrict 
the scattering properties of solitons and so provide the basis of our
understanding of quasi-integrability. We have also looked at the properties of the soliton field configurations numerically and have found  a good support of our claims. As a side result we have obtained some results on the forces involving two solitons in the NLS models and its deformations; namely that these forces are rather complicated and depend on the relative phase between the solitons (and for some values of this phase result in an attraction and for some other ones in a repulsion).

Clearly, our observations should be extended to other models, such as perturbed Toda models in (1+1) dimensions and, more importantly, to models in higher dimensions. Such extensions are under active considerations right now.

\newpage

\appendix

\section{Explicit expressions for quantities involved in the gauge transformation \rf{gaugetransfatilde}}
\label{sec:rotatedpot}
\setcounter{equation}{0} 

We give in this appendix the first few explicit expressions for the parameters $\zeta_i^{(-n)}$, $i=1,2$, introduced in \rf{abelg}, for the components $a_x$ of the connection defined in  \rf{gaugetransfatilde}, and the quantities $\alpha^{(j,-n)}$, $j=1,2,3$, introduced in \rf{rotateb0}. On the r.h.s. of the equations below we use the following notation:  (for partial derivatives w.r.t. $x$ and $t$)
\be
\star^{(n,m)}\equiv \pa_x^n\pa_t^m\, \star
\ee

The expressions for $\zeta_i^{(-n)}$ are:
\br
\zeta_1^{(-1)}&=&0,
\nonumber\\
\zeta_2^{(-1)}&=& 2 \sqrt{\mid \eta \mid } \sqrt{R },
\nonumber\\
\zeta_1^{(-2)}&=& \frac{i \sqrt{\mid \eta \mid } R^{(1,0)} }{\sqrt{R }},
\nonumber\\
\zeta_2^{(-2)}&=& \sqrt{\mid \eta \mid } \varphi ^{(1,0)}  \sqrt{R },
\nonumber\\
\zeta_1^{(-3)}&=& \frac{i \left(\sqrt{\mid \eta \mid } \varphi ^{(1,0)}  R^{(1,0)} +\sqrt{\mid \eta \mid } \varphi
   ^{(2,0)}  R \right)}{\sqrt{R }},
\\
\zeta_2^{(-3)}&=& \frac{16 \mid \eta \mid ^{3/2} \sigma R^3+3 \sqrt{\mid \eta \mid } \(\varphi^{(1,0)}\)^2 R^2-6
   \sqrt{\mid \eta \mid } R^{(2,0)}  R +3 \sqrt{\mid \eta \mid } \(R^{(1,0)}\)^2}{6 R ^{3/2}},
\nonumber\\
\zeta_1^{(-4)}&=& \frac{i}{12 R ^{5/2}}
 \left[64 \mid \eta \mid ^{3/2} \sigma R^{(1,0)}  R ^3+9 \sqrt{\mid \eta \mid } 
\(\varphi^{(1,0)}\)^2 R^{(1,0)}  R ^2+18 \sqrt{\mid \eta \mid } \varphi ^{(1,0)}  \varphi
   ^{(2,0)}  R ^3
   \right. \nonumber\\
 &-&  \left.  12 \sqrt{\mid \eta \mid } R^{(3,0)}  R ^2+18 \sqrt{\mid \eta \mid }
   R^{(1,0)}  R^{(2,0)}  R -9 \sqrt{\mid \eta \mid } \(R^{(1,0)}\)^3\right],
\nonumber\\
\zeta_2^{(-4)}&=& \frac{1}{4 R ^{3/2}}\left[
16 \mid \eta \mid ^{3/2} \sigma \varphi ^{(1,0)}  R ^3-6 \sqrt{\mid \eta \mid } \varphi ^{(2,0)} 
   R^{(1,0)}  R -6 \sqrt{\mid \eta \mid } \varphi ^{(1,0)}  R^{(2,0)}  R 
   \right. \nonumber\\
   &+& \left. 3
   \sqrt{\mid \eta \mid } \varphi ^{(1,0)}  \(R^{(1,0)}\)^2+\sqrt{\mid \eta \mid } \(\varphi ^{(1,0)}\)^3
   R ^2-4 \sqrt{\mid \eta \mid } \varphi ^{(3,0)}  R ^2\right].
\nonumber
\er
The components $a_x^{(3,n)}$ introduced in \rf{axcomp} are:
\br
a_x^{(3,0)}&=&\frac{1}{2} i \varphi ^{(1,0)}, 
\nonumber\\
a_x^{(3,-1)}&=&2 i \mid \eta \mid  \sigma R, 
\nonumber\\
a_x^{(3,-2)}&=&i \mid \eta \mid  \sigma \varphi ^{(1,0)}  R, 
\\
a_x^{(3,-3)}&=&\frac{i \mid \eta \mid  \left(4 \mid \eta \mid  R ^3+\sigma \(\varphi ^{(1,0)}\)^2 R ^2-2 \sigma R^{(2,0)} 
   R +\sigma \(R^{(1,0)}\)^2\right)}{2 R },
\nonumber\\
a_x^{(3,-4)}&=&\frac{i \mid \eta \mid}{4 R }  \left[12 \mid \eta \mid  \varphi ^{(1,0)}  R ^3-6 \sigma R  \left(\varphi
   ^{(2,0)}  R^{(1,0)} +\varphi ^{(1,0)}  R^{(2,0)} \right)+3 \sigma \varphi
   ^{(1,0)}  \(R^{(1,0)}\)^2
   \right. \nonumber\\
   &+& \left. \sigma \left(\(\varphi ^{(1,0)}\)^3-4 \varphi
   ^{(3,0)} \right) R ^2\right].
\nonumber
\er
The quantities $\alpha^{(j,-n)}$, introduced in \rf{rotateb0} are:
\br
\alpha^{(3,0)}&=&1,
\nonumber\\
\alpha^{(3,-1)}&=&0,
\nonumber\\
\alpha^{(3,-2)}&=&2 \mid \eta \mid  \sigma R, 
\lab{anomaly3}\\
\alpha^{(3,-3)}&=&2 \mid \eta \mid  \sigma \varphi ^{(1,0)}  R, 
\nonumber\\
\alpha^{(3,-4)}&=&6 \mid \eta \mid ^2 R ^2+\frac{3}{2} \mid \eta \mid  \sigma \(\varphi ^{(1,0)}\)^2 R -2 \mid \eta \mid  \sigma
   R^{(2,0)} +\frac{3 \mid \eta \mid  \sigma \(R^{(1,0)}\)^2}{2 R }
\nonumber
\er
and
\br
\alpha^{(1,0)}&=&0,
\nonumber\\
\alpha^{(1,-1)}&=&-2 \sqrt{\mid \eta \mid } \sqrt{R },
\nonumber\\
\alpha^{(1,-2)}&=&-\sqrt{\mid \eta \mid } \varphi ^{(1,0)}  \sqrt{R },
\\
\alpha^{(1,-3)}&=&-4 \mid \eta \mid ^{3/2} \sigma R ^{3/2}-\frac{1}{2} \sqrt{\mid \eta \mid } \(\varphi ^{(1,0)}\)^2
   \sqrt{R }-\frac{\sqrt{\mid \eta \mid } \(R^{(1,0)}\)^2}{2 R ^{3/2}}+\frac{\sqrt{\mid \eta \mid
   } R^{(2,0)} }{\sqrt{R }},
\nonumber\\
\alpha^{(1,-4)}&=&-6 \mid \eta \mid ^{3/2} \sigma \varphi ^{(1,0)}  R ^{3/2}-\frac{3 \sqrt{\mid \eta \mid } \varphi
   ^{(1,0)}  \(R^{(1,0)}\)^2}{4 R ^{3/2}}+\frac{3 \sqrt{\mid \eta \mid } \varphi
   ^{(1,0)}  R^{(2,0)} }{2 \sqrt{R }}
   \nonumber\\
   &+&\frac{3 \sqrt{\mid \eta \mid } \varphi
   ^{(2,0)}  R^{(1,0)} }{2 \sqrt{R }}-\frac{1}{4} \sqrt{\mid \eta \mid } 
   \(\varphi^{(1,0)}\)^3 \sqrt{R }+\sqrt{\mid \eta \mid } \varphi ^{(3,0)}  \sqrt{R }
\nonumber
\er
and
\br
\alpha^{(2,0)}&=&0,
\nonumber\\
\alpha^{(2,-1)}&=&0,
\nonumber\\
\alpha^{(2,-2)}&=&-\frac{i \sqrt{\mid \eta \mid } R^{(1,0)} }{\sqrt{R }},
\nonumber\\
\alpha^{(2,-3)}&=&-\frac{i \sqrt{\mid \eta \mid } \varphi ^{(1,0)}  R^{(1,0)} }{\sqrt{R }}-i \sqrt{\mid \eta \mid }
   \varphi ^{(2,0)}  \sqrt{R },
\\
\alpha^{(2,-4)}&=&-6 i \mid \eta \mid ^{3/2} \sigma \sqrt{R } R^{(1,0)} -\frac{3 i \sqrt{\mid \eta \mid } 
\(\varphi^{(1,0)}\)^2 R^{(1,0)} }{4 \sqrt{R }}-\frac{3}{2} i \sqrt{\mid \eta \mid } \varphi
   ^{(1,0)}  \varphi ^{(2,0)}  \sqrt{R }
   \nonumber\\
   &+&\frac{3 i \sqrt{\mid \eta \mid }
   \(R^{(1,0)}\)^3}{4 R ^{5/2}}-\frac{3 i \sqrt{\mid \eta \mid } R^{(2,0)} 
   R^{(1,0)} }{2 R ^{3/2}}+\frac{i \sqrt{\mid \eta \mid } R^{(3,0)} }{\sqrt{R }}.
\nonumber
\er

\section{The Hirota solutions}
\label{sec:hirota}
\setcounter{equation}{0} 

Here we construct the one and two bright soliton solutions of the integrable NLS theory \rf{nlseq} using the Hirota method. The one and two dark soliton solutions require a different procedure from the one described here. We introduce the Hirota tau-functions as 
\be
\psi_0=\frac{i}{\gamma}\,\frac{\tau_{+}}{\tau_0} \; ;\qquad \qquad \qquad \qquad \psib_0=-\frac{i}{{\bar \gamma}}\,\frac{\tau_{-}}{\tau_0},
\lab{taudef}
\ee
where $\eta=\gamma\,{\bar \gamma}$. The bright soliton solutions exist for  $\eta<0$ and so we need ${\bar \gamma}=-\gamma^*$, and then $\frac{\tau_{-}}{\tau_0}=-\(\frac{\tau_{+}}{\tau_0}\)^*$. Putting \rf{taudef} into the the NLS equation \rf{nlseq} and its complex conjugate we get the two Hirota equations
\br
&&\tau_0^2 \(i \partial_t \tau_{+} + \partial_x^2\tau_{+}\) - 
 2 \tau_0 \partial_x \tau_{+}\, \partial_x\tau_0 - 2 \tau_{+}^2 \,\tau_{-} - 
 \tau_0\, \tau_{+} \(i \partial_t \tau_0 + \partial_x^2 \tau_0\) + 2 \tau_{+} (\partial_x \tau_0)^2=0, \qquad\qquad
 \lab{taueq}\\
&&\tau_0^2 \(-i \partial_t \tau_{-} + \partial_x^2\tau_{-}\) - 
 2 \tau_0 \partial_x \tau_{-}\, \partial_x\tau_0 - 2 \tau_{-}^2 \,\tau_{+} - 
 \tau_0\, \tau_{-} \(-i \partial_t \tau_0 + \partial_x^2 \tau_0\) + 2 \tau_{-} (\partial_x \tau_0)^2=0.
 \nonumber
 \er

The one-soliton solution of \rf{taueq} is given by 
\br
\tau_0 &=& 1 + a_{+} a_{-}\,  \frac{z_1\, z_2}{(z_1 - z_2)^2}\, e^{i\Gamma\(z_1\)} e^{-i\Gamma\(z_2\)}, 
\nonumber\\
\tau_{+} &=& a_{-} \,z_2\,  e^{-i\Gamma\(z_2\)},
\nonumber\\
\tau_{-} &=& a_{+}\, z_1\, e^{i\Gamma\(z_1\)}
\er
with $a_{\pm}$, $z_1$ and $z_2$ being complex parameters and $\Gamma\(z_i\)= z_i^2\, t - z_i\, x$.  We choose $z_2=z_1^*$ and $a_{-}=-a_{+}^*$, which implies that $\tau_{-}=-\tau_{+}^*$, and $\tau_0$ is real. We then parametrize them as 
\be
a_{\pm}=i \, a\, e^{\pm i\,\theta},\qquad\ z_1=\frac{v}{2}+i\,\rho=\sqrt{\frac{v^2}{4}+\rho^2}\,\,e^{i\zeta},\qquad\ \gamma = i\sqrt{\mid \eta\mid}\,e^{i\phi},
\qquad {\bar \gamma} = i\sqrt{\mid \eta\mid}\,e^{-i\phi}
\ee
with $a>0$, and $v$ and $\rho$ both real. We replace $a$ by $x_0$ defined  as
\be
a \,\frac{\sqrt{\frac{v^2}{4}+\rho^2}}{2\,\mid \rho\mid}=e^{-\rho\,x_0}
\ee
and find from \rf{taudef} that 
\be
\psi_0= \frac{i\, e^{- i\,\(\theta+\zeta+\phi\)}}{\sqrt{\mid \eta\mid}}\,\mid \rho\mid\, 
\frac{e^{i\left[\(\rho^2-\frac{v^2}{4}\)\,t+ \frac{v}{2}\, x\right]}}{\cosh\left[\rho\,\(x-v\,t-x_0\)\right]}.
\ee

This expression, up to an overall constant phase factor (due to the symmetry \rf{phasetransf}) is the one-bright-soliton given in \rf{brightsol}.

The two-soliton solution of \rf{taueq} is given by 
\br
\tau_0 &=& 1 + a_{+} a_{-} \, \frac{z_1 z_2}{(z_1 - z_2)^2} \, e^{i\,\Gamma\(z_1\)} e^{-i\,\Gamma\(z_2\)}  + 
   b_{+} b_{-} \, \frac{w_1 w_2}{(w_1 - w_2)^2} \, e^{i\,\Gamma\(w_1\)} e^{-i\,\Gamma\(w_2\)}  
   \nonumber\\
   &+& 
   a_{+} b_{-} \, \frac{z_1 w_2}{(z_1 - w_2)^2} \, e^{i\,\Gamma\(z_1\)} e^{-i\,\Gamma\(w_2\)}  + 
   a_{-} b_{+} \, \frac{w_1 z_2}{(w_1 - z_2)^2} \, e^{-i\,\Gamma\(z_2\)} e^{i\,\Gamma\(w_1\)}  
   \nonumber\\
   &+& 
   a_{+} a_{-} b_{+} b_{-} \, \frac{z_1 z_2 w_1 w_2 (z_1 - w_1)^2 (z_2 - w_2)^2}{(z_1 - z_2)^2 (w_1 - w_2)^2 (z_1 - w_2)^2 (w_1 - z_2)^2}
   e^{i\,\Gamma\(z_1\)} e^{-i\,\Gamma\(z_2\)} e^{i\,\Gamma\(w_1\)} e^{-i\,\Gamma\(w_2\)}, 
   \nonumber\\
\tau_{+}&=& a_{-} z_2 e^{-i\,\Gamma\(z_2\)} + b_{-} w_2 e^{-i\,\Gamma\(w_2\)} + 
   a_{+} a_{-} b_{-} \, \frac{w_2 z_1 z_2(w_2 - z_2)^2  }{(w_2 - z_1)^2 (z_1 - z_2)^2}
   e^{i\,\Gamma\(z_1\)} e^{-i\,\Gamma\(z_2\)} e^{-i\,\Gamma\(w_2\)}  
   \nonumber\\
   &+&
   a_{-} b_{+} b_{-} \, \frac{w_1 w_2 z_2 (w_2 - z_2)^2 }{(w_1 - w_2)^2 (w_1 - z_2)^2}
   e^{-i\,\Gamma\(z_2\)} e^{i\,\Gamma\(w_1\)} e^{-i\,\Gamma\(w_2\)}, 
   \nonumber\\
\tau_{-}&=& a_{+} z_1 e^{i\,\Gamma\(z_1\)} + b_{+} w_1 e^{i\,\Gamma\(w_1\)} + 
   a_{+} a_{-} b_{+} \, \frac{w_1 z_1 z_2 (w_1 - z_1)^2  }{(w_1 - z_2)^2 (z_1 - z_2)^2}
   e^{i\,\Gamma\(z_1\)} e^{i\,\Gamma\(w_1\)} e^{-i\,\Gamma\(z_2\)}  
   \nonumber\\
   &+& 
   a_{+} b_{+} b_{-} \, \frac{w_1 w_2 z_1 (w_1 - z_1)^2 }{(w_1 - w_2)^2 (w_2 - z_1)^2}
   e^{i\,\Gamma\(z_1\)} e^{i\,\Gamma\(w_1\)} e^{-i\,\Gamma\(w_2\)},  
   \lab{rawtwosol}
\er
where $a_{\pm}$, $b_{\pm}$, $z_1$, $z_2$, $w_1$ and $w_2$ are arbitrary complex parameters, and as before, $\Gamma\(w_i\)=w_i^2\, t - w_i\, x$. The two-bright-soliton solution of the NLS theory \rf{nlseq}, corresponding to $\eta<0$, is obtained by taking $\tau_{-}=-\tau_{+}^*$, and $\tau_0$ real. One way of getting this involves putting
\be
z_2=z_1^*,\qquad\qquad w_2=w_1^*, \qquad\qquad a_{-}=-a_{+}^*, \qquad\qquad b_{-}=-b_{+}^*
\ee
and then parametrizing them as
\be
a_{\pm}=i \, a_1\, e^{\pm i\,\theta_1},\qquad\ b_{\pm}=i \, a_2\, e^{\pm i\,\theta_2},
\qquad\ \gamma = i\sqrt{\mid \eta\mid}\,e^{i\phi},
\qquad {\bar \gamma} = i\sqrt{\mid \eta\mid}\,e^{-i\phi}
\lab{thetadef}
\ee
with $a_i>0$, $i=1,2$, and
\be
z_1=\frac{v_1}{2}+i\,\rho_1=\sqrt{\frac{v_1^2}{4}+\rho_1^2}\,\,e^{i\zeta_1},
\qquad\qquad
w_1=\frac{v_2}{2}+i\,\rho_2=\sqrt{\frac{v_2^2}{4}+\rho_2^2}\,\,e^{i\zeta_2}.
\lab{zetaidef}
\ee

This gives us
\br
\Gamma\(z_1\)&=& z_1^2\,t-z_1\,x=
\(\frac{v_1^2}{4}-\rho_1^2\)\,t- \frac{v_1}{2}\, x -i\,\rho_1\,\(x-v_1\,t\),
\nonumber\\
\Gamma\(w_1\)&=& w_1^2\,t-w_1\,x=
\(\frac{v_2^2}{4}-\rho_2^2\)\,t- \frac{v_2}{2}\, x -i\,\rho_2\,\(x-v_2\,t\).
\er
Finally,  we replace  $a_i$ by $x_i^{(0)}$, $i=1,2$, defined as
\be
a_i \, \frac{\sqrt{\frac{v_i^2}{4}+\rho_i^2}}{2\mid\rho_i\mid}=e^{-\rho_i\, x_i^{(0)}}.
\ee
Putting all these expressions into \rf{rawtwosol} and into \rf{taudef} we obtain the final form of the two-bright-soliton solution:
\br
\psi_0=\frac{i\,2\,e^{-i\,\phi}}{\sqrt{\mid \eta\mid}}
\left[\frac{W_1\,e^{X_1} 
 + W_2\,e^{X_2}
 +\frac{\Lambda_{-}}{\Lambda_{+}}e^{\(X_1+X_2\)}\left[
 W_2 \, e^{X_1}\, e^{-i\,2\(\delta_{+}+\delta_{-}\)}
 +W_1  \,e^{X_2}\, e^{i\,2\(\delta_{+}-\delta_{-}\)}\,
\right]}{1 + e^{2\, X_1}  + e^{2\, X_2}   
+ \left[\frac{\Lambda_{-}}{\Lambda_{+}}\right]^2\,e^{2\, \(X_1+X_2\)}- 
   32\, \frac{\mid \rho_1\mid\,\mid\rho_2\mid}{\Lambda_{+}}\,\, 
   \cos\( \Omega_1-\Omega_2 -2\,\delta_{+}\)\,\,e^{\(X_1+X_2\)}}\right],
  \nonumber\\ 
   \lab{twobrightsoliton}
\er
where
\be
\Lambda_{\pm}=\(v_1-v_2\)^2+4\(\rho_1\pm\rho_2\)^2\; ; \qquad\qquad
\delta_{\pm}={\rm ArcTan}\left[\frac{2\,\(\rho_1\pm \rho_2\)}{\(v_1-v_2\)}\right]
\ee
and
\be
W_i=\mid\rho_i\mid \,e^{-i\,\Omega_i}
\ee
with
\be
\Omega_i =  \(\frac{v_i^2}{4}-\rho_i^2\)\,t- \frac{v_i}{2}\, x +\theta_i+\zeta_i,
\qquad\qquad 
X_i= \rho_i\,\(x-v_i\,t-x_i^{(0)}\)\qquad\qquad i=1,2.
\ee

\vspace{2cm}

{\bf Acknowledgements:} LAF and GL are grateful for the hospitality at the Department of Mathematical Sciences of Durham University where part of this work was carried out.
LAF and WJZ would like to thank the Royal Society for awarding them a grant which made their collaboration possible.  LAF is partially supported 
by CNPq (Brazil), and GL is supported by a scholarship from CNPq (Brazil).

\end{document}